\documentclass[11pt, a4paper]{article}
\pdfoutput=1
\usepackage{jheppub}
\usepackage{subcaption}
\usepackage[T1]{fontenc}
\usepackage{float}
\usepackage{bm}
\usepackage{latexsym, amsmath, amstext, amssymb, amsfonts, amscd, bm, array, multirow, amsbsy, mathrsfs}
\usepackage{amsthm}
\usepackage{t1enc}
\usepackage{indentfirst}
\usepackage{pb-diagram}
\usepackage{graphicx}
\usepackage{enumerate}
\usepackage{color}
\usepackage[all]{xy}
\usepackage{hyperref}
\hypersetup{colorlinks=false,pdfborderstyle={/S/U/W 0}}

\usepackage{url}

\DeclareMathOperator{\arccot}{arccot}

\theoremstyle{plain}
\newtheorem{thm}{Theorem}[section]

\theoremstyle{definition}
\newtheorem{definition}[thm]{Definition}

\theoremstyle{remark}

\newcommand{\A}{{\Bbb{A}}}

\newcommand{\bl}{\begin{Lemma}}
	\newcommand{\el}{\end{Lemma}}
\newcommand{\bt}{\begin{Theorem}}
	\newcommand{\et}{\end{Theorem}}
\newcommand{\bd}{\begin{definition}}
	\newcommand{\ed}{\end{definition}}

\newcommand{\Aut}{\mathrm{Aut}}

\newcommand{\eqdef}{\stackrel{{\rm def.}}{=}}

\DeclareFontFamily{U}{rsf}{}
\DeclareFontShape{U}{rsf}{m}{n}{<5> <6> rsfs5 <7> <8> <9> rsfs7 <10-> rsfs10}{}
\DeclareMathAlphabet\Scr{U}{rsf}{m}{n}

\def\N{\mathbb{N}}
\def\Z{\mathbb{Z}}
\def\C{\mathbb{C}}
\def\R{\mathbb{R}}
\def\Q{\mathbb{Q}}

\def\H{\mathbb{H}}

\def\SL{\mathrm{SL}}
\def\PSL{\mathrm{PSL}}

\def\dd{\mathrm{d}}

\def\vol{\mathrm{vol}}

\def\Int{\mathrm{Int}}
\def\Fr{\mathrm{Fr}}

\def\hJ{\hat{J}}

\def\bt{\mathbf{t}}

\def\U{\mathrm{U}}

\newcommand{\be}{\begin{equation*}}
\newcommand{\ee}{\end{equation*}}
\newcommand{\ben}{\begin{equation}}
\newcommand{\een}{\end{equation}}
\newcommand{\beqa}{\begin{eqnarray*}}
	\newcommand{\eeqa}{\end{eqnarray*}}
\newcommand{\beqan}{\begin{eqnarray}}
\newcommand{\eeqan}{\end{eqnarray}}

\newcommand{\nn}{\nonumber}

\newcommand{\Tr}{\mathrm{Tr}}
\newcommand{\tr}{\mathrm{tr}}

\def\Hess{\mathrm{Hess}}

\def\cC{{\mathcal C}}

\def\O{\mathrm{O}}

\def\hSigma{\hat{\Sigma}}

\def\cP{\mathcal{P}}
\def\cN{\mathcal{N}}
\def\cG{{\mathcal{G}}}

\def\cF{\mathcal{F}}
\def\cC{\mathcal{C}}
\def\cH{\mathcal{H}}

\def\SU{\mathrm{SU}}

\def\G_2{\mathrm{G_2}}

\def\cL{\mathcal{L}}

\def\fC{\mathfrak{C}}

\def\fF{\mathfrak{F}}
\def\fI{\mathfrak{I}}
\def\fD{\mathfrak{D}}

\def\P{\mathbb{P}}

\def\mD{\mathbb{D}}
\def\mA{\mathbb{A}}

\def\G{\mathrm{G}}

\def\Im{\mathrm{Im}}

\def\hPhi{\hat{\Phi}}
\def\bPhi{\bar{\Phi}}

\newcommand{\twopartdef}[4]
{
	\left\{
	\begin{array}{ll}
		#1 & \mbox{if } #2 \\
		#3 & \mbox{if } #4
	\end{array}
	\right.
}

\newcommand{\twopartdefmod}[4]
{
	\left\{
	\begin{array}{ll}
		#1 & \mbox{} #2 \\
		#3 & \mbox{} #4
	\end{array}
	\right.
}

\newcommand{\threepartdefmod}[6]
{
	\left\{
	\begin{array}{ll}
		#1 & \mbox{} #2 \\
		#3 & \mbox{} #4 \\
		#5 & \mbox{} #6
	\end{array}
	\right.
}

\def\mD{\mathbb{D}}
\def\fD{\mathfrak{D}}

\def\grad{\mathrm{grad}}

\def\bd{\boldsymbol{\dd}}

\def\rS{\mathrm{S}}

\def\bSigma{\boldsymbol{\Sigma}}

\def\area{\mathrm{area}}

\def\Crit{\mathrm{Crit}}
\def\rT{\mathrm{T}}

\def\Iso{\mathrm{Iso}}
\def\Re{\mathrm{Re}}
\def\Im{\mathrm{Im}}
\def\fJ{\mathfrak{J}}

\def\i{\mathbf{i}}
\def\sc{\mathrm{sc}}

\def\tPhi{{\tilde \Phi}}
\def\tvarphi{{\tilde \varphi}}

\def\cK{\mathcal{K}}

\def\PSU{\mathrm{PSU}}

\def\CP{\mathbb{C}\mathbb{P}}

\def\bSigma{\overline{\Sigma}}
\def\hSigma{\hat{\Sigma}}

\def\fc{\frak{c}}
\def\hG{{\hat G}}
\def\hDelta{{\hat \Delta}}
\def\rF{\mathrm{F}}
\def\rC{\mathrm{C}}

\def\ttvarphi{\tilde{\boldsymbol{\varphi}}}

\title{Generalized two-field $\alpha$-attractor models from geometrically finite hyperbolic surfaces}

\author{C. I. Lazaroiu$^1$ and C. S. Shahbazi$^2$}

\affiliation{$^1$ Center for Geometry and Physics, Institute for Basic
  Science, Pohang 37673, Republic of Korea \\ $^2$ Institut f\"ur Theoretische Physik, Leibniz
  Universit\"at Hannover, Germany.\\}

\emailAdd{calin@ibs.re.kr, carlos.shahbazi@itp.uni-hannover.de}

\abstract{We consider four-dimensional gravity coupled to a non-linear
sigma model whose scalar manifold is a non-compact geometrically
finite surface $\Sigma$ endowed with a Riemannian metric of constant
negative curvature.  When the space-time is an FLRW universe, such
theories produce a very wide generalization of two-field
$\alpha$-attractor models, being parameterized by a positive constant
$\alpha$, by the choice of a finitely-generated surface group
$\Gamma\subset \PSL(2,\R)$ (which is isomorphic with the fundamental
group of $\Sigma$) and by the choice of a scalar potential defined on
$\Sigma$. The traditional two-field $\alpha$-attractor models arise
when $\Gamma$ is the trivial group, in which case $\Sigma$ is the
Poincar\'{e} disk. We give a general prescription for the study of
such models through uniformization in the so-called ``non-elementary''
case and discuss some of their qualitative features in the gradient
flow approximation, which we relate to Morse theory.  We also discuss
some aspects of the SRST approximation in these models, showing that
it is generally not well-suited for studying dynamics near cusp ends.
When $\Sigma$ is non-compact and the scalar potential is
``well-behaved'' at the ends, we show that, in the {\em naive} local
one-field truncation, our generalized models have the same universal
behavior as ordinary one-field $\alpha$-attractors if inflation
happens near any of the ends of $\Sigma$ where the extended potential
has a local maximum, for trajectories which are well approximated by
non-canonically parameterized geodesics near the ends; we also discuss
spiral trajectories near the ends.  Generalized two field
$\alpha$-attractors illustrate interesting consequences of nonlinear
sigma models whose scalar manifold is not simply connected. They
provide a large class of tractable cosmological models with
non-trivial topology of the scalar field space.}
\begin{document}

\keywords{Non-linear sigma models, cosmology, hyperbolic geometry, uniformization, mathematical physics.}

\maketitle

\pagebreak


\section*{Introduction}
\label{Intro}

Inflation in the early universe can be described reasonably well by
so-called $\alpha$-attractor models
\cite{alpha1,alpha2,alpha3,alpha4,alpha5,Escher}. In their two-field
version (see, for example, \cite{alpha5,Escher}), such models arise
from cosmological solutions of four-dimensional gravity coupled to a
nonlinear sigma model whose scalar manifold $\Sigma$ (i.e. the target
manifold of the system of two real scalar fields) is the open unit
disk endowed with its unique complete metric $\cG$ (which determines
the kinetic energy term of the scalar fields) of constant Gaussian
curvature $K$ equal to $-\frac{1}{3\alpha}$, where $\alpha$ is a
positive constant. The ``universal'' behavior of such models in the
radial one-field truncation close to the conformal boundary of the
unit disk is a consequence of the hyperbolic character of $\cG$
\cite{alpha4,alpha5,
  Escher}.

While ordinary one-field\footnote{We count the number of {\em real}
  scalar fields present in the model.} models suffice to explain
current cosmological data \cite{Planck}, there are at least a few good
reasons to study the inflationary and post-inflationary dynamics of
two-field (and of more general multi-field) models, which form a
subject of active and continued interest
\cite{PT1,PT2,Welling,Sch1,Sch2,m1,m2,m3,m4,m5,m6,Gong,Dias1,Dias2,Mulryne,c1,BBJMV}.
First, it is possible that higher precision observations in the medium
future may detect deviations from one-field model predictions. Second,
it is considerably easier to produce multi-field models in fundamental
theories of gravity (such as string theory) than it is to produce
one-field models. Third, multi-field models are of theoretical
interest in themselves. In particular, it is well-known that such
models display behavior which is qualitatively new with respect to
that of one-field models; this happens due to the
higher-dimensionality of the target manifold of the system of real
scalar fields.

In this paper, we initiate a systematic study of two-field
cosmological models whose target manifold is an arbitrary borderless,
connected, oriented and non-compact two-dimensional smooth manifold $\Sigma$
endowed with a complete Riemannian metric $\cG$ of constant negative
curvature. Similar to the case of ordinary two-field
$\alpha$-attractor models, the Gaussian curvature $K$ of $\cG$ can be
parameterized by a positive constant $\alpha$ which is defined though
the relation $K=-\frac{1}{3\alpha}$. Since the open unit disk endowed
with its unique complete metric of constant and fixed negative
Gaussian curvature $K$ provides the simplest example of such a
Riemannian two-manifold, the cosmological models considered in this
paper can form an extremely wide generalization of ordinary two-field
$\alpha$-attractors, so we shall call them {\em two-field generalized
  $\alpha$-attractor models}.

Writing $\cG=3\alpha G$ produces a Riemannian metric $G$ on $\Sigma$
whose Gaussian curvature is constant and equal to $-1$. Thus
$(\Sigma,G)$ is an (oriented, connected, non-compact and borderless)
{\em hyperbolic surface} and the two-field cosmological model defined
by $(\Sigma,\cG)$ is equivalently parameterized by the real positive
number $\alpha$ and by $(\Sigma,G)$. The geometry and topology of
non-compact hyperbolic surfaces are extremely rich. For example, such
a surface can have infinite genus as well as a (countable) infinity of
Freudenthal ends \cite{Freudenthal,Peschke,Richards}; a simple example
of this phenomenon is provided by the surface $\Sigma=\C\setminus \Z$,
which has infinite genus and whose set of ends can be identified with
the set of integer numbers. A full topological classification of
oriented, borderless, connected and non-compact surfaces is provided
by the so-called Ker\'ekj\'art\'o-Stoilow model (see \cite{Stoilow}).
By the uniformization theorem of Poincar\'e and Koebe (see \cite{unif}
for a modern account), a hyperbolic surface $(\Sigma,G)$ is
parameterized up to isometry by the choice of the conjugacy class of a
finitely-generated surface group $\Gamma\subset \PSL(2,\R)$ (i.e. a
discrete subgroup of $\PSL(2,\R)$ without elliptic elements). Since
there exists a continuous infinity of such conjugacy classes, it
follows that the number of distinct choices of scalar manifold is
continuously infinite for any fixed constant $\alpha$. From this
perspective, traditional two-field $\alpha$-attractor models represent
just {\em one} type of a continuous infinity of geometrically distinct
models.

To simplify matters, in this paper we shall mostly focus on so-called
{\em geometrically finite}\footnote{Sometimes called {\em
topologically finite}.} hyperbolic surfaces, defined as those
hyperbolic surfaces $(\Sigma,G)$ whose fundamental group
$\pi_1(\Sigma)$ is finitely generated (but has an infinity of
elements); such surfaces necessarily have finite genus as well as a
finite number of ends. Even with this limitation, coupling
four-dimensional gravity to the scalar sigma model with target space
given by $(\Sigma,\cG)$ (while including an arbitrary smooth scalar
potential $\Phi:\Sigma\rightarrow \R$) produces a theory whose
cosmological solutions define a very wide generalization of ordinary
two-field $\alpha$-attractor models. 

When $(\Sigma,G)$ is geometrically finite, uniformization theory
implies that the local form of the hyperbolic metric $G$ on certain
canonically-defined neighborhoods of each end has an explicit form in
certain local coordinates defined on that neighborhood; the number of
such allowed local forms is small and determines the so-called
``hyperbolic type'' of the end (see, for example
\cite{Borthwick}). Using this fact, we show that two-field
cosmological models defined by geometrically finite hyperbolic
surfaces have the same universal behavior as ordinary
$\alpha$-attractors for certain special trajectories, in the leading
slow-roll approximation considered in the {\em naive} one-field
truncation near each end, provided that the scalar potential is
``well-behaved ``and ``locally maximal'' at that end in the sense that it extends
smoothly to a vicinity of the end considered in the
Ker\'ekj\'art\'o-Stoilow \cite{Richards,Stoilow} compactification of
$\Sigma$ and that its extension has a local maximum at the
corresponding ideal point. This further supports our choice of name
for such models. Given this similarity to ordinary two-field
$\alpha$-attractors, generalized models of this type may provide
interesting candidates for a description of inflation in the early
universe. More precisely, one finds (see \cite{elem,modular} for
explicit numerical analysis in certain examples) that inflationary
trajectories producing 50-60 e-folds can usually be found provided that
inflation takes place within the canonical neighborhood of an
end. This is of course not surprising given the naive universality
property mentioned above. However, unlike the ordinary models,
generalized two-field $\alpha$-attractors allow for much more
flexibility since $\Sigma$ can have more than one end (as well as
different types of ends).  Of course, potential interest in such
models is no way limited to their inflationary behavior. As for
ordinary one-field models, one must also consider post-inflationary
dynamics. It turns out that the cosmological trajectories of the
models considered in this paper display quite intricate behavior away
from the ends of $\Sigma$, which is partly due to the non-trivial
topology of the surface; this is illustrated qualitatively in Section
\ref{sec:Morse} and quantitatively in much more detail in the examples
analyzed in references \cite{elem,modular}.

The paper is organized as follows. In Section \ref{sec:ES}, we give
the global formulation of two-field cosmological models with arbitrary
Riemann surface targets and discuss a series of increasingly
restrictive approximations which are useful for analyzing them. The
so-called ``gradient flow approximation'' described in that section
will be used in Section \ref{sec:Morse} to give a qualitative picture of cosmological
trajectories and a relation to Morse theory. We stress that the
well-known SRST (slow-roll -- slow-turn) approximation \cite{PT1,PT2}
turns out to be too restrictive for a proper study of this class of
models (for example, it can fail near cusp ends, as we show in
Subsection \ref{subsec:SRSTEnds}).  Due to this fact, we discuss a
series of weaker approximations (which include, but are not limited
to, the gradient flow approximation).  Some of the approximations in
this series do not seem to have been considered in detail in the
literature, but they may turn out to be useful for deeper studies of
the models considered in the present paper. Section \ref{sec:genalpha}
discusses classical cosmological trajectories in generalized two-field
$\alpha$-attractor models with geometrically finite hyperbolic surface
targets. We show that the naive one-field truncation for certain
special trajectories has universal behavior in the slow-roll
approximation near all ends where the scalar potential is
``well-behaved'' and ``locally maximal''. We also discuss some aspects
of the SRST approximation near the ends (showing that it can fail near
cusp ends) and illustrate a form of spiral inflation which can occur
within canonical neighborhoods of the ends. In Section \ref{sec:lift},
we propose a general method for studying such models which relies on
lifting the cosmological equations to the Poincar\'e disk or to the
upper half plane\footnote{This relates a subclass of our models to the
  ``modular inflation models'' considered in \cite{Sch1, Sch2}. See
  \cite{modular} for a detailed discussion of the precise connection
  in the case of the hyperbolic triply-punctured sphere.}. This
approach assumes knowledge of a fundamental polygon for the
uniformizing surface group $\Gamma$, the general computation of which
is an open problem when $(\Sigma,G)$ has infinite hyperbolic area (in
the sense that no general stopping algorithm for computing such a
polygon is known for a general infinite area geometrically finite
hyperbolic surface). Section \ref{sec:Morse} gives a qualitative
description of inflationary trajectories in the gradient flow
approximation using the topological pants decomposition induced by
Morse theory. In Section \ref{sec:exphen}, we discuss some
phenomenological aspects of our models and certain examples which are
studied in detail in references \cite{elem,modular}. Finally, Section
\ref{sec:conclusions} concludes. Appendix \ref{app:SpecCoord}
summarizes some useful formulas in isothermal and semi-geodesic
coordinates, which are often used in various computations within this
paper. The remaining appendices summarize certain classical results of
uniformization theory (many of which go back to Poincar\'e). These are
used throughout the paper but some of them may be unfamiliar to the
cosmology community. Appendix \ref{sec:unif} recalls some fundamental
results on the uniformization of smooth (but not necessarily compact)
Riemann surfaces. Appendix \ref{sec:topfinite} summarizes relevant
classical results regarding topologically finite surfaces and their
Ker\'ekj\'art\'o-Stoilow compactification (which is a closed
surface). It also recalls some fundamental results regarding
topologically finite surfaces endowed with a conformal structure and
their conformal compactification (which generally is a surface with
boundary). The Ker\'ekj\'art\'o-Stoilow and conformal
compactifications are conceptually important for understanding the
behavior of our models near the ends of $\Sigma$. In Appendix
\ref{sec:geomfinite}, we recall some classical results on
geometrically finite hyperbolic surfaces, paying special attention to
certain canonical neighborhoods of their ends, on which the hyperbolic
metric can be brought to one of a few explicit forms. This paper
assumes some basic familiarity with the notion of Freudenthal end of a
manifold, for which we refer the reader to \cite{Freudenthal,Peschke}.
For the case of surfaces, the Freudenthal theory of ends reduces to
the classical theory of ``ideal boundary points'' developed by
Ker\'ekj\'art\'o and Stoilow (see \cite{Richards,Stoilow}).

\vspace{2mm}

\paragraph{Notations and conventions.}
All manifolds considered are smooth, connected, oriented and
paracompact (hence also second-countable). All homeomorphisms and
diffeomorphisms considered are orientation-preserving. By definition,
a Lorentzian four-manifold has ``mostly plus'' signature.

\section{Cosmological models with two real scalar fields minimally coupled to gravity}
\label{sec:ES}

In this section, we give the general description of cosmological
models with two real scalar fields minimally coupled to gravity,
allowing for scalar manifolds of non-trivial topology. Our formulation
is globally valid (in particular, it is coordinate free), since we
allow for arbitrary topology of the target manifold $\Sigma$. Some of
the local formulas are standard, but the reader should pay attention
to the mathematical aspects involved in the global approach. We also
discuss certain increasingly restrictive approximations which are
useful when studying inflation in such models, the strictest of which
is the well-known SRST approximation (which, as shown in Subsection
\ref{subsec:SRSTEnds}, turns out to be of limited usefulness for such
models).  The so-called ``gradient flow approximation'' will be used
in Section \ref{sec:Morse}. The reader may notice that some of the approximations 
discussed in this section are not usually considered in the cosmology literature; 
they may be useful for our class of models due to the fact that the SRST 
approximation fails near cusp ends.

\subsection{Two-dimensional scalar manifolds and scalar potentials}

\noindent Let $(\Sigma,\cG)$ be any oriented, connected and complete
two-dimensional Riemannian manifold without boundary (called the {\em
  scalar manifold}) and $\Phi:\Sigma\rightarrow \R$ be a smooth
function (called the {\em scalar potential}).

\vspace{2mm}

\paragraph{Remark.}
The condition that $\Sigma$ be orientable is physically unimportant
and can be relaxed; we make this assumption merely for simplicity.  We
do {\em not} assume that $\Sigma$ is compact; as we shall see below,
allowing $\Sigma$ to be non-compact is of direct relevance to
cosmological applications. We require that the scalar manifold metric
$\cG$ be complete in order to avoid problems with conservation of
energy.

\vspace{2mm}

With such weak assumptions, the topology of $\Sigma$ can be quite
involved. When $\Sigma$ is compact, its (oriented) diffeomorphism
class\footnote{The homeomorphism and diffeomorphism classifications of
  two-manifolds coincide as implied by Kirby's torus trick.} is
completely determined by its genus $g$. In that case, the fundamental
group $\pi_1(\Sigma)$ is finitely generated on $2g$ generators and its
Abelianization $H_1(\Sigma,\Z)$ is isomorphic with $\Z^{2g}$. The
topological classification is much more subtle when $\Sigma$ is
non-compact.  In that case, the fundamental group can be infinitely
generated\footnote{A simple example is provided by the planar
  non-compact surface $\Sigma=\C\setminus \Z$, which has infinitely
  generated fundamental group.} and $\Sigma$ can have an infinite
number of ends in the sense of Freudenthal \cite{Freudenthal,
  Peschke}. The ends correspond to the ``ideal points'' of the ``ideal
boundary'' of the so-called {\em end compactification} (a.k.a. {\em
  Ker\'ekj\'art\'o-Stoilow compactification}) $\hSigma$ of $\Sigma$
(see \cite{Richards}). The ideal boundary is a totally-disconnected,
compact and separable topological space, containing a closed subset
corresponding to ``non-planar'' ends.  Together with this subspace,
the ideal boundary determines the topology of $\Sigma$. In fact, every
pair of nested totally disconnected, compact and separable spaces
occurs as the ideal boundary of some orientable non-compact surface
\cite{Richards}. Moreover, $\Sigma$ admits a canonical {\em Stoilow
  presentation} (or {\em Stoilow model}) \cite{Stoilow} as the surface
obtained by removing from the Riemann sphere $\C\sqcup\{\infty\}\simeq
\CP^1$ a compact totally disconnected set $B$ of points lying on the
real axis (where $B$ corresponds to the ideal boundary) and a finite
or countable collection of pairs of mutually-disjoint disks which are
symmetric with respect to the real axis and whose bounding circles can
accumulate only on the set $B$, after which the bounding circles are
pairwise identified. In latter sections of the paper, we will focus
for simplicity on the case when $\Sigma$ is {\em topologically
  finite}, which means that its fundamental group $\pi_1(\Sigma)$ is
finitely-generated. In that case, the genus of $\Sigma$ is finite and
the ideal boundary consists of a finite number of points $p_1,\ldots,
p_n$, all of which correspond to planar ends; in this situation, the
end compactification $\hSigma$ is a compact oriented surface from
which $\Sigma$ is obtained by removing the points $p_j$ (see Appendix
\ref{sec:topfinite} for details). The case of compact $\Sigma$ arises when the
ideal boundary is empty, i.e. when $\Sigma$ has no ends; in that case,
one has $\hSigma=\Sigma$.

\subsection{The Einstein-Scalar theory defined by $(\Sigma,\cG,\Phi)$}

\noindent Any triplet $(\Sigma,\cG,\Phi)$ as above allows one to define an
Einstein-Scalar theory on any four-dimensional oriented manifold $X$
which admits Lorentzian metrics. This theory includes four-dimensional
gravity (described by a Lorentzian metric $g$ defined on $X$) and a
smooth map $\varphi:X\rightarrow \Sigma$ (which locally describes two
real scalar fields), with action:
\ben
\label{S}
S[g,\varphi]=\int_X \cL(g,\varphi) \vol_g~~,
\een
where $\vol_g$ is the volume form of $(X,g)$ and $\cL(g,\varphi)$ is
the Lagrange density:
\ben
\label{cL}
\cL(g,\varphi)\eqdef \frac{M^2}{2} \mathrm{R}(g)-\frac{1}{2}\Tr_g \varphi^\ast(\cG)-\Phi\circ \varphi~~.
\een
Here $\mathrm{R}(g)$ is the scalar curvature of $g$ and $M$ is the
reduced Planck mass. The notation $\varphi^\ast(\cG)$ denotes the
pull-back through $\varphi$ of the metric $\cG$ (this pull-back is a symmetric
covariant 2-tensor field defined on $X$), while
$\Tr_g\varphi^\ast(\cG)$ denotes the trace of the tensor field of type
$(1,1)$ obtained by raising one of the indices of $\varphi^\ast(\cG)$
using the metric $g$. The notation $\Phi\circ \varphi$ denotes the
smooth map from $X$ to $\R$ obtained by composing $\varphi$ with
$\Phi$. The formulation \eqref{cL} allows one to
define such a theory globally for any topology of the oriented scalar
manifold $\Sigma$ and any topology of the oriented space-time $X$. For
any fixed Lorentzian metric $g$, the Lagrange density:
\be
\cL_g(\varphi)\eqdef -\frac{1}{2}\Tr_g
\varphi^\ast(\cG)-\Phi\circ \varphi
\ee
defines the non-linear sigma model with source $(X,g)$,
target space $(\Sigma,\cG)$ and scalar potential $\Phi$.

\vspace{2mm}

\paragraph{Local expressions in real coordinates on $\Sigma$.} 
If $(U,(x^\mu)_{\mu=0\ldots 3})$ and $(V,
(y^\alpha)_{\alpha=1,2})$ are local coordinate systems on $X$ and
$\Sigma$ such that $\varphi(U)\subset V$, then the map $\varphi$ has
locally-defined components:
\be
\varphi^\alpha(x)=(y^\alpha\circ \varphi)(x)~~\mathrm{for}~~ x\in U
\ee
and the metrics $g$ and $\cG$ have squared line elements:
\be
\dd s_g^2=g_{\mu\nu}\dd x^\mu\dd x^\nu~~,~~\dd s_\cG^2=\cG_{\alpha\beta}\dd y^\alpha\dd y^\beta~~.
\ee
Moreover, we have:
\be
\Tr_g \varphi^\ast(\cG)= g^{\mu \nu} \cG_{\alpha\beta} \partial_\mu \varphi^\alpha\partial_\nu \varphi^\beta~~,~~(\Phi\circ\varphi)(x)=\Phi(\varphi^1(x),\varphi^2(x))~~.
\ee
Hence $\cL(g,\varphi)$ coincides locally with the usual expression for the Lagrange
density of two real scalar fields minimally coupled to gravity.

\vspace{2mm}

\paragraph{Local expressions in a $\cG$-compatible complex coordinate on $\Sigma$.}

By definition, a {\em conformal structure} on $\Sigma$ is a conformal
equivalence class of metrics on $\Sigma$. Since $\Sigma$ is a surface,
any almost complex structure $J$ on $\Sigma$ has vanishing Nijenhuis
tensor and hence is integrable (i.e. it is a complex
structure). Moreover, any two-form on $\Sigma$ is closed and hence any
$J$-Hermitian metric on $\Sigma$ is K\"ahler. Recall that the set of
conformal equivalence classes of Riemannian metrics on $\Sigma$ is in
bijection with the set of orientation-compatible complex structures on
$\Sigma$. This bijection takes the conformal equivalence class of a
Riemannian metric $\cG$ into the unique orientation-compatible complex
structure $J$ which has the property that $\cG$ is Hermitian (and
hence K\"ahler) with respect to $J$.  Endowing $\Sigma$ with the complex
structure $J$ determined by $\cG$, let $z$ be a local $J$-holomorphic
coordinate on $\Sigma$, defined on an open subset $V\subset
\Sigma$. Since $\cG$ is Hermitian with respect to $J$, we have $\dd
s_\cG^2|_V=\lambda(z,\bar{z})^2|\dd z|^2$ for some positive function
$\lambda(z,\bar{z})>0$ (See Appendix
\ref{app:SpecCoord}). Choosing a local chart $(U,(x^\mu))$ of $X$ such
that $\varphi(U)\subset V$ and setting $z(x)\eqdef z(\varphi(x))$, the
map $\varphi$ is described locally by the complex-valued scalar field
$z(x)$ and the Lagrange density takes the local form:

\ben
\label{cLcomplex}
\cL(g,z)=\frac{M^2}{2} \mathrm{R}(g)-\frac{1}{2}\lambda^2(z,\bar{z})
g^{\mu\nu}\partial_\mu z\partial_\nu \bar{z}- \Phi(z,\bar{z})~~.
\een

\subsection{Cosmological models defined by $(\Sigma,\cG,\Phi)$}
\label{subsec:FLRW}

\noindent By definition, a {\em cosmological model} defined by
$(\Sigma,\cG,\Phi)$ is a solution of the equations of motion of the
theory \eqref{S}-\eqref{cL} when $(X,g)$ is an FLRW universe and
$\varphi$ depends only on the cosmological time. We shall assume for
simplicity that the spatial section is flat and simply connected. With
these assumptions, the cosmological models of interest are defined by
the following conditions:
\begin{enumerate}
\itemsep 0.0em
\item $X$ is diffeomorphic with $\R^4$, with global coordinates
  $(t,x^1,x^2,x^3)$
\item The squared line element of $g$ has the form:
\ben
\label{FLRW}
\dd s^2_g=-\dd t^2+a(t)^2\sum_{i=1}^3{(\dd x^i)^2}~~,
\een
where $a(t)> 0$. In particular, the space-time metric $g$ is
determined by the single function $a(t)$.
\item $\varphi$ depends only on $t$.
\item $(a(t),\varphi(t))$ are such that $(g,\varphi)$ is a solution
of the equations of motion derived from the action functional
\eqref{S}.
\end{enumerate}
Let $||\cdot ||_\cG$ denote the norm induced by $\cG$ on the fibers of
the tensor and exterior powers of the tangent bundle $T\Sigma$ and of
the cotangent bundle $T^\ast \Sigma$. Since $\varphi$ depends only on
$t$, it describes a smooth curve $\varphi:\fI\rightarrow \Sigma$ in
$\Sigma$, where $\fI$ is a maximal interval of definition of the
solution.  Setting $\dot{~}\eqdef \frac{\dd}{\dd t}$, let $H\eqdef
\frac{\dot{a}}{a}$ denote the Hubble parameter and let
$\dot{\varphi}(t)\eqdef \frac{\dd \varphi(t)}{\dd t}\in
T_{\varphi(t)}\Sigma$ for $t\in \fI$. Let $\sigma$ be the proper
length along this curve measured starting from $t=t_0\in \fI$:
\be
\sigma(t) \eqdef \int_{t_0}^t {\dd t'}||\dot{\varphi}(t')||_\cG~\Longrightarrow~ \dot{\sigma}(t)=||\dot{\varphi}(t)||_\cG~~ (t\in \fI)~~.
\ee
We assume that $||\dot{\varphi}(t)||$ does not vanish for $t\in \fJ$.
The equations of motion derived from the action \eqref{S} when $g$ is
given by \eqref{FLRW} and $\varphi$ depends only on $t$ reduce to:
\beqan
\nabla_t \dot{\varphi}+3H \dot{\varphi}+(\grad_{\cG} \Phi)\circ \varphi &=&0~~,   \label{eom}\\
\frac{1}{3}\dot{H}+H^2 - \frac{\Phi\circ \varphi}{3M^2} &=& 0~~,\label{F1}\\
\dot{H} +\frac{\dot{\sigma}^2}{2M^2} &=& 0~~,\label{F2}
\eeqan
where:
\be
\nabla_t\eqdef \nabla_{\dot{\varphi}(t)}
\ee
is the covariant derivative with respect to $\dot{\varphi}(t)$.
Assuming $H(t)>0$, equations \eqref{F1} and \eqref{F2} give:
\ben
\label{Hvarphi}
H(t)=\frac{1}{\sqrt{6} M}\left[||\dot{\varphi}(t)||_\cG^2+2\Phi(\varphi(t))\right]^{1/2}~~.
\een
This allows us to eliminate $H(t)$, thus reducing the system
\eqref{eom}-\eqref{F2} to the single non-linear equation:
\ben
\label{eomsingle}
\nabla_t \dot{\varphi}(t)+\frac{1}{M} \sqrt{\frac{3}{2}} \left[||\dot{\varphi}(t)||_\cG^2+2\Phi(\varphi(t))\right]^{1/2}\dot{\varphi}(t)+ (\grad_{\cG} \Phi)(\varphi(t))=0~~.
\een
The initial conditions at cosmological time $t_0\in \fI$ take
the form:
\ben
\label{ini}
\varphi(t_0)=\varphi_0~~,~~\dot{\varphi}(t_0)=v_0\in T_{\varphi_0} \Sigma~~,
\een
where $\varphi_0$ is a point of $\Sigma$ and $v_0$ is a vector tangent
to $\Sigma$ at that point. Together with the initial conditions
\eqref{ini}, equation \eqref{eomsingle} determines the function
$\varphi(t)$, which also determines $H(t)$ through relation
\eqref{Hvarphi}.

\subsection{Conditions for inflation}

\noindent Equation \eqref{F2} shows that $\dot{H}$ is negative. As usual, define the
{\em first slow-roll parameter} through:
\ben
\label{epsilon_def}
\epsilon(t)\eqdef -\frac{\dot{H}(t)}{H(t)^2}>0~~.
\een
Then:
\be
\frac{\ddot{a}}{a}=\dot{H}+H^2=H^2(1-\epsilon)
\ee
and the condition for inflation $\ddot{a}>0$ amounts to
$\epsilon(t)<1$. By definition, inflation occurs for time intervals
during which $H>0$ and $\epsilon<1$, i.e. for intervals where $a$
is a (strictly) convex and increasing function of the cosmological
time $t$. Using \eqref{epsilon_def}, equation \eqref{F1} becomes:
\be
H(t)^2(1-\frac{\epsilon(t)}{3})=\frac{1}{3M^2}\Phi(\varphi(t))
\ee
and the conditions for inflation read:
\ben
\label{infcond}
H>0~~\mathrm{and}~~\Phi > 2 M^2 H^2~\Longleftrightarrow~\Phi(\varphi(t))>0~~\mathrm{and}~~0 < H(t) < \frac{1}{M}\sqrt{\frac{\Phi(\varphi(t))}{2}}~~.
\een
Using \eqref{Hvarphi}, these amount to $H(t)>0$ together with the condition:
\be
||\dot{\varphi}(t)||_\cG^2 < \Phi(\varphi(t))~~.
\ee
When $H(t)>0$, we have:
\be
\epsilon(t)\ll 1 ~~\mathrm{iff}~~||\dot{\varphi}(t)||_\cG^2 \ll \Phi(\varphi(t))~~.
\ee

\subsection{The gradient flow approximation}

\noindent In this subsection, we discuss an approximation in which
cosmological trajectories of the model are replaced by
reparameterized gradient flow lines of the scalar potential
$\Phi$. This allows one to derive qualitative features of the model
using the well-known properties of gradient flows on Riemann surfaces,
which is especially useful when $\Phi$ is a Morse function (see
Section \ref{sec:Morse} for an application of this approximation). We
stress that the approximation discussed in this subsection is much
less restrictive than the well-known SRST approximation \cite{PT1,
  PT2}. The latter is used traditionally when studying cosmological
perturbations in two-field models but turns out to be ill-suited for
understanding deeper aspects of generalized two-field $\alpha$-attractors (see
Subsection \ref{subsec:SRSTEnds} for how the SRST approximation can
fail near cusp ends).

Assuming $H(t)>0$, define the {\em gradient flow vector
  parameter} through:
\ben
\label{etadef}
 \eta(t)\eqdef -\frac{1}{H\dot{\sigma}}\nabla_t \dot{\varphi}=-\frac{1}{H} \frac{\nabla_t \dot{\varphi}}{||\dot{\varphi}||_\cG}=-M \sqrt{6} \left[||\dot{\varphi}(t)||_\cG^2+2\Phi(\varphi(t))\right]^{-1/2}\frac{\nabla_t \dot{\varphi}}{||\dot{\varphi}||_\cG}~~,
\een
i.e.:
\ben
\label{eta1}
\eta(t)=3\frac{\left[||\dot{\varphi}(t)||_\cG^2+
2\Phi(\varphi(t))\right]^{1/2}\dot{\varphi}(t)+
M\sqrt{\frac{2}{3}} (\grad_{\cG} \Phi)(\varphi(t))}{||\dot{\varphi(t)}||_\cG \left[||\dot{\varphi}(t)||_\cG^2+2\Phi(\varphi(t))\right]^{1/2}}~~.
\een
Using \eqref{etadef}, equation \eqref{eom} becomes:
\ben
\label{eom_eta}
3H||\dot{\varphi}||_\cG (\vartheta-\frac{\eta}{3})+\grad_{\cG}\Phi=0~~,
\een
where:
\ben
\label{taudef}
\vartheta(t)\eqdef \frac{\dot{\varphi}(t)}{||\dot{\varphi}(t)||_\cG}
\een
is the unit tangent vector to the curve $\varphi$ at time $t$. Since
$||\vartheta||_\cG=1$, the term proportional to $\eta$ can be neglected in
this equation when the {\em kinematic gradient flow condition}:
\ben
\label{gradflowcond}
||\eta||_\cG\ll 1
\een
is satisfied. When \eqref{gradflowcond} holds, equation
\eqref{eom_eta} reduces to:
\ben
\label{appgradflow}
\dot{\varphi}(t)\simeq -\frac{1}{3H(t)}(\grad_{\cG}\Phi)(\varphi(t))~\Longleftrightarrow~
 \frac{\dd \varphi(q)}{\dd q}\simeq -(\grad_{\cG}\Phi)(\varphi(q))~~,
\een
which shows that $\varphi$ can be approximated by a gradient flow
trajectory $\varphi_\bullet(q)$ of the potential $\Phi$ with respect
to the metric $\cG$ and the parameter:
\ben
\label{qdef}
q(t)\eqdef q_0+\int_{t_0}^{t}\dd t'\frac{1}{3H(t')}~~.
\een
In this expression, we assumed that $t_0$ is chosen within a time
interval on which the gradient flow approximation holds and we set
$q(t_0)=q_0$. Using \eqref{appgradflow}, relation \eqref{Hvarphi}
reduces to the algebraic equation:
\be
H^4 -\frac{\Phi}{3M^2} H^2-\frac{||\dd \Phi||_\cG^2}{54 M^2}=0~~,
\ee
which gives:
\ben
\label{Hgradflow}
H(t(q))=\frac{1}{M\sqrt{6}}\Bigg(\Phi(\varphi_\bullet(q)) 
+\sqrt{\Phi(\varphi_\bullet(q))^2+
\frac{2}{3} M^2 ||(\dd \Phi)(\varphi_\bullet(q))||_\cG^2}~~\Bigg)^{1/2}~~,
\een
where we assumed $H(t)>0$. Thus \eqref{qdef} reduces to:
\ben
\label{tq}
t-t_0=\frac{1}{M}\sqrt{\frac{3}{2}}\int_{q_0}^q \dd q  \Bigg(\Phi(\varphi_\bullet(q))+ \sqrt{\Phi(\varphi_\bullet(q))^2+\frac{2}{3} M^2 ||(\dd \Phi)(\varphi_\bullet(q))||_\cG^2} ~~\Bigg)^{1/2}~~.
\een
This relation can be used to determine $t$ as a function of $q$ given
the approximating gradient flow trajectory $\varphi_\bullet(q)$.

Let us assume that $\Phi(\varphi(t))>0$, as required by the first inflation
condition in \eqref{infcond}. Then the second inflation condition
amounts to the following inequality in the gradient flow approximation:
\be
M\frac{||(\dd \Phi)(\varphi_\bullet(q))||_\cG}{\Phi(\varphi_\bullet(q))}<\frac{3}{\sqrt{2}}~~.
\ee
Together with \eqref{tq}, the last inequality can be used to determine
the inflationary time intervals within the gradient flow
approximation. Notice from \eqref{Hgradflow} that $H$ is determined by
the point $\varphi_\bullet(q)\in \Sigma$ in this
approximation.

A gradient flow trajectory is determined uniquely by the gradient flow equation:
\ben
\label{gradflow}
\frac{\dd \varphi_\bullet(q)}{\dd q}= -(\grad_{\cG}\Phi)(\varphi_\bullet(q))
\een
(which is a {\em first} order differential equation), together with the
initial condition:
\be
\varphi_\bullet(q_0)=\varphi_0~~.
\ee
Validity of the gradient flow approximation at $t=t_0$ requires
$\frac{\dd\varphi}{\dd q}(q_0)=-(\grad_{\cG}\Phi)(\varphi_0)$,
i.e.:
\ben
\label{v0gradflow}
v_0=-\frac{(\grad_\cG\Phi)(\varphi_0)}{3H_0}~~,
\een
where $H_0\eqdef H(t_0)$ is determined from \eqref{Hgradflow} as:
\ben
\label{Hgradflow0}
H_0=\frac{1}{M\sqrt{6}}\left(\Phi(\varphi_0)+\sqrt{\Phi(\varphi_0)^2+\frac{2}{3}M^2||(\dd \Phi)(\varphi_0)||_\cG^2}\right)^{1/2}~~.
\een
In particular, the gradient flow approximation constrains the initial
velocity $v_0$ in terms of the initial value $\varphi_0$, as expected
from the fact that, in this approximation, the second order equation
\eqref{eomsingle} is replaced by the first order equation \eqref{gradflow}.

\subsection{The potential gradient flow condition}

\noindent In the gradient flow approximation, we have
$||\dot{\varphi}||_\cG \!=\frac{1}{3H} ||\dd \Phi||_\cG$ and equation \eqref{F2}
gives:
\be
\dot{H}\simeq -\frac{1}{18 M^2 H^2}||\dd \Phi||_\cG^2~~.
\ee
Relation \eqref{qdef} implies:
\be
\nabla_t =\nabla_{\dot{\varphi}(t)}\simeq \frac{1}{3H}\nabla_{\frac{\dd\varphi_\bullet(q)}{\dd q}}=-\frac{1}{3H}\nabla_{\grad_\cG \Phi}~~.
\ee
Thus:
\beqa
 \nabla_t \dot{\varphi}&=&\!\nabla_t \!\left[\frac{1}{3H}\grad_{\cG}\Phi\right]=-\frac{\dot{H}}{3H^2}\grad_\cG \Phi+\!\frac{1}{3H} (\nabla_t\, \grad_\cG \Phi) \simeq \nn\\
&\simeq& \frac{1}{9H^2}\left[\frac{1}{6 M^2 H^2}||\dd \Phi||_\cG^2\, \grad_\cG \Phi -\nabla_{\grad_\cG\Phi}\, \grad_\cG \Phi \right]~~
\eeqa
and:
\be
\eta=-\frac{1}{H} \frac{\nabla_t \dot{\varphi}}{||\dot{\varphi}||_\cG}\simeq -3\frac{\nabla_t \dot{\varphi}}{||\dd \Phi||_\cG}\simeq
\frac{1}{3H^2} \frac{\nabla_{\grad_\cG \Phi}\,\grad_\cG \Phi}{||\dd\Phi||_\cG}-\frac{1}{18 M^2 H^4}||\dd\Phi||_\cG \,\grad_\cG \Phi~~.
\ee
This gives:
\ben
\label{etasq}
||\eta||_\cG^2 \simeq \frac{1}{9 H^4} \frac{||\nabla_{\grad_\cG \Phi}\,\grad_\cG \Phi||_\cG^2}{||\dd \Phi||_\cG^2}- \frac{1}{27 M^2 H^6} 
\cG({\grad_\cG\Phi}, \nabla_{\grad_\cG\Phi}\,\grad_\cG\Phi)+\frac{||\dd\Phi||_\cG^4}{324 M^4 H^8}~~.
\een
Since $\cG$ is covariantly constant with respect to its Levi-Civita connection $\nabla$,
the following relation holds for any vector field $X$ defined on $\Sigma$\,:
\be
\partial_X ||X||_\cG^2=2\cG(X, \nabla_X X)~~.
\ee
Using this relation for $X=\grad_\cG \Phi$, equation \eqref{etasq} becomes:
\be
 ||\eta||_\cG^2 \simeq \!\frac{1}{9 H^4} \frac{||\nabla_{\grad_\cG \Phi}\,\grad_\cG \Phi||_\cG^2}{||\dd \Phi||_\cG^2}-\frac{1}{54 M^2 H^6} \partial_{\grad_\cG\Phi}||\dd\Phi||_\cG^2+\frac{||\dd\Phi||_\cG^4}{324 M^4 H^8}~~.
\ee
Hence consistency of the gradient flow approximation requires
that the {\em potential gradient flow condition}:
\ben
\label{potgradflow}
\frac{1}{9 H^4} \frac{||\nabla_{\grad_\cG \Phi}\,\grad_\cG \Phi||_\cG^2}{||\dd \Phi||_\cG^2}-\frac{1}{54 M^2 H^6} \partial_{\grad_\cG\Phi}||\dd\Phi||_\cG^2 +\frac{||\dd\Phi||_\cG^4}{324 M^4 H^8}\ll 1
\een
is satisfied. In this relation, $H^2$ can be expressed in terms of
$\Phi$ and $||\dd \Phi||_\cG$ using relation \eqref{Hgradflow}, thus
obtaining a condition involving only the metric $\cG$ as well as
$\Phi$ and its first and second order derivatives:
\beqa
\!\!\!\!\!\!\frac{\left[\!\!
	\begin{array}{l}
	12\left\Vert \dd\Phi \right\Vert_\cG ^{6}M^{4}-12\left\Vert \dd\Phi \right\Vert_\cG
	^{2}M^{4}\left( \Phi +\sqrt{\Phi ^{2}+\frac{2}{3}M^{2}\left\Vert \dd\Phi
		\right\Vert_\cG ^{2}}\right) \partial _{\grad_{\mathcal{G}}\Phi
	}\left\Vert \dd\Phi \right\Vert_\cG ^{2} +\\
	\\
	+8M^{4}\left[ M^{2}\left\Vert \dd\Phi \right\Vert_\cG ^{2}+3\Phi \left( \Phi +
	\sqrt{\Phi ^{2}+\frac{2}{3}M^{2}\left\Vert \dd\Phi \right\Vert_\cG ^{2}}\right)
	\right] \left\Vert \nabla _{\grad_{\mathcal{G}}\,\Phi }\grad_{
		\mathcal{G}}\Phi \right\Vert_\cG ^{2}
	\end{array}
	\!\! \right] }{3\left\Vert \dd\Phi \right\Vert_\cG ^{2}\left( \Phi +\sqrt{\Phi ^{2}+
		\frac{2}{3}M^{2}\left\Vert \dd\Phi \right\Vert_\cG ^{2}}\right) ^{4}} \!&\ll &\!1~.
\nonumber \\
&& 
\eeqa
The potential gradient flow condition is necessary for validity of
the gradient flow approximation.

\subsection{Decomposition in a Frenet frame}

\noindent The equations of Subsection \ref{subsec:FLRW} define globally a two-field
cosmological model with non-canonical kinetic term. 
As usual in the theory of curves, it is convenient to choose a {\em framing} of
$\varphi$, which in this case is a unit norm smooth vector field $n\in
\Gamma(\fI, \varphi^\ast(T\Sigma))$ which is everywhere orthogonal to
the unit tangent vector \eqref{taudef}:
\be
n(t)\perp \vartheta(t)=0~~,~\mathrm{i.e.}~~\cG(n(t),\vartheta(t))=0~~,~~\forall t\in \fI~~.
\ee
We have $\vartheta(t),n(t)\in T_{\varphi(t)} \Sigma$ for all $t\in \fI$. Since
$\Sigma$ is oriented, we can choose $n$ so that $(\vartheta(t),n(t))$ is a
positively-oriented orthonormal basis of $T_{\varphi(t)}\Sigma$ for
all $t$. The Frenet-Serret equations of $\varphi$
read:
\ben
\label{FS}
\nabla_t\vartheta (t) = \dot{\sigma}\, \kappa\, n~~,~~\nabla_t n(t) =-\dot{\sigma}\,\kappa\, \vartheta~~,
\een
where $\kappa(t)$ is the extrinsic curvature of $\varphi$.
Using the first of equations \eqref{FS} and the relation
$\dot{\varphi}=\dot{\sigma} \vartheta$ gives:
\ben
\label{nabla_phi}
\nabla_t \dot{\varphi}=\ddot{\sigma}\vartheta+\dot{\sigma}^2 \kappa n~~.
\een
Thus:
\ben
\label{etadec}
\eta=\eta_\parallel \vartheta+\eta_\perp n~~,
\een
where
\beqa
\label{etaparperp}
&& \eta_\parallel\eqdef \cG(\vartheta,\eta)=-\frac{\ddot{\sigma}}{H\dot{\sigma}}~~,~~\eta_\perp\eqdef \cG(n,\eta)= -\frac{\dot{\sigma}}{H}\kappa=\frac{\partial_n\Phi}{H\dot{\sigma}}
\eeqa
are known \cite{Welling} as the {\em second slow-roll parameter}
($\eta_\parallel$) and the {\em first slow-turn parameter}
($\eta_\perp$). Using \eqref{nabla_phi} and the relation
$\dot{\varphi}=\dot{\sigma}\vartheta$ shows that \eqref{eom} is equivalent
with the system:
\beqan
&& \ddot{\sigma}+3H \dot{\sigma}+\partial_\vartheta \Phi=0~~,\label{eom1}\\
&& \kappa=-\frac{1}{\dot{\sigma}^2}\partial_n\Phi~~. \label{eom2}
\eeqan
Notice that $\dot{\Phi}\eqdef \frac{\dd}{\dd t} (\Phi\circ
\varphi)=(\dd \Phi)(\dot{\varphi})=\dot{\sigma}\partial_\vartheta \Phi$.
Using \eqref{etadec}, the system becomes:
\beqan
\label{etasys}
&& 3(1-\frac{\eta_\parallel}{3}) H \dot{\sigma}=-\partial_\vartheta \Phi~~,\nn\\
&& \eta_\perp H \dot{\sigma}=\partial_n\Phi~~.
\eeqan
Since $||\eta||_\cG=\sqrt{\eta_\parallel^2+\eta_\perp^2}$, the
gradient flow condition $||\eta||_\cG\ll 1$ amounts to
$|\eta_\parallel|\ll 1$ and $|\eta_\perp|\ll 1$. 

\subsection{The slow gradient flow approximation}

\noindent In this subsection, we discuss an approximation which is more
restrictive than the gradient flow approximation in that it implies
the latter. This consists of approximating cosmological trajectories
by gradient flow lines and further assuming that the motion along
such lines is ``slow''.

\vspace{2mm}

Formally, the {\em slow gradient flow approximation} is
defined by the {\em kinematic slow gradient flow conditions}:
\ben
\label{slowgradcond}
\epsilon\ll 1~~\mathrm{and}~~||\eta||_\cG\ll 1~~.
\een
Notice that these are more restrictive than the kinematic gradient flow condition.

\vspace{2mm}

\paragraph{Proposition.}
When \eqref{slowgradcond} hold, we have 
$\vartheta\simeq \frac{\grad_\cG \Phi}{||\dd \Phi||_\cG}$
and the scalar potential $\Phi$ satisfies the {\em potential slow
  gradient flow conditions}:
\beqan
M|\partial_n \log \Phi| &\ll & 1 \label{slowgrad1}~~,\\
\left|\frac{\partial_n\Phi}{\partial_\vartheta\Phi}\right| &\ll & 1~\label{slowgrad2}~,~\\
M^2\frac{\Hess(\Phi)(\vartheta,\vartheta)}{\Phi} &\ll & 1~\label{slowgrad3}~,
\eeqan~
where $\Hess(\Phi)$ is the Hessian of $\Phi$.

\vspace{2mm}

\paragraph{Proof.}
Using the first slow-roll parameter, equations \eqref{F1} and \eqref{F2} become:
\ben
\label{F1F2epsilon}
\dot{\sigma}=M H\sqrt{2\epsilon}~~\mathrm{and}~~\Phi=3M_p^2 H^2\left(1+\frac{\epsilon}{3}\right)~~.
\een
Assume that $\eta_\parallel\ll 1$. Then the first term of equation \eqref{eom1} can be neglected and
\eqref{eom1} becomes:
\be
H\dot{\sigma}\simeq -\frac{1}{3}\partial_\vartheta \Phi~\Longrightarrow ~\eta_\perp=-\frac{3\partial_n\Phi}{\partial_\vartheta\Phi}~~.
\ee
Differentiating this relation with respect to $t$ and dividing the resulting equation by $H^2$ gives:
\ben
\label{eqetapar}
\eta_\parallel+\epsilon=\frac{1}{3H^2}\nabla_\vartheta \partial_\vartheta \Phi=M^2\left(1+\frac{\epsilon}{3}\right)\frac{\nabla_\vartheta\partial_\vartheta \Phi}{\Phi}~~,
\een
where in the last equality we used \eqref{F1F2epsilon}. Let us further
assume that $\epsilon\ll 1$. Then \eqref{eqetapar} gives
$\eta_\parallel\simeq M^2\frac{\nabla_\vartheta\partial_\vartheta
  \Phi}{\Phi}$. In this case, the full gradient flow condition
$||\eta||_\cG\ll 1$ implies \eqref{slowgrad2} and
\eqref{slowgrad3}.

Now let us assume that the full conditions \eqref{slowgradcond} hold.
Then \eqref{gradflow} applies, giving:
\be
\dot{\sigma}\simeq ||\dot{\varphi}||_\cG=\frac{|\partial_n \Phi|}{3H}~~.
\ee
Substituting this into the relation
$\epsilon=\frac{\dot{\sigma}^2}{2H^2M^2}$ gives:
\ben
\label{ep}
\epsilon\simeq \frac{|\partial_n \Phi|^2}{18 H^4 M^2}~~.
\een
Since $\epsilon$ is small, the second relation in \eqref{F1F2epsilon}
implies $H^2\simeq \frac{\Phi}{3M^2}$.  Substituting this into
\eqref{ep} gives:
\ben
\epsilon\simeq \frac{M^2}{2}|\partial_n \log \Phi|^2~~.
\een
Hence the slow gradient flow approximation also implies \eqref{slowgrad1}.

\subsection{The SRST approximation}

\noindent Following \cite{m1,m2,m3, m4, m5, m6, PT1, PT2, Gong} (see
\cite{Welling} for a recent review), we end this section by briefly
recalling the traditional SRST approximation, which is considerably
more restrictive that the approximations discussed above\footnote{As we shall
see below (see, for example, Subsection \ref{subsec:SRSTEnds}), this popular
approximation is insufficient for the study of generalized
two-field $\alpha$-attractor models.}. Define the {\em third slow-roll parameter}
$\xi_\parallel$ and the {\em second slow-turn parameter} $\xi_\perp$
through:
\beqa
&& \xi_\parallel\eqdef -\frac{\dddot{\sigma}}{H\ddot{\sigma}}~~,~~\xi_\perp\eqdef -\frac{\dot{\eta_\perp}}{H\eta_\perp}~~.
\eeqa
The {\em Hubble slow-roll parameters} are defined through:
\beqa
&& \epsilon_H\eqdef \epsilon ~~,\\
&& \eta_H\eqdef \eta_\parallel-\epsilon~~,\\
&& \xi_H\eqdef -\frac{\dot{\eta_H}}{H\eta_H}=\frac{\eta_\parallel(\xi_\parallel-\eta_\parallel-3\epsilon)+2\epsilon^2}{\eta_\parallel-\epsilon}~~.
\eeqa
With these definitions, the {\em kinematical slow-roll (SR)
  approximation} corresponds to $|\epsilon_H|$,$|\eta_H|$, $|\xi_H|\ll 1$
while the {\em kinematical slow-turn (ST) approximation} corresponds
to $|\eta_\perp|,|\xi_\perp|\ll 1$ (see \cite{Welling}). The {\em
  kinematical slow-roll -- slow-turn (SRST) approximation} corresponds
to imposing all of these conditions simultaneously.

As explained in \cite{Welling}, the kinematical SRST approximation implies the
{\em potential SRST conditions}\,:
\beqan
M ||\dd \log \Phi||_\cG &\ll & 1 \label{grad}~~,\\
M^2 \left |\frac{\Hess(\Phi)(\vartheta,\vartheta)}{\Phi}\right | &\ll & 1 \label{Hess1}~~,\\
M^2 \left |\frac{\Hess(\Phi)(n,\vartheta)}{\Phi}\right | &\ll & 1\label{Hess2}~~.
\eeqan
Since the conditions $\epsilon_H,\eta_H,\eta_\perp \ll 1$ imply
$||\eta||_\cG\ll 1$, the SRST approximation implies the slow gradient
flow approximation and hence also the gradient flow approximation.
However, the gradient flow approximation (which will be used in
Section \ref{sec:Morse}) is much less restrictive than the SRST
approximation, since it requires only $|\eta_\parallel|\ll 1$ and
$|\eta_\perp|\ll 1$, conditions which constrain only the second
slow-roll parameter and the first slow-turn parameter. Similarly, the
slow gradient flow approximation is considerably less restrictive than
the SRST approximation. We stress that experience with explicit
examples shows that the SRST approximation is generally quite
ill-suited for a study of generalized two-field $\alpha$-attractor models, for
example if one wishes to study cosmological dynamics near cusp ends
or within the compact core of $\Sigma$.

\subsection{The strong potential SRST conditions}

\noindent The potential SRST conditions are implied by the stronger conditions:
\ben
\label{spSRST}
M ||\dd \log \Phi||_\cG \ll  1~~,~~M^2 \frac{||\Hess(\Phi)||_\cG}{|\Phi|} \ll 1~~,
\een
which we shall call the {\em strong potential SRST conditions}. These
conditions are convenient since they are easy to test. When
these conditions are satisfied, it is reasonable to expect that the
SRST approximation is valid \cite{PT1, PT2, Welling}.

\section{Generalized two-field $\alpha$-attractor models}
\label{sec:genalpha}

In this section, we introduce two-field generalized $\alpha$-attractor
models and discuss some of their fundamental properties.  This section
makes free use of mathematical results and terminology which are
summarized in Appendices \ref{sec:unif}, \ref{sec:topfinite} and
\ref{sec:geomfinite}.

\vspace{2mm}

Let $(\Sigma, G)$ be a geometrically finite hyperbolic surface with
$n$ ends. For any real positive number $\alpha>0$, consider the
rescaled hyperbolic metric:
\be
\cG\eqdef 3\alpha G~~,
\ee
which has Gaussian curvature $K=-\frac{1}{3\alpha}$. Let $\Phi$ be a
smooth real-valued function defined on $\Sigma$. By definition, a
{\em generalized two-field $\alpha$-attractor model} defined by $\alpha$ and
$(\Sigma,G,\Phi)$ is a cosmological model associated to the triplet
$(\Sigma,\cG,\Phi)$ as in Section \ref{sec:ES}.

When $\Sigma$ is simply connected, the hyperbolic surface $(\Sigma,
G)$ is isometric with the Poincar\'e disk. In this case, $G$ is
uniquely determined and coincides with the Poincar\'e metric, hence
the generalized two-field $\alpha$-attractor model reduces to the two-field
model discussed in \cite{Escher}. When $\Sigma$ is not simply
connected, isometry classes of hyperbolic metrics on $\Sigma$ with
fixed type of ends form a moduli space. When $\Sigma$ is
non-elementary, this moduli space has real dimension equal to
$6g-6+3n_f+2n_c$ (see \cite{Borthwick}). In particular, generalized
two-field $\alpha$-attractor models form an {\em uncountably infinite} family of
generalizations of the hyperbolic disk model of \cite{Escher}.

\subsection{Symmetries}

\noindent Isometries of the scalar manifold $(\Sigma,G)$ which
preserve the scalar potential $\Phi$ induce symmetries of the
equations of motion of the generalized two-field $\alpha$-attractor
model. Restricting to orientation preserving isometries gives the
symmetry group\footnote{Note that these are only the `obvious' symmetries of the 
model and not the most general Noether symmetries.}:
\be
\Aut^+(\Sigma,\cG,\Phi)=\Aut^+(\Sigma,G,\Phi)=\{h\in \Iso^+(\Sigma,G)~|~\Phi\circ h=\Phi \}~.
\ee
The results summarized in Appendix \ref{subsec:Iso} imply:
\begin{enumerate}[A.]
\itemsep 0.0em
\item In the elementary case, the group $\Aut^+(\Sigma,G,\Phi)$ is
  infinite (namely, isomorphic with $\U(1)$) iff $\Phi$ is invariant
  under $\Iso^+(\Sigma,G)\simeq \U(1)$. This happens iff $\Phi$
  depends only on $|u|$ when $\Sigma$ is conformal to the disk $\mD$,
  the hyperbolic punctured disk $\mD^\ast$ or an annulus $\A(R)$. When $\Phi$
  depends on both $|u|$ and $\arg u$, the group
  $\Aut^+(\Sigma,G,\Phi)$ is a finite cyclic group, which is trivial
  in the generic case.
\item In the non-elementary case, the symmetry group
  $\Aut^+(\Sigma,G,\Phi)$ is necessarily finite, being a normal
  subgroup of the finite group $\Aut(\Sigma,G)\simeq
  N(\Gamma)/\Gamma$, where $\Gamma$ is the uniformizing Fuchsian group
  of $\Sigma$ (see Apppendix \ref{subsec:Iso}).
\end{enumerate}

\noindent In particular, generalized two-field $\alpha$-attractors based on
non-elementary hyperbolic surfaces {\em cannot} admit D-term
embeddings \cite{FS,FFS1,FFS2} in gauged $\cN=1$ supergravity with a
single chiral multiplet, since the construction of the latter involves
gauging a {\em continuous} group of isometries of $(\Sigma,\cG)$.
However, such models do admit $F$-term embeddings in $\cN=1$
supergravity provided that the scalar potential is induced by
a holomorphic superpotential.

\subsection{Well-behaved scalar potentials}
\label{subsec:wb}

\noindent Let $\hSigma$ be the end compactification of $\Sigma$ 
(see Appendix \ref{sec:topfinite}). A
scalar potential $\Phi:\Sigma\rightarrow \R$ is called {\em
  well-behaved at the ideal point (or end) $p\in \hSigma\setminus
  \Sigma$} if there exists a smooth function
$\hPhi_p:\Sigma\sqcup\{p\}\rightarrow \R$ such that:
\be
\Phi=\hPhi_p|_{\Sigma}~~.
\ee
In this case, $\hPhi_p$ is uniquely determined by $\Phi$ through
continuity and is called the {\em extension of $\Phi$} to $p$.
The potential $\hPhi$ is called {\em globally well-behaved} if there
exists a globally-defined smooth function $\hPhi:\hSigma\rightarrow
\R$ such that:
\be
\Phi=\hPhi|_\Sigma~~.
\ee
In this case, $\hPhi$ is uniquely determined by $\Phi$ through
continuity and is called the {\em global extension of $\Phi$} to
$\hSigma$. Notice that $\Phi$ is globally well-behaved iff it is
well-behaved at each end of $\Sigma$. As we shall see below,
potentials which are well-behaved at $p$ lead to the same universal
expression for the spectral index and tensor to scalar ratio as
ordinary $\alpha$-attractor models in the leading order of the
slow-roll approximation if an appropriate one-field truncation is
performed in the vicinity of $p$.

Consider semi-geodesic coordinates $(r,\theta)$ near $p$  
in which the hyperbolic metric $G$ has the asymptotic form
\eqref{asmetric}. Setting:
\be
\phi\eqdef 2\arccot(r)\in (0,\pi)~\Longleftrightarrow~ r=\cot\left(\frac{\phi}{2}\right)\in (0,+\infty)~~,
\ee
the change of coordinates from $(r,\theta)$ to $(\phi,\theta)$ induces
a diffeomorphism $\mu_p:V_p\rightarrow \rS^2\setminus \{\nu,\sigma\}$ which
identifies the punctured semi-geodesic coordinate neighborhood
$V_p\subset \Sigma$ of $p$ with the unit sphere with the north and south
poles $\nu$ and $\sigma$ removed. Here $\phi$ and $\theta$ are viewed
as spherical coordinates on $\rS^2=\{(\sin\phi\cos\theta,
\sin\phi\sin\theta, \cos\phi)\,|\,\phi\in [0,\pi],\theta\in
                  [0,2\pi)\}\subset \R^3$. 
The limit $r\rightarrow +\infty$ corresponds to $\phi\rightarrow 0$,
i.e. to the north pole $\nu\in \rS^2$, while the limit $r\rightarrow
0$ corresponds to $\phi\rightarrow \pi$, i.e. to the south pole
$\sigma \in \rS^2$. In particular, the north pole of $\rS^2$
corresponds to the ideal point $p$. A potential $\Phi:\Sigma
\rightarrow \R$ is well-behaved at $p$ iff there exists a smooth
function $\bPhi_p:\rS^2\setminus \{\sigma\}\rightarrow \R$ such that
$\Phi=\bPhi_p\circ \mu_p$, which gives:
\ben
\label{PhibPhi}
\Phi(r,\theta)=\bPhi_p(2\arccot(r),\theta)~~. 
\een 
The condition that $\bPhi_p$ is smooth at the north pole implies that
$\bPhi_p(\phi,\theta)$ has a finite limit for $\phi\rightarrow 0$
(which equals $\hPhi_p(p)$) and hence $\Phi(r,\theta)$ has a
$\theta$-independent limit for $r\rightarrow +\infty$. Moreover,
\eqref{PhibPhi} gives:
\beqan
\label{parPhi}
&& \partial_r \Phi=-\frac{2}{1+r^2}\partial_\phi \bPhi_p~~,\nn\\
&&\partial_r^2 \Phi=\frac{4r}{(1+r^2)^2}\partial_\phi \bPhi_p+\frac{4}{(1+r^2)^2}\partial_\phi^2 \bPhi_p~~,\nn\\
&& \partial_\theta \Phi=\partial_\theta \bPhi_p~~,~~\partial_\theta^2\Phi=\partial_\theta^2 \bPhi_p~~,\nn\\
&&\partial_r \partial_\theta \Phi=-\frac{2}{1+r^2}\partial_\phi\partial_\theta \bPhi_p~~.
\eeqan
These relations imply the following asymptotics for $r\gg 1$:
\beqan
\label{asparPhi}
&& \partial_r \Phi\simeq -\frac{2}{r^2} \partial_\phi \bPhi_p~~,~~\partial_r^2 \Phi\simeq \frac{4}{r^3}\partial_\phi \bPhi_p\nn\\
&& \partial_\theta \Phi=\partial_\theta \bPhi_p~~,~~\partial_\theta^2\Phi=\partial_\theta^2 \bPhi_p~~,~~\partial_r \partial_\theta \Phi\simeq -\frac{2}{r^2}\partial_\phi\partial_\theta \bPhi_p~~.
\eeqan

\paragraph{The Laplace expansion of $\hPhi$.}
In this paragraph, we discuss a natural expansion for the extension of
well-behaved scalar potentials, which gives a systematic way to
approximate such potentials in our class of models. Some applications
of this type of expansion can be found in references \cite{elem,modular}. Let $\fc$ be
the conformal equivalence class of the hyperbolic metric $G$ on
$\Sigma$ and let ${\hat \fc}$ be its prolongation to $\hSigma$ (see
Appendix \ref{subsec:conf_compactif}). Let $\hG$ denote the canonical
complete metric in the conformal class ${\hat \fc}$, which was defined
in the same subsection. Notice that the restriction of $\hG$ to
$\Sigma$ differs from $G$ (in particular, this restriction is {\em
not} a complete metric on $\Sigma$ !). Since $\hSigma$ is compact, the
sign opposite $-\hDelta$ of the Laplacian $\hDelta$ of $\hG$ is a
positive operator with purely discrete spectrum, whose distinct
eigenvalues $0=\lambda_0<\lambda_1<\lambda_2<\ldots $ accumulate only
at $+\infty$. The eigenspace $\cH_l$ of the eigenvalue $\lambda_l$ is
finite-dimensional and consists of all smooth functions
$f:\hSigma\rightarrow \C$ which satisfy:
\be
\hDelta f=-\lambda_l f~~.
\ee
For $l=0$, all eigenfunctions are constant and we have $\cH_0=\C 1$,
where $1$ denotes the constant unit function defined on $\hSigma$.
The Hilbert space $L^2(\hSigma,\hG)$ of complex-valued
square-integrable functions (modulo equivalence almost everywhere)
with respect to the Lebesgue measure defined by $\hG$ has an
orthogonal direct sum decomposition:
\be
L^2(\hSigma,\hG)=\overline{\bigoplus}_{l=0}^\infty \cH_l~~
\ee
where $\overline{\bigoplus}_{l=0}^\infty$ denotes the completed direct
sum. Let $M_l\eqdef \dim_\C \cH_l$ denote the multiplicity of the
eigenvalue $\lambda_l$ and let $u_{l1},\ldots, u_{l M_l}$ denote an
orthonormal basis of $\cH_l$ for each $l\in \N$, where $M_0=1$ and
$u_{01}=1$. Since $\lambda_l$ are real, the complex conjugates
$\overline{u_{lm}}$ also form an orthonormal basis of $\cH_l$, so we
have:
\be
\overline{u_{lm}}=\sum_{m'=1}^{M_l} q^{(l)}_{mm'}u_{lm'}~~\forall {m=1,\ldots, M_l}~~,
\ee
for some $q^{(l)}_{m m'}\in \C$, where $Q^{(l)}\eqdef (q^{(l)}_{m
  m'})_{m,m'=1\ldots M_l}$ is a unitary and symmetric
matrix\footnote{We have $[Q^{(l)}]^t=Q^{(l)}$ and
  $Q^{(l)}\overline{Q^{(l)}}=1$.} and $q^{(0)}_{11}=1$.

Let $\Phi$ be a globally well-behaved scalar potential on $\Sigma$ and
let $\hPhi$ be its global extension to $\hSigma$. Then $\hPhi$ can be expanded
uniquely as:
\ben
\label{hPhiExpGen}
\hPhi=\sum_{l=0}^\infty \sum_{m=1}^{M_l} C_{lm} u_{lm}~~,
\een
where $C_{lm}$ are complex constants given by: 
\be
C_{lm}=\int_{\hSigma}(\hPhi\,\overline{u_{lm}})\vol_{\hG}
\ee
and the series in the right hand side converges to the left hand side
in the $L^2$ norm determined by $\hG$. Since $\hPhi$ is real-valued,
we have:
\be
\overline{C_{lm}}q^{(l)}_{mm'}=C_{lm'}~~.
\ee
The expansion \eqref{hPhiExpGen} gives a systematic method to
approximate globally well-behaved potentials by truncating away all
modes with $l$ larger than some given cutoff.

When $\Sigma$ is a planar surface (i.e. when its genus $g$ vanishes), the end
compactification $\hSigma$ is a sphere $\rS^2$ and the canonical
complete metric $\hG$ determined by the conformal prolongation ${\hat
  \fc}$ of the conformal class $\fc$ of $G$ is the round metric of
radius one. In this case, we have $\lambda_l=l(l+1)$ with $l\in \N$
and $u_{lm}$ can be taken to be the spherical harmonics $Y_{lm}$. Thus
\eqref{hPhiExpGen} reduces to the classical Laplace-Fourier expansion
on the sphere; some models of this type are discussed in \cite{elem,modular}. 
When $\Sigma$ has genus one, the end compactification
$\hSigma$ endowed with the metric $\hG$ and the complex structure
$\hJ$ is a complex torus $\rT^2=\C/L$, where $L=\Z\oplus \Z\tau$ is
the lattice generated by $1$ and $\tau\in \H$, with $|\tau|\geq 1$ and
$\Re(\tau)\leq \frac{1}{2}$ (see Appendix
\ref{subsec:conf_compactif}). In this case, an orthonormal basis of
eigenfunctions of $-\hDelta$ is given by $u_w(z)=e^{2\pi \i \langle
  w,z\rangle}$, where $z$ is the complex coordinate induced from $\C$
and the 2-dimensional vector $w\in \R^2\equiv \C$ runs over the
elements of the dual lattice $L^\vee$. The eigenvalue of $u_w$ is
$\lambda_w=4\pi^2||w||^2$. When the genus $g$ of $\Sigma$ is greater
than one, the metric $\hG$ is hyperbolic. In that case, a basis of
eigenfunctions of $-\hDelta$ is sensitive to the hyperbolic geometry
of $(\hSigma,\hG)$ and cannot in general be determined
explicitly. However, results from the well-developed spectral theory
of compact hyperbolic surfaces \cite{Buser} can in principle be
applied to extract information about the spectrum and eigenfunctions
of $-\hDelta$.

\subsection{Naive local one-field truncations and ``universality'' for certain special trajectories near the ends}

\label{subsec:univ}

\noindent In this subsection, we briefly discuss ``universality'' of generalized
two-field $\alpha$-attractor models in the naive classical truncation near the
ends. We stress that one-field truncations are far from sufficient
when studying two-field generalized two-field $\alpha$-attractors, which in our view
cannot be understood properly unless one treats them systematically 
as two-field models. We also stress that we do not consider in detail the quantum perturbations 
of such models, since this subject lies outside the scope of the present paper. 

\vspace{2mm}

\paragraph{Universality in one-field models.}
As explained in \cite{alpha4}, $\alpha$-attractor models with a {\em
  single} real scalar field have universal behavior\footnote{Near $x=0$.} when:
\begin{itemize}
\itemsep 0.0em
\item The scalar manifold is diffeomorphic with an open interval $I$,
  thus admitting a global real coordinate $x:I\stackrel{\sim}{\rightarrow}
  (0,1)$.
\item The coefficient $s(x)$ of the field space metric $\dd s^2
  =s(x)\dd x^2$ has asymptotic form:
\ben
\label{Gexp}
s(x)=_{x\ll 1}\frac{a}{x^{p}} (1+\O(x))
\een
\item The one-field scalar potential $W(x)$ has an asymptotic
  expansion:
\ben
\label{Vexp}
W(x)=_{x\ll 1} W_0\left(1-c x + \O(x^2)\right)~~,
\een
\end{itemize}
where $a>0$ and $c>0$ are constants. In this case, the spectral index
$n_s$ and the tensor to scalar ratio $r$ in the leading order of the
one-field slow-roll approximation in the region $x\ll 1$ are given by
\cite{alpha4}:
\ben
\label{nsr}
n_{s}\approx 1-\frac{p}{p-1}\frac{1}{N(t)},~~~r\approx \frac{8c^{\frac{p-2}{p-1}
	}a^{\frac{1}{p-1}}}{(p-1)^{\frac{p}{p-1}}}\frac{1}{N^{\frac{p}{p-1}}}~~,
\een
where:
\be
N\eqdef \log \frac{a(t_1)}{a(t_0)}
\ee
is the number of e-folds ($t_0$ and $t_1$ being the cosmological times
at which inflation starts and ends).

\vspace{2mm}

\paragraph{Universality in the truncated model on the hyperbolic disk.}

Consider the {\em two-field} $\alpha$-attractor model based on the
Poincar\'e disk:
\be
\mD=\{u\in \C|~|u|<1\}~~.
\ee
At the classical level, this can be truncated to a one-field model by taking
$\varphi(t)=(x(t),\theta_0)$, with $\theta_0$ independent of
$t$. Setting $u=(1-x)e^{\i\theta}$ (with $x\in (0,1)$ and $\theta\in
\R/(2\pi \Z)$), the rescaled Poincar\'e metric on $\mD$ takes the
form:
\be
\dd s_\cG^2=\frac{3 \alpha}{2}\frac{4}{x^2(2-x)^2} [\dd x^2+(1-x)^2\dd \theta^2]\simeq_{x\ll 1} \frac{3 \alpha}{2}\frac{1}{x^2} (\dd x^2+\dd \theta^2)
\ee
and the {\em reduced} field $\varphi(t)=(x(t),\theta_0)$ ``feels''
only the radial metric:
\be
\dd s_\cG^2=\frac{3 \alpha}{2}\frac{4}{x^2(2-x)^2} \dd x^2=_{x\ll 1}\frac{3 \alpha}{2}\left[\frac{1}{x^2}+\O(x^2)\right]\dd x^2~~,
\ee
which has the form \eqref{Gexp} with $p=2$ and
$a=\frac{3\alpha}{2}$. Let us assume that $\Phi(x,\theta)$ has an
asymptotic expansion of the form:
\ben
\label{PhiExp}
\Phi(x,\theta)=_{x\ll 1} a-b(\theta)x+\O(x^2)~~,
\een
where $a$ is a non-zero constant and $b(\theta)> 0$ is a smooth
function defined on the circle $\R/(2\pi \Z)$. Then $W(x)\eqdef
\Phi(x,\theta_0)$ has an expansion of the form \eqref{Vexp} for $x\leq
1$, where $W_0=a$ and $c\eqdef \frac{b(\theta_0)}{a}$ (the latter of
which generally depends on $\theta_0$). In this case, the argument of
\cite{alpha4} shows that, to leading order of the slow-roll
approximation for the {\em truncated} model, the quantities $n_s$ and
$r$ are given by relation \eqref{nsr} with $p=2$:
\ben
\label{nsr2}
n_{s}\approx 1-\frac{2}{N},~~~r\approx \frac{12\alpha}{N^2}~~,
\een
which fit well current observational data.

Notice that the right hand sides of these expressions have no explicit
dependence of the constant $c$ (and hence of the choice of
$\theta_0$). However, the number $N$ of e-folds realized during
inflation can depend implicitly on $c$ and hence on the choice of
$\theta_0$. A limit argument (based on taking $W_0$ to zero) shows
that the same relations \eqref{nsr} are obtained when $a=0$.

\

\paragraph{Universality for certain special trajectories in the naive local truncation of generalized two-field $\alpha$-attractor models
near the ends.}

A similar argument can be applied to generalized two-field $\alpha$-attractor
models in a punctured neighborhood of each ideal point $p\in
\hSigma\setminus \Sigma$. On a sufficiently small such neighborhood,
the hyperbolic metric $G$ has a $\U(1)$ symmetry, which in
semi-geodesic coordinates $(r,\theta)$ acts by shifts of $\theta$. 
Moreover, the hyperbolic metric can be approximated by
\eqref{asmetric} near the end, where the ideal point corresponds to
$r\rightarrow +\infty$. The change of coordinates $r=-\log x$ (with
$x=e^{-r}>0$) places the ideal point at $x=0$ and brings the rescaled
asymptotic metric $\cG$ to the form:
\be
\dd s_\cG^2=_{x\ll 1}\twopartdef{\frac{3\alpha}{2} \frac{1}{x^2}\left[\dd x^2+\left(\frac{c_p}{4\pi}\right)^2 \dd \theta^2\right]}{\epsilon_p=+1~
\mathrm{(plane,~ horn,~ funnel ~end)}}
{\frac{3\alpha}{2} \frac{1}{x^2} \left[\dd x^2+\left(\frac{c_p}{4\pi}\right)^2 \!\! x^4 \dd \theta^2\right]}{\epsilon_p=-1~\mathrm{(cusp~end)}}~~
\ee
Suppose that $\Phi$ is well-behaved at the end $p$. In this case,
considering a small enough disk $D_p$ centered at $p$, we have:
\be
\Phi(r,\theta)=\hPhi_p(e^{-r},\theta)~~\mathrm{for}~~r\gg 1
\ee
for some smooth function $\hPhi_p:D_p\rightarrow \R$.  Using the
Taylor expansion of $\hPhi_p$ around $x=0$, this implies that $\Phi$
has the asymptotic form \eqref{PhiExp}, where
$b(\theta)=-(\partial_{x_1}\hPhi_p|_{x=0})\cos\theta-(\partial_{x_2}\hPhi_p|_{x=0})\sin\theta$
(with $x_1\!=\!x\cos\theta$ and $x_2\!=\!x\sin\theta$). On such a small
neighborhood of $p$, one can truncate the two-field model by setting
$\varphi(t)\!=\!(x(t),\theta_0)$ with $\theta_0$ independent of $t$. Such
a truncation is justified when the inflationary trajectories near the
end are well-approximated by non-canonically parameterized geodesics
flowing from the end toward the compact core\footnote{see Appendix
\ref{sub:isom} for a definition of the compact core.} of $(\Sigma,G)$.  The
truncated model ``feels'' the metric:
\be
\dd s^2 \simeq_{x\ll 1} \frac{3\alpha}{2} \frac{\dd x}{x^2}~~,
\ee
which has the same form for all types of ends. Assuming
$b(\theta_0)>0$ (so that inflation proceeds from the end toward the
interior), the same argument as above shows that $n_s$ and $r$ are
given by \eqref{nsr2} in the leading order of the slow-roll
approximation for the {\em truncated} model. Thus:

\vspace{2mm}

\begin{itemize}
\item {\em Suppose that $\Phi$ is well-behaved at an end of
$(\Sigma,G)$ where its extension has a local maximum and that
inflation takes place near that end and proceeds away from the
end. Then the generalized two-field $\alpha$-attractor model admits a
{\em naive} local truncation to a one-field model in a punctured
vicinity of the corresponding ideal point for which the universal
relations \eqref{nsr2} apply in the leading order of the one-field
slow-roll approximation, for the special trajectories described
above}.
\end{itemize}

\vspace{2mm}

\noindent This shows that generalized two-field $\alpha$-attractor models have
the same kind of universal behavior as the disk models of
\cite{Escher} near each such end, at least in the naive one-field truncation
discussed above.  

Of course, the classical local truncation to a
one-field model used in the argument of \cite{Escher} (as well as
above) is somewhat simplistic. In particular, one may worry that
quantum effects could destabilize the one-field trajectory and
change the predictions for radiative corrections. For ordinary
$\alpha$-attractors (those models for which $\Sigma$ is the hyperbolic
disk $\mD$), a resolution of this problem can be  obtained when restricting 
to those $\alpha$-attractor models which admit embeddings in $\cN=1$
supergravity coupled to one chiral multiplet. For such models, it was
argued in \cite{alpha5} that the naive truncation can be stabilized in
a universal way provided that the superpotential is sufficiently
well-behaved. As explained later on in Section \ref{sec:lift}, any
generalized two-field $\alpha$-attractor model can be lifted to a model defined
on the hyperbolic disk by using the uniformization map
$\pi_\mD:\mD\rightarrow \Sigma$.  This lift is an ordinary
$\alpha$-attractor whose scalar potential is invariant under the action
of the uniformizing surface group $\Gamma\subset \PSL(2,\R)$. When the
generalized two-field $\alpha$-attractor model admits an F-term embedding in
$\cN=1$ supergravity coupled to a single chiral multiplet with
superpotential $W:\Sigma\rightarrow \C$, the lifted model also admits
such an embedding, with superpotential given by the lift ${\tilde
  W}\eqdef W\circ \pi_\mD$ of $W$, which is $\Gamma$-invariant. The
cosmological trajectories of the model defined by $\Sigma$ are the
$\pi_\mD$-images of the cosmological trajectories of this lifted
model. The two models are related by the field redefinition induced by
$\pi_\mD$, which implements the quotient of the lifted model through
the discrete group $\Gamma$. Provided that this discrete symmetry is
preserved when quantizing perturbations of the lifted model around a
classical solution, the effect of quantum perturbations of the model
defined by $\Sigma$ can be obtained from those of the lifted model by
projection through the uniformization map (which implements the
quotient through $\Gamma$). This implies that cosmological solutions
of the generalized two-field $\alpha$-attractor model have the same local 
properties as those of the lifted model, to which stabilization
arguments such as those of \cite{alpha5} apply for radial trajectories
on the Poincar\'e disk. The trajectories involved in the naive
truncations near the ends of $\Sigma$ lift to small portions of radial
trajectories on $\mD$ which are located near the conformal boundary of
$\mD$. In models which have appropriate F-term embeddings in
supergravity one can thus expect that, at least near flaring ends, the
local one-field truncations discussed above can be stabilized by
constructions similar to those used for ordinary $\alpha$-attractors,
provided that the discrete $\Gamma$-symmetry of the lifted model is
unbroken by such corrections. To make this argument rigorous, one must
of course first classify F-term embeddings of generalized two-field
$\alpha$-attractors into $\cN=1$ supergravity, a subject which we plan
to discuss in a separate publication. Here, we mention only that
generalized two-field $\alpha$-attractors can admit F-term embeddings
which are `twisted' by a unitary character of the fundamental group of
$\Sigma$, an aspect which must be taken into account when studying the
corresponding supergravity models. We mention that universality in
two-field $\alpha$-attractors defined on the hyperbolic disk was
recently studied for curved trajectories in reference \cite{univ} 
(such trajectories do not coincide with the lift to the hyperbolic
disk of the very special trajectories considered in this subsection). 

Notice, however, that generalized two-field $\alpha$-attractors are
intrinsically {\em two-field} models, since they are interesting
precisely due to the non-trivial topology of the 2-manifold
$\Sigma$. A proper understanding of the dynamics of such models simply
cannot be attained by considering one-field truncations, which in our
opinion are of limited interest in this context (in particular, since
such truncations cannot make sense globally when $\Sigma$ is
topologically non-trivial). In the next subsection, we briefly discuss
the classical two-field dynamics of such models in the SRST
approximation near the ends, showing that, in this approximation, the
models have different behavior near cusp and flaring ends. In
particular, we show that the SRST approximation can fail near cusp
ends and hence (as already mentioned before) this approximation is
insufficient for the study of generalized two-field
$\alpha$-attractors.

\subsection{The strong potential SRST conditions near the ends}
\label{subsec:SRSTEnds}

\noindent Let $\Phi:\Sigma\rightarrow \R$ be a scalar potential which
is well-behaved at an ideal point $p\in \hSigma\setminus \Sigma$ and
assume that the smooth extension $\hPhi_p:\Sigma\sqcup\{p\}\rightarrow
\R$ of $\Phi$ does not vanish at $p$. Consider semi-geodesic
coordinates $(r,\theta)$ in a neighborhood of $p$. Using relation
\eqref{asmetric} of Appendix \ref{subsec:endmetric}, we find the
following asymptotic expressions for $r\gg 1$:
\beqan
\label{gradhess}
&&M ||\dd \log \Phi||_\cG \simeq_{r\gg 1} \frac{M}{\sqrt{3\alpha}}\frac{\sqrt{(\partial_r\Phi)^2+\left(\frac{4\pi}{c_p}\right)^2e^{-2\epsilon_p r} (\partial_\theta\Phi)^2}}{|\Phi|} \nn\\
&&\frac{M_p^2}{|\Phi|}||\Hess(\Phi)||_\cG \simeq_{r\gg1} \\
&&\frac{M^2}{3\alpha} \frac{\sqrt{(\partial_r^2\Phi)^2+2\left(\frac{4\pi}{c_p}\right)^2(\partial_r\partial_\theta \Phi)^2e^{-2\epsilon_p r} +
\left(\frac{4\pi}{c_p}\right)^4 e^{-4\epsilon_p r} (\partial_\theta^2 \Phi)^2}}{|\Phi|}~.\nn
\eeqan
The quantities $c_p$ and $\epsilon_p$ are defined in equation \eqref{Cdef} of Appendix \ref{subsec:endmetric}. 
Below, we use the terminology and results of that appendix.  

\vspace{2mm}

\paragraph{Flaring ends.}

When $p$ corresponds to a flaring end, we have $\epsilon_p=+1$ and the
exponential terms appearing in the right hand side of \eqref{gradhess}
are suppressed when $r\gg 1$, since we assume that $\Phi$ is
well-behaved at $p$. In this case, we find:
\beqa
\label{gradhess1}
&&M ||\dd \log \Phi||_\cG \simeq_{r\gg 1} \frac{M}{\sqrt{3\alpha}}\left|\frac{\partial_r\Phi}{\Phi}\right|\\
&&\frac{M^2}{|\Phi|}||\Hess(\Phi)||_\cG \simeq_{r\gg 1} \frac{M^2}{3\alpha} \, \left|\frac{\partial_r^2\Phi}{\Phi}\right|~~
\eeqa
and the strong potential SRST conditions are satisfied on a pointed
neighborhood of $p$ provided that:
\ben
\label{cond1}
\frac{M}{\sqrt{3\alpha}}\left|\frac{\partial_r\Phi}{\Phi}\right|\ll 1~~\mathrm{and}~~ \frac{M^2}{3\alpha} \left|\frac{\partial_r^2\Phi}{\Phi}\right| \ll 1~~\mathrm{for}~~r\gg 1~~.
\een
Using relations \eqref{asparPhi} shows that \eqref{cond1} amount to
the conditions:
\ben
\frac{2M}{r^2\sqrt{3\alpha}}\left|\frac{(\partial_\phi\bPhi_p)(\nu)}{\bPhi_p(\nu)}\right|\ll 1~~\mathrm{and}~~ \frac{4M^2}{3\alpha r^3} \left|\frac{(\partial_\phi^2\bPhi_p)(\nu)}{\bPhi_p(\nu)}\right|\ll 1~~,
\een
where $r=\cot(\frac{\phi}{2})$. These conditions are satisfied for $r\gg r_p$, where:
\be
r_p\eqdef \max\left[ \left(\frac{2M}{\sqrt{3\alpha}}\left|\frac{(\partial_\phi\bPhi_p)(\nu)}{\bPhi_p(\nu)}\right|\right)^{1/2} \!\! , \left(\frac{4M^2}{3\alpha}\left|\frac{(\partial_\phi^2\bPhi_p)(\nu)}{\bPhi_p(\nu)}\right|\right)^{1/3}\right]~~.
\ee
In particular, the strong SRST conditions are always satisfied on a
small enough vicinity of a flaring end (when the scalar potential is
well-behaved at that end) and one can use the SRST approximation
there.

\vspace{2mm}

\paragraph{Cusp ends.}

When $p$ corresponds to a cusp end, we have $\epsilon_p=-1$ and
$c_p=2$. In this case, the strong potential SRST conditions hold on a
pointed neighborhood of $p$ provided that, beyond \eqref{cond1}, the
following conditions are also satisfied for $r\gg 1$:
\ben
\label{cond2}
\frac{2\pi M}{\sqrt{3\alpha}}  e^r \left|\frac{\partial_\theta\Phi}{\Phi}\right|\ll 1~~,~~
\frac{(2\pi M) ^2}{3\alpha} e^{2r} \left|\frac{\partial_\theta^2\Phi}{\Phi}\right|\ll 1~~,
~~\frac{2^{3/2} \pi M^2}{3\alpha} e^r \left|\frac{\partial_r \partial_\theta\Phi}{\Phi}\right|\ll 1~~.
\een
As before, conditions \eqref{cond1} are automatically satisfied on a
small enough pointed neighborhood of $p$. On the other hand,
conditions \eqref{cond2} require fine-tuning of the potential $\Phi$
in a neighborhood of the cusp. For example, these conditions are
satisfied in the {\em non-generic} case when $\bPhi_p$ is independent
of $\theta$ on some neighborhood of the north pole of $\rS^2$.

In particular, the strong SRST conditions fail to hold near a cusp end
when $\Phi$ is a generic potential which is well-behaved at that end
and hence the SRST approximation can fail near such ends. This shows
that the SRST approximation is generally not well-suited for studying
cosmological dynamics near cusp ends.

\subsection{On computing power spectra}

\noindent When the SRST approximation holds during the inflationary
period (for example, if inflation takes place near a flaring end), the
power spectrum can be approximated as explained in \cite{m1, m2, m3,
  m4, m5, m6, PT1, PT2} (see \cite{Welling, Gong} for recent
reviews). However, a detailed study of cosmological perturbations in
our models requires going beyond the SRST approximation (in
particular, because the latter can fail near cusp ends).  In general,
this requires the use of a numerical approach such as that developed
in \cite{Dias1, Dias2, Mulryne}. Moreover, for certain classes of
trajectories one might be able to find an effective one-field
description using the approach of \cite{m3}. These subjects lie outside
the scope of the present paper and we hope to address them in future
work.

\subsection{Spiral trajectories near the ends}

\noindent In semi-geodesic coordinates near an end $p\in
\hSigma\setminus \Sigma$, we have:
\be
\dd s_\cG^2\simeq_{r\gg 1} 3\alpha \left[\dd r^2+\left( \frac{c_p}{4\pi }\right) ^{2}e^{2\epsilon_p r} \dd\theta^2\right]~~,
\ee
where $c_p$ and $\epsilon_p$ are given in equation \eqref{Cdef} and the ideal
point $p$ corresponds to $r\rightarrow +\infty$. Using this asymptotic
form, equations \eqref{Christoffel} give:
\be
\Gamma _{\theta \theta }^{r} =-3\epsilon \alpha \left( \frac{c_p}{4\pi }
\right) ^{2}e^{2\epsilon_p r}~~, ~~
\Gamma _{r\theta }^{\theta } = \Gamma^\theta_{\theta r}=\epsilon_p~~.
\ee
Substituting in \eqref{eom}, we find that the equation of motion
in a vicinity of the end reduces to the system:
\beqan
&&\ddot{r}-3\epsilon \alpha \left( \frac{c_p}{4\pi }\right) ^{2}e^{2\epsilon_p r}
\dot{\theta}^{2}+3H\dot{r}+\frac{1}{3\alpha }\partial _{r}\Phi =0~~,\label{1} \\
&&\ddot{\theta}+2\epsilon_p \dot{r}\dot{\theta}+3H\dot{\theta}+\frac{1}{3\alpha }
\left(\frac{4\pi }{c_p}\right) ^{2}e^{-2\epsilon_p r}\partial _{\theta }\Phi
=0~~.  \label{2}
\eeqan
The generic solution of this system has $\dot{r}\neq 0$ and
$\dot{\theta}\neq 0$, thus being a portion of a spiral which ``winds''
around the ideal point. This gives a form a spiral inflation in our
class of models. As already clear from Subsection
\ref{subsec:SRSTEnds}, cosmological dynamics depends markedly on the
type of end under consideration, i.e. on the values of $\epsilon_p$
and $c_p$ allowed by equation \eqref{Cdef}. Note that this type of
spiral inflation differs from that discussed in
\cite{Barenboim,Barenboim-V} (since, unlike the present paper,
loc. cit. considers two-field cosmological models with flat target
space). With the exception of the single case when $(\Sigma,G)$ is the
Poincar\'e disk $\mD$, the type of spiral inflation discussed here
also differs from that studied in references
\cite{hspiral1,hspiral2,hspiral3}, where only the Poincar\'e disk was
considered. As explained in Appendix \ref{subsec:endmetric}, the
Poincar\'e disk $\mD$ is the {\em only} geometrically finite
hyperbolic surface which admits a plane end. Hence the case $c_p=2\pi$
with $\epsilon_p=+1$ in equations \eqref{1} and \eqref{2} (which makes
those equations consistent with the equations of motion considered in
references \cite{hspiral1,hspiral2,hspiral3}) arises {\em only} for
those spiral trajectories on the Poincar\'e disk $\mD$ which are close
to the conformal boundary of $\mD$.

\

\noindent Let us consider the system \eqref{1}-\eqref{2} for various types of ends.

\begin{itemize}
\itemsep 0.0em
\item For flaring ends, we have $\epsilon_p=+1$ and the equations of
  motion take the form:
\beqan
&& \ddot{r}-3\alpha \left( \frac{c_p}{4\pi }\right) ^{2}e^{2r}\dot{\theta}^{2}+3H
\dot{r}+\frac{1}{3\alpha }\partial _{r}\Phi  =0 \label{1-1} ~~,\\
&& \ddot{\theta}+2\dot{r}\dot{\theta}+3H\dot{\theta} \simeq 0~~,
\label{2-1}
\eeqan
where:
\begin{equation*}
H = \frac{1}{\sqrt{6} M} \sqrt{\frac{3 \alpha}{2}\left[\dot{r}^2 + \left(\frac{c_p}{4\pi}\right)^2 e^{2 r} \dot{\theta}^{2}\right] + 2 \Phi(r,\theta)}~~.
\end{equation*}
Assuming $\dot{\theta}\neq 0$, we can write \eqref{2-1} as:
\begin{equation*}
\frac{\dd}{\dd t}\ln \dot{\theta} =-\left( 2\dot{r}+3H\right)~~.
\end{equation*}

\item For cusp ends, we have $\epsilon_p=-1$ and $c_p=2$, thus
  the equations of motion become:
\beqan
\ddot{r}+3H\dot{r}+\frac{1}{3\alpha}\partial _{r}\Phi  &\simeq &0  \label{1-2}~~, \\
\ddot{\theta}-2\dot{r}\dot{\theta}+3H\dot{\theta}+\frac{4\pi ^{2}}{3\alpha }
e^{2r}\partial _{\theta }\Phi  &=&0~~,  \label{2-2}
\eeqan
where:
\begin{equation*}
H = \frac{1}{\sqrt{6} M} \sqrt{\frac{3 \alpha}{2}\left[\dot{r}^{2} + \frac{e^{-2 r}}{(2\pi)^2}  \dot{\theta}^2\right] + 2 \Phi(r,\theta)}~~~.
\end{equation*}
\end{itemize}

\paragraph{Remark.}
As already mentioned above, certain types of spiral trajectories in
two-field cosmological models whose scalar manifold $(\Sigma,G)$ is
the Poincar\'e disk $\mD$ were discussed in references
\cite{hspiral1,hspiral2,hspiral3}, which consider various explicit
scalar potentials and/or make certain special assumptions. Unlike the
present paper, none of those references discusses spiral trajectories
near the ends of a hyperbolic surface which differs from the
Poincar\'e disk, hence the results of loc. cit. cannot apply without
modification\footnote{Reference \cite{hspiral1} considered spiral
  trajectories on the Poincar\'e disk $\mD$ in the so-called {\em
    angular inflation} regime, an approximation in which the radial
  coordinate $r$ is fixed while the slow-roll approximation is assumed
  to apply to the angular coordinate $\theta$; notice that such an
  approximation can only apply to special portions of certain
  particular spiral trajectories nearby the conformal boundary of
  $\mD$. Reference \cite{hspiral2} studies so-called ``sidetracked
  inflation'' in two-field models, an inflationary scenario which is
  highly sensitive to the precise form of the target space metric
  $\cG$ and of the scalar potential $\Phi$ of the model. For the
  special case of the Poincar\'e disk and using a rather special form
  for the scalar potential, loc. cit. shows that the background
  dynamics can be described by a certain effective single field
  model. Those conclusions of \cite{hspiral2} are quite
  model-dependent and it is unclear to what extent they could apply to
  the much more general situation considered herein, given that our
  scalar potentials are considerably more general than those
  considered in loc. cit. and since the local form of the hyperbolic
  metric close to an end of a geometrically finite hyperbolic surface
  which differs from the Poincar\'e disk depends on whether that end
  is a cusp, a funnel or a horn end (see Appendix
  \ref{subsec:endmetric}) and differs from the hyperbolic metric on
  $\mD$. Finally, the brief note \cite{hspiral3} points out some
  general features of inflation in models whose target space is the
  Poincar\'e disk $\mD$, features which are a direct consequence of the
  special behavior of auto-parallel curves on $\mD$. It is well-known
  that auto-parallel curves on $\mD$ behave quite differently from
  auto-parallel curves on more general hyperbolic surfaces.}
for the cusp, funnel or horn ends (see Appendix
\ref{subsec:endmetric}) of a geometrically finite hyperbolic surface
$(\Sigma,G)$ which differs from $\mD$. Since the Poincar\'e disk is
the {\em only} geometrically finite hyperbolic surface which has a
plane end, all other cases of spiral inflation considered above have
not yet been discussed in the literature. We stress once again that
the class of hyperbolic surfaces considered in this paper is vastly
more general than the Poincar\'e disk and that the geometry of such
surfaces differs markedly from that of $\mD$ --- as illustrated by the
presence of the parameters $c_p$ and $\epsilon_p$ in the equations of
motion \eqref{1} and \eqref{2} above and by the different behavior of
the SRST conditions at cusp and flaring ends. In particular, the
parameters $c_p$ and $\epsilon_p$ (which are defined in equation
\eqref{Cdef}) depend on the hyperbolic type of the end of $\Sigma$
near which the trajectory spirals and they lead to different
cosmological dynamics near each type of end. It would be interesting
to adapt some of methods and ideas of references
\cite{hspiral1,hspiral2,hspiral3} to the wider setting of spiral
trajectories nearby cusp, funnel and horn ends of a general hyperbolic
surface and to apply them to the special types of trajectories and
scalar potentials to which they might be relevant. More conceptually,
one can ask what is the class of trajectories in two-field generalized
$\alpha$-attractor models to which one could apply the methods of
\cite{m3} (which in certain cases allow one to build an effective
one-field description of two-field models) --- a problem which we hope
to address in future work. However (given that all approximation
methods which are currently available for two-field models depend on
various assumptions) we are of the opinion that a complete
understanding of general trajectories and of the corresponding
cosmological perturbations for the class of two-field models
considered in the present paper requires a numerical approach.

\subsection{An example of spiral trajectory near a cusp end}

\begin{figure}[H]
\centering
\begin{minipage}{.44\textwidth}
\centering
\includegraphics[width=.77\linewidth]{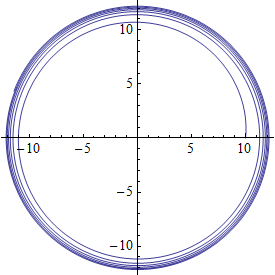}
\captionof{figure}{Numerical plot of the pair $(r(t)\cos\theta(t), r(t)\sin\theta(t))$ 
(with the initial conditions indicated in the text) for $t\in [0,50]$.}
\label{fig:spiralcusp1}
\end{minipage}\hfill
\begin{minipage}{.45\textwidth}
\centering
\includegraphics[width=.77\linewidth]{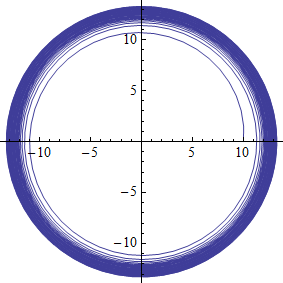}
\caption{Numerical plot of solution $(r(t)\cos\theta(t), r(t)\sin\theta(t))$ 
(with the initial conditions indicated in the text) for $t\in [0,500]$.}
\label{fig:spiralcusp2}
\end{minipage}
\end{figure}

\noindent In Figures \ref{fig:spiralcusp1} and \ref{fig:spiralcusp2},
we show a numerical solution of the system \eqref{1-2}-\eqref{2-2},
which corresponds to a spiral trajectory near a cusp end (for such an
end, one has $c_p=2$ and $\epsilon_p=-1$ by equation
\eqref{Cdef}). For simplicity, we took $\Phi=0$ and
$\alpha=\frac{1}{3}$, $r(0) = 10$, $\dot{r}(0)= 1$, $\theta(0) = 0$
and $\dot{\theta}(0)=1$. Both figures represent the same solution, but
in Figure \ref{fig:spiralcusp1} we follow the trajectory for a shorter
time for reasons of clarity of the plot.

\vspace{-3mm}

\begin{figure}[H]
\centering
\begin{minipage}{.44\textwidth}
\centering \includegraphics[width=0.9\linewidth]{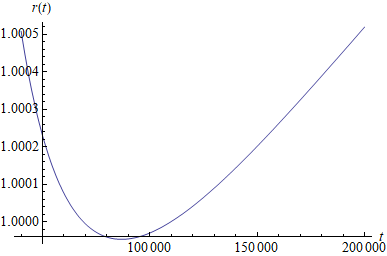}
\captionof{figure}{Plot of $\frac{1}{6.8580}\frac{r(t)}{(\ln
    t)^{0.3411}}$ as a function of $t$ for $t/10^4\in [4, 20]$. The
  plot shows that $r(t)$ increases at least as fast as $(\ln
  t)^{0.3411}$ with a $0.5\%$ discrepancy.}
\label{fig:rlog}
\end{minipage}
\hfill
\begin{minipage}{0.44\textwidth}
\vskip 1em
\!\!\!\!\!\!\!\!\!\!\includegraphics[width=\linewidth]{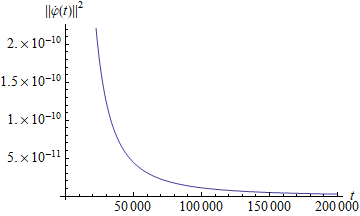}
\caption{Plot of $||\dot{\varphi}||_\cG^{2}$ as a function of $t$. The
  plot shows that the norm of $\dot{\varphi}$ decreases toward zero
  as the trajectory spirals around the cusp end.}
\label{fig:speedtotal}
\end{minipage}
\end{figure}

\noindent Figures \ref{fig:rlog} and \ref{fig:speedtotal} show the increase of $r$ and 
the decrease of the norm of $\dot{\varphi}$ with the cosmological time $t$,
for the trajectory with the initial conditions given above. 

\paragraph{Remark.}
Other examples of spiral trajectories (including their global behavior away
from cusp, funnel and plane ends) in certain explicit generalized
two-field $\alpha$-attractor models can be found in references
\cite{elem,modular}.

\section{Lift of generalized two-field $\alpha$-attractor models to the disk and upper half plane}
\label{sec:lift}

This section makes free use of mathematical terminology and material summarized in Appendix B. 

\subsection{Lift to the Poincar\'e disk}

\noindent Let $\Delta\subset \PSU(1,1)$ be the uniformizing group of $(\Sigma,G)$ 
to the disk and $\pi_\mD:\mD\rightarrow \Sigma=\mD/\Delta$
denote the holomorphic covering map. Let ${\tilde \Phi}\eqdef \Phi\circ
\pi_\mD:\mD\rightarrow \R$ denote the lift of the scalar potential to $\mD$, 
which is $\Delta$-invariant:
\be
{\tilde \Phi}(\delta u)={\tilde \Phi}(u)~~,~~\forall u\in \mD\, , \quad\forall \delta \in \Delta~~.
\ee
Let $||\cdot||_\mD$ denote the fiberwise norm induced by the Poincar\'e
metric on the tangent bundle $T\mD$ and $\grad_\mD\tPhi$ denote the
gradient of $\tPhi$ with respect to the Poincar\'e metric.

Let ${\tilde \varphi}_0\in \mD$ be a lift of the initial
value $\varphi_0\in \Sigma$, i.e. a point of the unit disk which
satisfies:
\be
\pi_\mD(\tvarphi_0)=\varphi_0~~.
\ee
Since $\pi_\mD$ is a local diffeomorphism, the differential
$\dd_{{\tilde \varphi}_0}\pi_\mD:T_{{\tilde \varphi}_0}\mD\rightarrow
T_{\varphi_0} \Sigma$ is bijective. Hence there exists a unique
tangent vector ${\tilde v}_0\in T_{{\tilde \varphi}_0}\mD$ which
satisfies:
\be
(\dd_{{\tilde \varphi}_0}\pi_\mD)({\tilde v}_0)=v_0~~.
\ee

\paragraph{Proposition.}
Let $\cG=3\alpha G$. Then the solution $\varphi:\fI\rightarrow \Sigma$
of \eqref{eomsingle} satisfying the initial conditions \eqref{ini} can
be written as $\varphi=\pi_\mD\circ \tvarphi$, where
$\tvarphi:\fI\rightarrow \mD$ is a solution of the equation:
\ben
\label{eomsinglel}
\nabla_t \dot{\tvarphi}(t)+\frac{1}{M} \sqrt{\frac{3}{2}} \left[3\alpha ||\dot{\tvarphi}(t)||_\mD^2+2\tPhi(\tvarphi(t))\right]^{1/2}\dot{\tvarphi}(t) +\frac{1}{3\alpha}(\grad_\mD \tPhi)(\tvarphi(t))=0
\een
which satisfies the initial conditions:
\ben
\label{inil}
\tvarphi(t_0)={\tilde \varphi}_0~~,~~\dot{\tvarphi}(t_0)={\tilde v}_0~~.
\een
The solution $\tvarphi:\fI\rightarrow \mD$ of \eqref{eomsinglel}
is called the {\em lift} of the solution $\varphi:\fI\rightarrow
\Sigma$ through the point $\tvarphi_0\in \pi_\mD^{-1}(\{\varphi_0\})$.

\vspace{2mm}

\paragraph{Proof.}
Since $\pi_\mD$ is a universal covering map, the smooth curve
$\varphi:\fI\rightarrow \Sigma$ which satisfies
$\varphi(t_0)=\varphi_0$ lifts uniquely to a curve ${\tilde
  \varphi}:\fI\rightarrow \mD$ satisfying $\pi_\mD\circ {\tilde
  \varphi}=\varphi$ and ${\tilde \varphi}(t_0)={\tilde
  \varphi}_0$. Moreover, the condition $\dot{\varphi}(t_0)=v_0$ is
equivalent with $\dot{\tilde \varphi}(t_0)={\tilde v}_0$ since
$\pi_\mD$ is a local diffeomorphism. The lift $\tilde{\varphi}$
satisfies equation \eqref{eomsinglel} since $\varphi$ satisfies
\eqref{eomsingle} while $\pi_\mD$ is a local isometry between
$(\mD,3\alpha G_\mD)$ and $(\Sigma,\cG)$, where $G_\mD$ is the
Poincar\'e metric of $\mD$.

\vspace{2mm}

\paragraph{Remark.}
Since $\Delta$ is an infinite discrete group, there exists a countable
infinity of choices of lifts $\tvarphi_0$ of $\varphi_0$. These lifts
form an orbit of $\Delta$ on $\mD$. The set of accumulation points of
this orbit coincides with the limit set $\Lambda_\mD\subset
\partial_\infty \mD$ of $\Delta$.

\vspace{2mm}

In semi-geodesic coordinates $u=\tanh(\frac{r}{2})e^{i\theta}$ on
$\mD$, the Poincar\'e metric has the form \eqref{semig}
with $H(r)=\sinh(r)$. Using \eqref{Christoffel}, we find:
\be
\Gamma^{r}_{\theta\theta}=-\frac{1}{2}\sinh(2 r)~~,~~\Gamma^{\theta}_{r\theta}=\Gamma^{\theta}_{\theta r}=\coth(r)~~,
\ee
which gives:
\be
\nabla_t \dot{r}=\ddot{r}-\frac{1}{2}\sinh(2r)\dot{\theta}^2~~,~~\nabla_t \dot{\theta}=\ddot{\theta}+2\coth(r)\dot{r}\dot{\theta}~~.
\ee
Thus \eqref{eomsinglel} takes the following form in semi-geodesic
coordinates on $\mD$:
\beqan
\label{eldisk}
&& \ddot{r}-\frac{1}{2}\sinh(2r)\dot{\theta}^2 +\frac{1}{M} \sqrt{\frac{3}{2} \left[ 3\alpha(\dot{r}^2+\sinh^2(r)\dot{\theta}^2)+2\tPhi(r,\theta)\right]}\dot{r} +\frac{1}{3\alpha}\partial_r\tPhi(r,\theta)=0~~,\nn\\
&& \ddot{\theta}+2\coth(r)\dot{r}\dot{\theta}+\frac{1}{M} \sqrt{\frac{3}{2} \left[ 3\alpha(\dot{r}^2+\sinh^2(r)\dot{\theta}^2)+2\tPhi(r,\theta)\right]}\dot{\theta}+\frac{1}{3\alpha}\frac{\partial_\theta\tPhi(r,\theta)}{\sinh^2(r)}=0~.
\eeqan

\subsection{Lift to the upper half plane}

\noindent One can also lift to the upper half plane using a holomorphic covering
map $\pi_\H:\H\rightarrow \Sigma=\H/\Gamma$ (recall that
$\pi_\mD=\pi_\H\circ f$ and $\Delta=Q^{-1}\Gamma Q$, where $f$ is the
Cayley map \eqref{f} and $Q$ is the Cayley element \eqref{Qdef}). The
Poincar\'e metric of $\H$ in Cartesian coordinates $x=\Re \tau$ and
$y=\Im \tau$ takes the form \eqref{sHCart} and has Christoffel
symbols:
\be
\Gamma_{xx}^y=\frac{1}{y}~~,~~\Gamma_{yy}^y=\Gamma_{xy}^x=\Gamma_{yx}^x=-\frac{1}{y}~~,
\ee
which gives:
\be
\nabla_t \dot{x}=\ddot{x}-\frac{2}{y}\dot{x}\dot{y}~~,~~\nabla_t{\dot y}=\ddot{y}+\frac{1}{y}(\dot{x}^2-\dot{y}^2)~~.
\ee
Thus \eqref{eomsinglel} takes the following form in Cartesian coordinates on $\H$:
\beqan
\label{elplane}
&& \ddot{x}-\frac{2}{y}\dot{x}\dot{y} +\frac{1}{M} \sqrt{\frac{3}{2}} \left[3\alpha \frac{\dot{x}^2+\dot{y}^2}{y^2}+2\tPhi(x,y)\right]^{1/2}\dot{x}+\frac{1}{3\alpha} y^2 \partial_x\tPhi(x,y)=0 ~~,\\
&& \ddot{y}+\frac{1}{y}(\dot{x}^2-\dot{y}^2)+\frac{1}{M} \sqrt{\frac{3}{2}} \left[3\alpha \frac{\dot{x}^2+\dot{y}^2}{y^2}+2\tPhi(x,y)\right]^{1/2}\dot{y}+ \frac{1}{3\alpha} y^2\partial_y\tPhi(x,y)=0~~.\nn
\eeqan

\vspace{2mm}

\paragraph{Remark.}
Given that every generalized two-field $\alpha$-attractor model can be lifted to
$\H$, it may naively appear that we have gained nothing by considering
general geometrically finite hyperbolic surfaces. Notice, however,
that the covering map $\pi_\H$ is quite non-trivial, because
$\Gamma\subset \PSL(2,\R)$ is an {\em infinite} discrete group, whose
orbits have accumulation points on the set $\Lambda_\H\subset
\partial_\infty \H$. In practice, the effect of the projection through
$\pi_\H$ can be described by computing a fundamental
polygon\footnote{For example, a Dirichlet domain for $\Delta$
  centered at the origin of $\mD$, which is known as a Ford domain.}
for the action of $\Gamma$ on $\H$ (see below). Determining a
fundamental polygon for a given Fuchsian group is a classical but
highly non-trivial problem. When $\Sigma$ has finite hyperbolic area
(i.e., when $\Gamma$ is of the first kind) and when the elements of
$\Gamma$ have entries belonging to an algebraic extension of $\Q$, a
stopping algorithm for finding a Ford domain was given in
\cite{Voight}. When $\Sigma$ has infinite hyperbolic area, no general
stopping algorithm for computing a fundamental polygon of $\Gamma$ is
currently known. Some models for which a fundamental polygon (as well as the 
uniformizing map) is explicitly known are discussed in detail in references 
\cite{elem, modular}.

\subsection{Describing the projection using a fundamental polygon}
\label{subsec:liftalg}

\noindent Let $\Delta\subset \PSU(1,1)$ be the uniformizing group of a
geometrically finite hyperbolic surface $(\Sigma,G)$. Provided that a
fundamental polygon $\fD_\mD\subset \mD$ of $\Delta$ is known, the
integral curves of \eqref{eomsingle} for $\cG=3\alpha G$ can be
determined by computing their lift to the interior of $\fD_\mD$ as
follows:
\begin{enumerate}
\itemsep 0.0em
\item The $\Delta$-orbit of the lifted initial value $\tvarphi_0\in
  \mD$ intersects $\fD_\mD$ in exactly one point. Hence by applying a
  $\Delta$-transformation if necessary, we can assume without loss of
  generality that $\tvarphi_0$ belongs to $\fD_\mD$. By slightly
  changing the initial time $t_0$, we can assume that $\tvarphi_0$
  belongs to the interior $\Int \fD_\mD$ of $\fD_\mD$.
\item Assuming $\tvarphi_0\in \fD_\mD$, compute a solution $\ttvarphi$
  of equation \eqref{eomsinglel} with initial conditions
  \eqref{inil}. Stop the computation at the first value $t_1>t_0$ for
  which $\ttvarphi(t)$ meets the relative frontier $\Fr\fD_\mD$ of
  $\fD_\mD$ in $\mD$.
\item Since $\fD_\mD$ is a fundamental polygon, there exists a unique element
  $\delta_1\in \Delta$ which maps the point $\ttvarphi(t_1)\in \Fr \fD_\mD$
  into a point $\tvarphi_1\eqdef \delta_1 \ttvarphi(t_1)$ which lies on
  $\Fr\fD_\mD$. Since $\delta_1$ acts as an isometry of $(\mD,\dd
  s^2_\mD)$, the velocity vector $\dot{\ttvarphi}(t_1)\in
  T_{\ttvarphi(t_1)}\mD$ is transported by $\delta_1$ to a congruent
  vector $\tilde{v}_1\eqdef (\delta_1)_\ast (\dot{\ttvarphi}(t_1))\in
  T_{\tvarphi_1}\mD$ with origin at the point $\tvarphi_1$.
\item Now compute a solution $\ttvarphi(t)$ of equation
  \eqref{eomsinglel} with initial conditions
  $\ttvarphi(t_1)=\tvarphi_1$ and $\dot{\ttvarphi}(t_1)=\tilde{v}_1$ and
  follow it to the first time $t_2>t_1$ for which $\ttvarphi(t_2)$ lies
  in the relative frontier of $\fD_\mD$. Then repeat the process above.
\end{enumerate}

\noindent This algorithm produces an infinite sequence of times
$t_0<t_1<t_2<t_3<\ldots $ and a map $\ttvarphi:[t_0,a)\setminus
T\rightarrow \fD_\mD$ (where $T\eqdef \{t_i|i\geq 1\}$ and $a\in
\R\sqcup\{+\infty\}$ satisfies $a\geq \sup T$) which is smooth inside
each interval $(t_i,t_{i+1})$ ($i\geq 0$) and satisfies
$\ttvarphi(t_0)=\tvarphi_0$ and
$\dot{\ttvarphi}(t_0)=\tilde{v}_0$. Moreover, it produces a sequence
of elements $\delta_i\in \Delta$ (with $i\geq 1$) such that
$\delta_i\ttvarphi(t_i-)=\ttvarphi(t_i+)$ and $(\delta_i)_\ast
(\dot{\ttvarphi}(t_i-))=\dot{\ttvarphi}(t_i+)=\tilde{v}_i$ for all
$i\geq 1$, where $\ttvarphi(t_i-)$ and $\ttvarphi(t_i+)$ denote the
lower and upper limits of $\ttvarphi$ at the point $t_i$, with a
similar notation for the partial limits of $\dot{\ttvarphi}$. It is
clear from the above that $\ttvarphi$ provides a lift of the forward
solution $\varphi:\fI\cap [t_0,+\infty)\rightarrow \Sigma$ to the
interior of the fundamental polygon $\fD_\mD$. In general, the
algorithm must be continued indefinitely to obtain a full lift to
$\Int\fD_\mD$, because the tessellation of $\mD$ by $\Delta$-images of
the fundamental polygon has an infinite number of tiles. In
particular, the errors in a numerical implementation of this algorithm
will accumulate as $i$ grows and the precision of the lifted
trajectory provided by a numerical solver will decrease for large $t$.

A different but equivalent approach is to compute a single full
trajectory $\tvarphi$ of \eqref{eomsinglel} on $\mD$ and then
determine its projection to $\Sigma$ by computing a tessellation of
$\mD$ through $\Delta$-images of the fundamental domain
$\fD_\mD$. Since $\Delta$ is an infinite group, the tessellation has
an infinite number of tiles, which accumulate toward the conformal
boundary $\partial_\infty\mD$. As a consequence, the error made in
determining the projection $\varphi(t)$ of $\tvarphi(t)$ to $\Sigma$
increases as $\tvarphi(t)$ approaches $\partial_\infty\mD$, i.e when
$\varphi(t)$ approaches an ideal point of $\Sigma$. In this region,
one can use instead the explicit form of the hyperbolic metric on a
small punctured neighborhood of the ideal point (see Appendix
\ref{subsec:endmetric}), which allows one to compute an approximate
trajectory on such a neighborhood, thus avoiding the problem of
accumulating tiles. Of course, similar algorithms can be formulated
for the lift to the upper half plane.

\subsection{Characterization of potentials which are well-behaved at a cusp or funnel end}

\noindent Let $\Phi:\Sigma\rightarrow \R$ be a smooth function defined
on $\Sigma$ and $\tPhi:\H\rightarrow \R$ be its $\Gamma$-invariant
lift to $\H$. Let $\fD_\H\subset \H$ be a fundamental polygon for the
uniformizing group $\Gamma\subset \PSL(2,\R)$. The construction of
Appendix \ref{subsec:cfcoords} implies:
\begin{itemize}
\itemsep 0.0em
\item $\Phi$ is well-behaved at a cusp ideal point $p\in
  \hSigma\setminus \Sigma$ iff there exists a smooth function
  $\hPhi_p:D\rightarrow \R$ defined on the disk $D\eqdef\{z\in
  \C|\,|z|<e^{-2\pi}\}$ such that, for some (equivalently, any) ideal
  vertex $v$ of $\fD_\H$ which corresponds to $p$, the restriction of
  the lift $\tPhi$ to a relative cusp neighborhood $\fC_v$ of $v$ in
  $\fD_\H$ has the form:
\be
\tPhi(\tau)=\hPhi_p(z_p(\tau))~~(\tau \in \fC_v)~~,
\ee
where $z_p(\tau)$ is the cusp coordinate \eqref{cuspcoord} near $p$.
\item $\Phi$ is well-behaved at a funnel ideal point $p\in
  \hSigma\setminus \Sigma$ iff there exists a smooth function
  $\hPhi_p:\mD\rightarrow \R$ defined on the unit disk $\mD=\{z\in
  \C||z|<1\}$ such that, for some (equivalently, any) free side $E$ of
  $\fD_\H$ which corresponds to $p$, the restriction of the lift
  $\tPhi$ to a relative funnel neighborhood $\fF_E$ of $E$ in $\fD_\H$
  has the form:
\be
\tPhi(\tau)=\hPhi_p\left(\frac{|z_p(\tau)|-\frac{1}{R_p}}{(1-\frac{1}{R_p})|z_p(\tau)|}z_p(\tau)\right)~~,
\ee
where $z_p(\tau)$ is the funnel coordinate \eqref{funnelcoord}. Here,
$R_p=e^{\frac{\pi^2}{\ell_p}}$ is the displacement radius \eqref{Relldef}, where
$\ell_p$ is the width of the hyperbolic funnel corresponding to $p$.
\end{itemize}
The condition for the cusp ideal points is obvious, while that for
funnel ideal points follows from the fact that, for each $R>1$, the
map $z\rightarrow \frac{|z|-\frac{1}{R}}{(1-\frac{1}{R})|z|}z$ is a
diffeomorphism from the annulus $\A(R)$ of inner radius $1/R$ and outer
radius $1$ to the punctured unit disk $\mD^\ast$.

\section{Morse theory and multi-path inflation}
\label{sec:Morse}

In this section, we give a qualitative general discussion of the
cosmological model in the gradient flow approximation (see Section
\ref{sec:ES}) for the case when the scalar potential is globally
well-behaved and the scalar manifold is geometrically finite and
non-elementary (i.e. it is not a disk, a punctured disk or an
annulus). In the gradient flow approximation, the qualitative features
of the model are determined by the critical points and level sets of
the scalar potential $\Phi$. Let $(\Sigma,G)$ be a geometrically
finite non-elementary surface and $\hSigma=\Sigma\sqcup\{p_1,\ldots,
p_n\}$ be its end compactification (see Appendix
\ref{sec:topfinite}). This section assumes some familiarity with
elementary results in Morse theory. These are briefly recalled in the
text for convenience of the reader.

\subsection{Compactly Morse potentials}

\noindent Recall that a smooth function $f:\hSigma\rightarrow \R$ is
called {\em Morse} if all its critical points have non-degenerate
Hessian. In this case, the critical points are isolated and they must
form a finite set since $\hSigma$ is compact. Morse functions are
generic in the space $\cC^\infty(\hSigma,\R)$ of all smooth functions
from $\hSigma$ to $\R$ (when the latter is endowed with the
compact-open topology) in the sense that they form an open and dense
subset of that topological space. Hence any smooth function
$f:\hSigma\rightarrow \R$ can be deformed to a Morse function by an
infinitesimally small perturbation.

We say that a smooth function $\Phi:\Sigma\rightarrow \R$ is {\em
  compactly Morse} if there exists a smooth Morse function
$\hPhi:\hSigma\rightarrow \R$ such that $\Phi=\hPhi|_\Sigma$ and such
that $p_1,\ldots, p_n$ are critical points of $\hPhi$. In this case,
the remaining critical points of $\hPhi$ lie inside $\Sigma$ and they
form the critical set $\Crit \Phi$ of $\Phi$. We have:
\be
\Crit \hPhi=\Crit\Phi \sqcup \{p_1,\ldots,p_n\}~~,
\ee
where $\Crit\hPhi$ denotes the critical set of $\hPhi$. Notice that
compactly Morse potentials are globally well-behaved in the sense of
Subsection \ref{subsec:wb}.  In particular, the extension $\hPhi$ of
$\Phi$ to $\hSigma$ is uniquely determined by $\Phi$ through
continuity. Since any smooth real-valued function defined on $\hSigma$
can be infinitesimally deformed to a Morse function, any well-behaved
scalar potential on $\Sigma$ can be infinitesimally deformed to a
compactly Morse potential. Since infinitesimal deformations cannot be
detected observationally, one can safely assume that a
physically-realistic well-behaved scalar potential is compactly Morse.

\subsection{Reconstructing the topology of $\Sigma$}

\noindent Let $\Phi$ be a compactly Morse potential and $\hPhi$ be its extension
to $\hSigma$. Recall that a critical point $c\in \Crit \hPhi$ is either:
\begin{itemize}
\itemsep 0.0em
\item a local minimum of $\hPhi$, when the Hessian of $\hPhi$ at $c$
  is positive-definite, i.e. when $c$ has Morse index $0$.
\item a saddle point of $\hPhi$, when the Hessian operator of $\hPhi$
  at $c$ has one positive eigenvalue and one negative eigenvalue, i.e
when $c$ has Morse index $1$.
\item a local maximum of $\hPhi$, when the Hessian of $\hPhi$ at $c$
  is negative-definite, i.e. when $c$ has Morse index $2$.
\end{itemize}
The local minima, saddles and local maxima of $\Phi$ coincide with
those local minima, saddles and local maxima of $\hPhi$ which lie in
$\Sigma$. Let ${\hat N}_m$, ${\hat N}_M$ and ${\hat N}_s$ denote the
number of local minima, local maxima and saddles of $\hPhi$ and $N_m$,
$N_M$ and $N_s$ denote the number of local minima, local maxima and
saddles of $\Phi$. Let $n_m$, $n_M$ and $n_s$ be the numbers of ideal
points $p_1,\ldots, p_n$ which are local minima, local maxima and
saddles of $\hPhi$. We have:
\ben
\label{nNrels}
 n=n_m+n_M+n_s~~,~~{\hat N}_m=N_m+n_m~~,~~ {\hat N}_M=N_M+n_M~~,~~
 {\hat N}_s=N_s+n_s~~.
\een
Morse's theorem applied to $\hPhi$ shows that the Euler characteristic
of $\hSigma$ is given by:
\ben
\label{chiMorse}
\chi(\hSigma)=2-2g={\hat N_m}+{\hat N}_M-{\hat N}_s=N_M+N_m-N_s+n_M+n_m-n_s~~,
\een
where $g$ is the genus of $\Sigma$ (which by definition equals the
genus of $\hSigma$). In particular, the quantity $2+{\hat N}_s-{\hat
  N}_M-{\hat N}_m$ is an even non-negative integer and the genus of
$\hSigma$ is given by:
\ben
\label{gMorse}
g=1+\frac{{\hat N}_s-{\hat N}_M-{\hat N}_m}{2}=1+\frac{N_s-N_M-N_m+n_s-n_M-n_m}{2}~~.
\een
Recall that $\Sigma$ is determined up to diffeomorphism by its genus
$g$ and by the number $n=n_M+n_m+n_s$ of its ends. Hence
\eqref{gMorse} implies that the topology of $\Sigma$ can be
reconstructed from knowledge of the numbers $n_M,n_m,n_s$ and
$N_M,N_m$ and $N_s$. In principle, these numbers could be determined
observationally by comparing cosmological correlators computed in the
model with observational data, thus allowing one to reconstruct the
topology of the scalar manifold $\Sigma$.

\subsection{The topological decomposition induced by a compactly Morse potential}

\noindent
A compactly Morse potential $\Phi$ induces a topological
decomposition:
\be
\hSigma=Y_1\cup \ldots \cup Y_{2g-2+n}~~,
\ee
where $Y_j$ are (possibly degenerate) topological pairs of pants,
i.e. surfaces diffeomorphic with a sphere with three holes, where any
of the holes is allowed to degenerate to a point. For any regular
value $a\in \hPhi(\hSigma)-\hPhi(\Crit \hPhi)\subset \R$, the preimage
$\hPhi^{-1}(\{a\})$ is a finite disjoint union of closed connected
curves, each of which is diffeomorphic with a circle. When $a$ is a
critical value such that $\hPhi^{-1}(\{a\})$ contains a single
critical point, one has two possibilities:
\begin{itemize}
\itemsep 0.0em
\item One of the circles degenerates to a single point $p\in
  \hSigma$, which is a local minimum or local maximum of $\hPhi$.
\item Two of the circles touch at a single point $p\in \hSigma$, which
  is a saddle point of $\hPhi$.
\end{itemize}
Two pants of a given pair of pants $Y_j$ meet at a saddle point and
each saddle point lies on a single pair of pants. Moreover, every
local maximum or local minimum of $\hPhi$ corresponds to a degenerate
bounding circle of a pair of pants.

One can isolate the local maxima and local minima of $\hPhi$ by
cutting out small open disks $D_1,\ldots, D_{{\hat N}_M+{\hat N}_m}$
around the local maximum and local minimum points, such that the
closure of each disk $D_k$ does not meet the closure of any other disk
or any other critical point.  Then $Z\eqdef \hSigma\setminus
(\cup_{k=1}^{{\hat N}_M+{\hat N}_m} D_k)$ is a compact surface with
boundary and the restriction of $\hPhi$ to $Z$ is a Morse function
whose critical points are saddles and lie in the interior of
$Z$. Using the Morse decomposition for surfaces with boundary, one can
decompose $Z$ as a union:
\be
Z=Z_1\cup\ldots \cup Z_{2g-2+n}
\ee
of a number $|\chi(\hSigma)|=2g-2+n$ of pairs of pants $Z_j$, all of
whose bounding circles are non-degenerate. This gives a
decomposition:
\be
\hSigma=Z_1\cup \ldots \cup Z_N\cup D_1\cup \ldots \cup D_{{\hat N}_M+{\hat N}_m}~~.
\ee
Notice that this decomposition is entirely determined by the topology of $\Sigma$ and by 
the scalar potential $\Phi$ (since $D_i$ and $Z_j$ are determined by that data) and hence is 
independent of the hyperbolic metric on $\Sigma$ and on the parameter $\alpha$ of the 
cosmological model. 

\subsection{Qualitative features and multipath inflation in the gradient flow approximation}

\noindent The gradient flow approximation allows one to give a
qualitative picture of cosmological trajectories when the potential $\Phi$
is compactly Morse. If the scalar field $\varphi$ starts rolling from
an initial value near a point $p\in \hSigma$ which is a local maximum
of $\hPhi$, then to first approximation the model has inflationary
behavior as long as $\varphi$ lies inside some neighborhood $D$ of
$p$.  In general, inflation ends after $\varphi$ exits $D$, after
which the field generally evolves in a complicated manner radiating
its energy and either settles in one of the local minima of $\Phi$ or
evolves further toward an ideal point which is a local minimum
of $\hPhi$. The full evolution can be quite complicated in general,
since $\Phi$ can have many critical points.

\begin{figure}[H]
\centering \includegraphics[width=60mm]{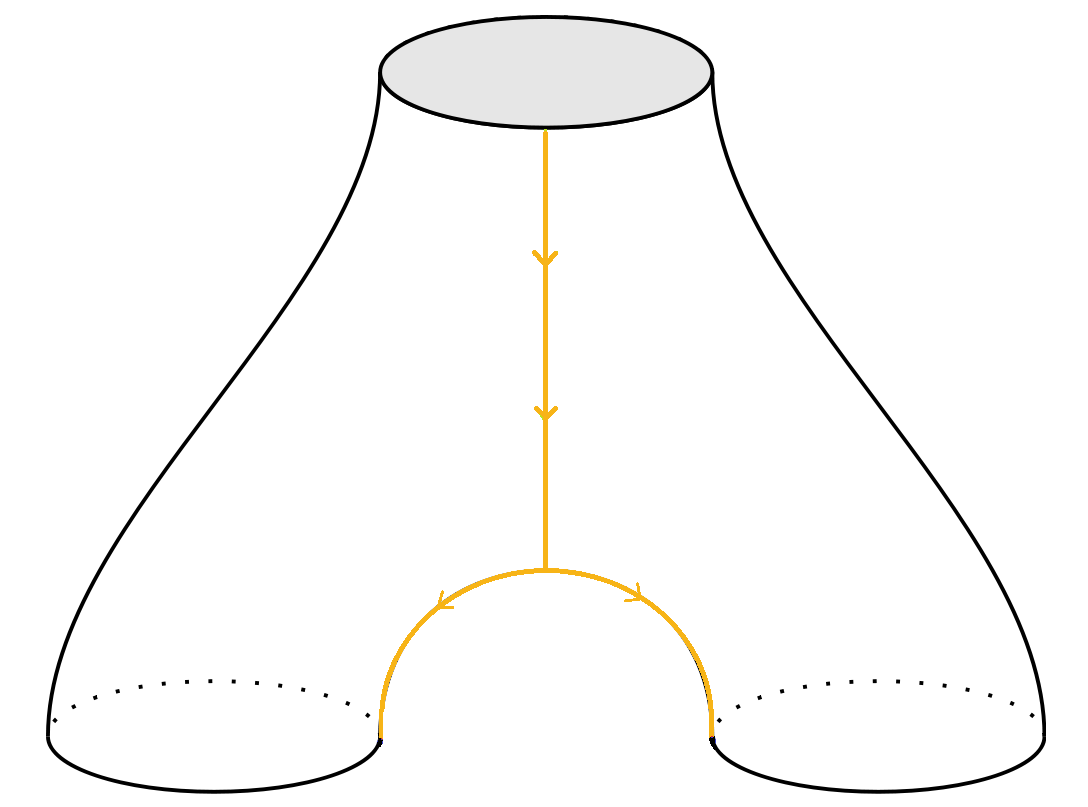}
\caption{Example of a gradient flow trajectory (the orange line) in the
  gradient flow approximation, shown within a topological pair of
  pants determined by the scalar potential
  $\Phi$. }
\label{fig:PantsBifurc}
\end{figure}

The gradient flow of $\Phi$ bifurcates at the saddle points, where the
level sets of $\Phi$ consist of two simple closed curves meeting at
the saddle (see Figure \ref{fig:PantsBifurc}). In the gradient flow
approximation, this implies that certain integral curves will
bifurcate within the topological pants components $Z_j$ determined by
$\Phi$, entering through one of the boundary components of $Z_j$,
hitting the unique saddle point contained inside $Z_j$, then following
one of the two trajectories which start from that saddle point to exit
through either of the other two boundary components of $Z_j$. In
particular, this gives a realization of multipath inflation in our
class of models.

\paragraph{Remark.}
For a detailed study beyond the gradient flow approximation, one could
instead use a {\em hyperbolic} pants decomposition (see Appendix
\ref{subsec:hyppants}).  If a hyperbolic pants decomposition
$N=X_1\cup \ldots \cup X_{2g-2+n}$ of $(\Sigma,G)$ is known, then one
can study the dynamics of the cosmological model separately inside
each pair of hyperbolic pants $X_j$. The latter can be computed by
presenting $X_j$ as the result of gluing two hyperbolic hexagons,
which reduces the problem to studying the model on a hyperbolic
hexagon and making appropriate identifications.

\section{Examples and phenomenological implications}
\label{sec:exphen}
  
Unlike one-field $\alpha$-attractors
\cite{alpha1,alpha2,alpha3,alpha4}, the models considered in this
paper are genuine two-field cosmological models. As such, they allow
for deviations from the traditional paradigm of inflationary
cosmology, which assumes the inflaton to be a single real scalar
field.  Current cosmological data are well accounted for by various
one-field models, but they can also be fit using multi-field
models. It is generally believed that improved measurements in the
future might detect deviations from one-field predictions, a
possibility which lead to renewed interest in multi-field models
\cite{m2,m3,m4,m5,m6,Gong} and in particular to interest in numerical
methods for computing cosmological perturbations in such models
\cite{Dias1,Dias2,Mulryne} beyond the limitations of the SRST
approximation \cite{PT1,PT2}. See \cite{c1} for an investigation of
constraints imposed on two-field models by Planck 2015 data
\cite{Planck}.

As explained in Subsection \ref{subsec:univ}, a generalized two-field
$\alpha$-attractor model based on a non-compact geometrically finite
hyperbolic surface $(\Sigma,G)$ and with a well-behaved scalar
potential $\Phi$ has universal behavior near each end of $\Sigma$
where the extended scalar potential has a local maximum, in a certain
`radial' one-field truncation near that end. Within the slow-roll
approximation for `radial' trajectories close to such an end of
$\Sigma$, these models predict the same spectral index $n_s$ and
tensor to scalar ratio $r$ as ordinary one-field $\alpha$-attractors,
namely (see Section \ref{subsec:univ}):
\be
n_s\approx 1-\frac{2}{N}~~,~~r\approx \frac{12\alpha}{N^2}~~,
\ee
where $N$ denotes the number of e-folds. Hence generalized two-field
$\alpha$-attractors are as promising observationally as ordinary
one-field $\alpha$-attractors, which are known to be in good agreement
with current cosmological data \cite{alpha1,alpha2,alpha3,alpha4}.

In references \cite{elem} and \cite{modular}, we study certain
explicit examples of generalized two-field $\alpha$-attractor models,
which differ topologically from the oft-studied case of the Poincar\'e
disk. Namely, reference \cite{elem} considers generalized two-field
$\alpha$-attractors based on the punctured hyperbolic disk $\mD^\ast$
and on hyperbolic annuli $\mA(R)$ of modulus $\mu=2\ln R>0$, for
certain natural choices of globally well-behaved scalar potentials. On
the other hand, reference \cite{modular} studies generalized two-field
$\alpha$-attractor models based on the hyperbolic triply-punctured
sphere (which is also known as the modular curve $Y(2)$), for a few
natural choices of scalar potential.  References \cite{elem,modular}
give numerous examples of numerically computed cosmological
trajectories for the models considered therein, showing that
generalized two-field $\alpha$-attractors display intricate
cosmological dynamics due to the delicate interplay between the
effects of the hyperbolic metric, the scalar potential and the
topology of the scalar manifold $\Sigma$. The last effect was not
considered previously in the literature --- given that, until now,
studies of two-field $\alpha$-attractors were limited to models based
on the Poincar\'e disk.

In the models of reference \cite{elem}, the universal behavior of
Subsection \ref{subsec:univ} arises for the radial trajectories on the
punctured disk $\mD^\ast$ and on the annulus $\mA(R)$ when inflation
happens close to any of the two components of the conformal boundary.
In fact, reference \cite{elem} gives explicit examples of such
trajectories which produce between $50-60$ e-folds, thereby showing
that generalized two-field $\alpha$-attractor models based on
$\mD^\ast$ or on $\mA(R)$ are compatible with current observational
data. In the same reference, we also give examples of non-geodesic
cosmological trajectories with $50-60$ e-folds. The latter
trajectories are {\em not} of the special type considered in
Subsection \ref{subsec:univ}; this shows that generalized two-field
$\alpha$-attractor models based on the hyperbolic surfaces $\mD^\ast$
and $\mA(R)$ can fit current cosmological data even for trajectories
with genuine two-field behavior.

Generalized two-field $\alpha$-attractor models based on the
hyperbolic triply-punctured sphere $Y(2)$ are discussed in reference
\cite{modular}. As shown in loc. cit., such models are again
compatible with current cosmological constraints. In that reference,
we again display cosmological trajectories producing between $50-60$
e-folds. This shows that generalized two-field $\alpha$-attractor
models based on $Y(2)$ are also compatible with current observational
constraints.

We remark that generalized two-field $\alpha$-attractors are
interesting for studies of post-inflationary dynamics, a problem for
which generic two field models have low predictive power (since they
involve the choice of an arbitrary metric for the scalar manifold
$\Sigma$). Indeed, generalized $\alpha$-attractors form a
mathematically {\em natural} class of two-field models with universal
behavior in the `radially truncated' inflationary regime near the
ends. Moreover, they show remarkable dynamical complexity beyond that
regime, as illustrated in detail in references \cite{elem,
  modular}. Such models could form a useful testing ground for
two-field model technology.

\section{Conclusions and further directions}
\label{sec:conclusions}

We proposed a wide generalization of ordinary two-field
$\alpha$-attractor models obtained by promoting the Poincar\'e disk to
a geometrically finite hyperbolic surface as well as a general
procedure for studying such models through uniformization techniques;
we also discussed certain general aspects of such models, using the
well-developed machinery of uniformization and two-dimensional
hyperbolic geometry. Our generalized models are parameterized by a
positive constant $\alpha$, by the choice of a surface group
$\Gamma\subset \PSL(2,\R)$ and by the choice of a smooth scalar
potential. Despite being extremely general, we showed that such models
have the same universal behavior as ordinary $\alpha$-attractors for
certain special trajectories, in a {\em naive} one-field truncation
near each end and in the slow-roll approximation for such
trajectories, provided that the scalar potential is well-behaved near
that end. We also discussed certain qualitative features of such
models in the gradient flow approximation and commented briefly on
some phenomenological aspects and on some examples which are studied
in detail in references \cite{elem,modular}. The present work sets the
ground for more detailed investigations of such models. There exist
numerous directions for further study of such models, of which we
point out a few.

First, one could investigate cosmological perturbation theory in such
models, adapting the numerical approach developed in \cite{Dias1,
  Dias2, Mulryne}; using numerical methods seems to be necessary
within the compact core of $\Sigma$ as well as near cusp ends (where
the SRST approximation can fail, as showed in Subsection
\ref{subsec:SRSTEnds}). In particular, the algorithm proposed in
Section \ref{sec:lift} could be combined with existing code in order
to perform numerical studies of cosmological correlators. 

Second, one could study in detail the particular case when $\Sigma$ is
a modular curve. When $\Gamma$ is an arithmetic subgroup of
$\PSL(2,\Z)$, an algorithm for finding a fundamental polygon was given
in \cite{Voight} (and was implemented in {\tt Sage} as part of the
{\em KFarey} package \cite{Sage}). In this case, the hyperbolic
surface $(\Sigma,G)$ has only cusp ends and the scalar potential can
be expanded in Maass wave forms \cite{Bump}. In general, this special
subclass of two-field generalized $\alpha$-attractor models leads to
connections with Teichm\"{u}ller theory and number theory and relates
to the framework of automorphic inflation which was developed in
\cite{S1, S2}. A simple example of this type (namely the model whose
scalar manifold is the hyperbolic triply punctured sphere (also known
as the modular curve $Y(2)$) is studied in reference \cite{modular};
as explained there, the generalized two-field $\alpha$-attractor model
having $Y(2)$ as its scalar manifold is related to --- but does {\em not}
coincide with -- the ``modular invariant $j$-model'' considered in
\cite{Sch1,Sch2}.

One could also study the F-term embedding of our theories into $\cN=1$
supergravity with a single chiral multiplet. When $\Gamma$ is an
arithmetic group, this leads to connections with the theory of
automorphic forms. It would be interesting to
study the extension of our models to the case when $\Sigma$ is a
non-orientable surface. In that situation, the uniformizing surface
group $\Gamma$ is replaced by a non-Euclidean crystallographic group
\cite{NEC1, NEC2}. It would also be interesting to consider the case
when $\Sigma$ is not topologically finite.

We refer the interested reader to reference \cite{elem} for a detailed
study of cosmological trajectories in two-field generalized $\alpha$-attractor
models based on elementary hyperbolic surfaces and to \cite{modular}
for a similar study in the model based on the hyperbolic triply
punctured sphere. As shown in those references, it is quite easy in
such models to obtain the observationally favored range of $50-60$
e-folds for various inflationary trajectories. This shows that the
class of models proposed in the present paper could be of
phenomenological interest.  Unlike ordinary two-field
$\alpha$-attractor models (which are based on the Poincar\'e disk),
generalized two-field $\alpha$ attractors can have much richer
post-inflationary dynamics, as illustrated qualitatively in Section
\ref{sec:Morse} of this paper and quantitatively in \cite{elem,
  modular} through numerical computation of various trajectories. The
surprising complexity of such trajectories indicates rich
possibilities which may be of interest to cosmological model building.
While much work remains to be done before ascertaining the
phenomenological viability of such models, it may well be worth
stepping into the world of general hyperbolic surfaces as a natural
special class of scalar manifolds for cosmological two-field models.
Indeed, hyperbolic surfaces form a family of two-dimensional
Riemannian manifolds to which two-field model technology can be
applied in a non-generic manner and usefully combined with the
powerful results of uniformization theory.

\acknowledgments{\noindent The work of C.I.L. was supported by grant
  IBS-R003-S1. C.S.S. thanks the CERN Theory Division for
  hospitality during the preparation of this work and R. Kallosh for
  comments. C.I.L and C.S.S. thank A. Marrani for participation during
  the early stages of the project and for interesting discussions.}

\appendix

\section{Isothermal and semi-geodesic coordinates on Riemann surfaces}
\label{app:SpecCoord}

\noindent In this paper, we often use two special kinds of local coordinate systems on $\Sigma$:

\

\begin{itemize}
\item {\bf Isothermal coordinates}. Let $x=\Re z$ and $y=\Im z$ be the
  real coordinates defined by a local holomorphic coordinate $z$ on a
  complex curve $(\Sigma,J)$. In such coordinates, any Hermitian (and
  hence K\"ahler) metric $\cG$ on $(\Sigma, J)$ takes the local form:
\ben
\label{isothermal}
\dd s^2=\lambda^2(z,\bar{z})|\dd z|^2=\lambda^2(\dd x^2+\dd y^2)~~,~\mathrm{where}~~~ \lambda(z,\bar{z})>0
\een
and the Gaussian curvature of $\cG$ is given by:
\be
K=-\Delta_\cG \log \lambda~~,
\ee
where:
\be
\Delta_\cG=\frac{1}{\lambda^2}\left(\frac{\partial^2}{\partial x^2}+\frac{\partial^2}{\partial y^2}\right)=\frac{4}{\lambda^2}\frac{\partial^2}{\partial z \partial \bar{z}}
\ee
is the Laplacian of $\cG$. The metric $\cG$ is hyperbolic iff $K=-1$,
which amounts to the condition:
\ben
\label{pcond}
\Delta_\cG\log\lambda =1~\Longleftrightarrow ~4 \frac{\partial^2\log
  \lambda}{\partial z \partial \bar{z}}=\lambda^2~\Longleftrightarrow~ \left(\frac{\partial^2}{\partial x^2}+\frac{\partial^2}{\partial
  y^2}\right)\log \lambda=\lambda^2~~.
\een
Any Riemannian metric $\cG$ on an oriented surface $\Sigma$ determines
a unique orientation-compatible complex structure $J$ on $\Sigma$ such
that $\cG$ is K\"ahler with respect to $J$; this complex structure depends only
on the conformal class of $\cG$. In this case, the isothermal
coordinates $x,y$ determined by a local $J$-holomorphic coordinate $z$
defined on an open subset $U\subset \Sigma$ are called {\em local
  isothermal coordinates} for $\cG$. The smooth function
$\lambda:U\rightarrow \R_{>0}$ defined through \eqref{isothermal} in
such coordinates is called the {\em local density of $\cG$} with
respect to $z$. When $\cG$ is hyperbolic, $\lambda$ satisfies
\eqref{pcond} and is called the {\em Poincar\'e density} of $\cG$.

\

\item {\bf Semi-geodesic coordinates.}
Any point of a Riemann surface $(\Sigma,\cG)$ admits an open
neighborhood $U$ supporting semi-geodesic coordinates, in which the
metric takes the form:
\ben
\label{semig}
\dd s^2=\dd r^2+ H(r,\theta)^2 \dd \theta^2~~,~~\mathrm{where}~~H(r,\theta)>0~~.
\een
Such coordinates are characterized by the property that the curves
$\theta\!=$\,constant are geodesics (with natural parameter $r$) which are
orthogonal to the surfaces $r\!=$\,constant. The Gaussian curvature is given by:
\be
K=-\frac{1}{H}\frac{\partial^2 H}{\partial r^2}~~,
\ee
while the non-vanishing Christoffel symbols are \cite[p. 389]{NT}:
\beqan
\Gamma _{\theta \theta }^{r} &=&-H \partial_r H \label{Christoffel}~~,\nn\\
\Gamma _{r\theta }^{\theta } &=& \Gamma^\theta_{\theta r} =\partial_r \log H ~~,\\
\Gamma^\theta_{\theta\theta}&=&\partial_\theta \log H~~.\nn
\eeqan
The metric $\cG$ is hyperbolic when $K=-1$, which amounts to the condition:
\be
\frac{\partial^2 H}{\partial r^2}=H~~.
\ee
\end{itemize}

\

\section{Uniformization of smooth Riemann surfaces}
\label{sec:unif}

\noindent In this appendix, we summarize some well-known results of
uniformization theory and of the theory of hyperbolic surfaces which
are used in various sections of the paper. Some standard references
for this classical subject are
\cite{Borthwick,Beardon,Katok,FrenchelNielsen}.

\subsection{Hyperbolic surfaces}
\label{subsec:unif_basics}

\noindent Let $\Sigma$ be an oriented surface, which need not be compact.  A
complete Riemannian metric $G$ on $\Sigma$ is called {\em hyperbolic}
if its Gaussian curvature $K(G)$ equals $-1$. In this case, the pair
$(\Sigma,G)$ is called a {\em hyperbolic surface}. Each conformal
class $\fc$ on $\Sigma$ contains at most one hyperbolic metric. The
uniformization theorem of Koebe and Poincar\'e \cite{unif} implies
that such a hyperbolic metric exists in $\fc$ iff $(\Sigma,J)$ is
covered holomorphically by the unit disk, where $J$ is the
orientation-compatible complex structure determined by $\fc$.  A
compact oriented surface admits a hyperbolic metric iff its genus $g$ is
strictly greater than one; in that case, there exists a continuous
infinity of hyperbolic metrics which form a moduli space of complex
dimension $3g-3$ (real dimension $6g-6$). The case when $\Sigma$ is
non-compact is discussed below.

\subsection{The Poincar\'e upper half plane}

\begin{figure}[H]
\centering
\includegraphics[width=.47\columnwidth]{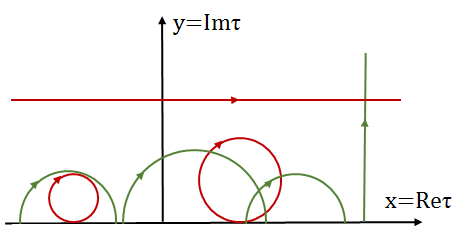}
\caption{Some geodesics (red) and horocycles (green) on the upper half-plane. }
\label{fig:geohoroH}
\end{figure}

\noindent The open upper half plane:
\be
\H\eqdef \{\tau\in \C|\Im \tau >0\}
\ee
with complex coordinate $\tau$ admits a unique complete metric of
constant Gaussian curvature $-1$, known as the {\em Poincar\'e plane
metric}. This metric is given by:
\ben
\label{sH}
\dd s^2_\H=\lambda_\H(\tau,\bar{\tau})^2|\dd \tau|^2~~\mathrm{with}~~\lambda_\H(\tau,\bar{\tau})=\frac{1}{\Im\tau}~~.
\een
The Cartesian coordinates $x=\Re \tau$ and $y=\Im \tau$ are isothermal
(see Appendix \ref{app:SpecCoord}) and we have:
\ben
\label{sHCart}
\dd s^2_\H=\frac{\dd x^2+\dd y^2}{y^2}~~.
\een
Notice that $\H$ is a non-compact manifold {\em without boundary} (an
open manifold), which has infinite area with respect to the metric
\eqref{sH}. However, $(\H,\dd s^2_\H)$ (which is isometric with the
Poincar\'e disk $(\mD,\dd s^2_\mD)$ discussed below) has a {\em
  conformal compactification} $\bar{\H}$ (see Appendix
\ref{subsec:conf_compactif}). The latter is a manifold with boundary
which is diffeomorphic with the closed unit disk $\bar{\mD}$. Thus
$\H$ has a {\em conformal boundary} $\partial_\infty \H$ which is
diffeomorphic with the unit circle $\rS^1$. The latter can be obtained
from the real axis $\Im \tau=0$ by adding the point at infinity:
\be
\partial_\infty \H\simeq \rS^1\simeq \{\tau\in \C\,|\,\Im \tau=0\}\sqcup \{\infty\}~~.
\ee
The group of orientation-preserving isometries of $\H$ is isomorphic
with $\PSL(2,\R)$, acting on $\H$ from the left by fractional linear
transformations:
\ben
\label{fraclin}
\tau\rightarrow A\tau\eqdef \frac{a\tau+b}{c\tau+d}~~,~~\forall A=\left[\begin{array}{cc} a & b \\ c & d \end{array}\right]\in \SL(2,\R)~~.
\een
The geodesics of $\H$ are as follows (see Figure \ref{fig:geohoroH}):
\begin{enumerate}[A.]
\item The Euclidean half-circles contained in $\H$ and having center
  on the real axis. The canonical orientation of such a geodesic is
  from left to right when the imaginary axis of $\H$ points upwards.
\item The Euclidean half-lines parallel to the imaginary axis. The
  canonical orientation of such a geodesic is from bottom to top when
  the imaginary axis of $\H$ points upwards.
\end{enumerate}
A {\em horocycle} of $\H$ is either a Euclidean circle which is
tangent to a point of the real axis or a Euclidean line parallel to
the real axis. In the second case, we say that the horocycle is
tangent to $\partial_\infty \H$ at the point $\infty$. The 
  canonical orientation of a horocycle is from left to right when the
imaginary axis of the Poincar\'e half-plane points upwards.

\subsection{The Poincar\'e disk}

\begin{figure}[H]
\centering
\includegraphics[width=.37\columnwidth]{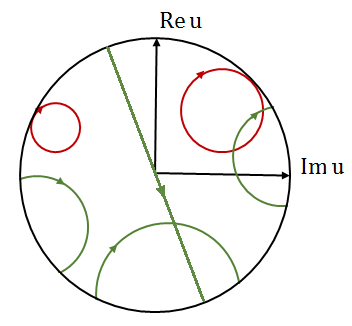}
\caption{Some geodesics (red) and horocycles (green) on the Poincar\'e disc. }
\label{fig:geohoroD}
\end{figure}

\noindent The open unit disk:
\be
\mD\eqdef \{u\in \C~|~|u|<1\}
\ee
with complex coordinate $u$ admits a unique complete metric of
constant Gaussian curvature $-1$. This is known as the {\em Poincar\'e disk
metric} and is given by:
\be
\dd s^2_\mD=\lambda^2_\mD(u,\bar{u})|\dd u|^2~~\mathrm{with}~~\lambda_\mD(u,\bar{u})=\frac{2}{1-|u|^2}~~.
\ee
In particular, $\Re u$ and $\Im u$ are isothermal coordinates. The
hyperbolic disk is isometric with the hyperbolic plane through the
biholomorphic transformation (the {\em Cayley map}) $f:\mD\rightarrow
\H$ given by:
\ben
\label{f}
\tau=f(u)\eqdef \frac{u+\i}{\i u+1}~\Rightarrow~ u=\frac{\i-\tau}{\i\tau -1}~~,
\een
which is the action of the Cayley element:
\ben
\label{Qdef}
Q\eqdef \frac{1}{2}\left[\begin{array}{cc} 1 & \i \\ \i & 1 \end{array}\right]\in \SL(2,\C)
\een
on $\tau$ by a fractional linear transformation. This extends to a
diffeomorphism of manifolds with boundary between the conformal
compactifications $\bar{\mD}$ and $\bar{\H}$, which takes the center
$u=0$ of $\mD$ into the point $\tau=\i$ and takes the conformal
boundary point $u=\i\in \partial_\infty \mD$ into the conformal
boundary point $\tau=\infty\in \partial_\infty \H$. In particular, $f$
identifies the conformal boundaries $\partial_\infty \mD\simeq \rS^1$
and $\partial_\infty \H$. The group of orientation-preserving
isometries of $\mD$ is $\PSU(1,1)$, where:
\be
\SU(1,1)=\left\{A=\left[\begin{array}{cc} \eta & \sigma \\ \bar{\sigma} 
& \bar{\eta} \end{array}\right] \Big|~\eta,\sigma\in \C~,~ |\eta|^2-|\sigma|^2=1\right\}
\ee
acts on $\mD$ through:
\be
u\rightarrow \frac{\eta u+\sigma}{\bar{\sigma} u+\bar{\eta}}~~.
\ee
The subgroups $\SU(1,1)$ and $\SL(2,\R)$ of $\SL(2,\C)$ are conjugate through
the Cayley element:
\be
Q\SU(1,1)Q^{-1}=\SL(2,\R)~~,
\ee
which implies a similar conjugacy relation between the subgroups
$\PSU(1,1)$ and $\PSL(2,\R)$ of $\PSL(2,\C)$. In particular,
$\PSU(1,1)$ is isomorphic with $\PSL(2,\R)$. Moreover, the subgroups
$\Gamma$ of $\PSL(2,\R)$ can be identified with the subgroups $\Delta$
of $\PSU(1,1)$ through the {\em Cayley correspondence}:
\ben
\label{DeltaGamma}
\Delta=Q^{-1}\Gamma Q~~.
\een
The geodesics of $\mD$ are those Euclidean half-circles contained in
$\mD$ which are orthogonal to $\partial_\infty\mD$, together with the
diameters of $\mD$. The canonical orientation of geodesics is induced
from that of geodesics in the Poincar\'e half-plane model. In the
Poincar\'e disk model, horocycles correspond to those Euclidean
circles contained in the closed unit disk which are tangent to
$\partial_\infty \mD$ at some point of the conformal boundary; their
canonical orientation is induced from that of horocycles in the
Poincar\'e half-plane model; the point of tangency is not part of the
horocycle (see Figure \ref{fig:geohoroD})

\subsection{Classification of elements of $\PSL(2,\R)$}
\label{subsec:el_classif}

\noindent An element $A=\left[\begin{array}{cc} a & b \\ c & d \end{array}\right]\in \PSL(2,\R)$ is called:
\begin{itemize}
\itemsep 0.0em
\item elliptic, if $|\tr(A)|<2$~,
\item parabolic, if $|\tr(A)|=2$~,
\item hyperbolic, if $|\tr(A)|>2$~.
\end{itemize}
The fixed points of $A\in \PSL(2,\R)$ in $\C$ are the solutions $\tau_\pm=\frac{(a-d) \pm \sqrt{(a+d)^2-4}}{2c}$
of the equation:
\be
c\tau^2+(d-a)\tau-b=0
\ee
(which has discriminant $\tr(A)^2-4$) and satisfy the Vieta relations:
\be
\tau_++\tau_-=\frac{a-d}{c}~~,~~~\tau_+\tau_-=-\frac{b}{c}~~,
\ee
as well as the relation:
\be
\tau_+-\tau_-=\frac{\sqrt{\tr(A)^2-4}}{c}~~.
\ee
The element $A$ is non-elliptic iff the roots belong to
$\partial_\infty \H=\R\sqcup \{\infty\}$, being parabolic or
hyperbolic iff the roots are respectively equal or distinct. We
concentrate on parabolic and hyperbolic elements, since smooth
surfaces are uniformized by Fuchsian groups without elliptic elements.

\vspace{2mm}

\paragraph{Parabolic elements and parabolic cyclic groups.}
The fixed point $\tau_P\in \partial_\infty \H$ of a parabolic element
$P\in \PSL(2,\R)$ can be moved to any position on $\partial_\infty \H$
by conjugating $P$ inside $\PSL(2,\R)$. Any horocycle tangent to
$\partial_\infty \H$ at $\tau_P$ is stabilized by the action of $P$ on
$\H$. The {\em unit horocycle} $c_P$ of $P$ is the unique horocycle
tangent to $\partial_\infty\H$ at $\tau_P$ such that the quotient
closed curve $c_P/\langle P \rangle$ has hyperbolic length equal to
one.

\vspace{2mm}

The subgroup of $\PSL(2,\R)$ consisting of all elements fixing a given
point $\tau_0\in \partial_\infty \H$ is an infinite cyclic group
$Z_{\tau_0}\subset \PSL(2,\R)$ consisting of parabolic elements
(called a {\em parabolic cyclic group}). All parabolic cyclic groups
are mutually conjugate inside $\PSL(2,\R)$. Moreover, the map
$\tau_0\rightarrow Z_{\tau_0}$ is a bijection between the points of
$\partial_\infty\H$ and the set of all parabolic cyclic subgroups of
$\PSL(2,\R)$. All elements $P\in Z_{\tau_0}$ satisfy $\tau_P=\tau_0$
and can be written as $P=P_{\tau_0}^m$ for some integer $m\in \Z$,
where $P_{\tau_0}$ is a generator of $Z_{\tau_0}$, which is called
{\em a parabolic generator} determined by $\tau_0$. There are exactly
two parabolic generators, namely $P_{\tau_0}$ and
$P_{\tau_0}^{-1}$. The {\em positive parabolic generator} is that parabolic
generator which maps each point of the unit horocycle tangent to
$\partial_\infty \H$ at $\tau_0$ to a point lying forward on that horocycle with
respect to the canonical orientation.

\vspace{2mm}

\paragraph{Hyperbolic elements and hyperbolic cyclic groups.}
The two fixed points $\tau_H^\pm$ of a hyperbolic element $H\in
\PSL(2,\R)$ can be moved to any distinct positions on
$\partial_\infty\H$ (for example, to the points $0$ and $\infty$) by
conjugating $H$ inside $\PSL(2,\R)$. The {\em axis} of a hyperbolic
element $H$ is the unique hyperbolic geodesic $c_H$ connecting the two
fixed points of $H$; this geodesic is stabilized by the action of $H$
on $\H$. Notice that the axis is completely determined by the two
fixed points of $H$. The positive number $\ell(H)>0$ defined through
the relation:
\ben
\label{elldef}
\tr(H)=2 \cosh\left(\frac{\ell(H)}{2}\right)
\een
is called the {\em displacement length} of $H$. It coincides with the
hyperbolic distance between any point $\tau\in c_H$ and its image
$H\tau\in c_H$. The {\em displacement radius} of $H$ is defined through:
\ben
\label{Relldef}
R_H \eqdef e^{\frac{~\pi^2}{\ell(H)}} >1~~.
\een
The subgroup of $\PSL(2,\R)$ consisting of all elements fixing two
given distinct points $\tau_1,\tau_2\in \partial_\infty \H$ is an
infinite cyclic group $Z_{\tau_1,\tau_2}\subset \PSL(2,\R)$ consisting
of hyperbolic elements (called a {\em hyperbolic cyclic group}). All
hyperbolic cyclic groups are mutually conjugate inside
$\PSL(2,\R)$. Moreover, the map $\{\tau_1,\tau_2\}\rightarrow
Z_{\tau_1,\tau_2}$ is a bijection between the set of two-element
subsets of $\partial_\infty\H$ and the set of all parabolic cyclic
subgroups of $\PSL(2,\R)$. All elements $H\in Z_{\tau_1,\tau_2}$
satisfy $\{\tau_H^+,\tau_H^-\}=\{\tau_1,\tau_2\}$ and can be written
as $H=H_{\tau_1,\tau_2}^m$ for some integer $m\in \Z$, where
$H_{\tau_1,\tau_2}$ is a generator of $Z_{\tau_1,\tau_2}$, which is
called a {\em hyperbolic generator} determined by the set
$\{\tau_1,\tau_2\}$. There are exactly two such generators, namely
$H_{\tau_1,\tau_2}$ and $H_{\tau_1,\tau_2}^{-1}$. The {\em positive
  hyperbolic generator} is that hyperbolic generator which maps each
point of the geodesic connecting $\tau_1$ and $\tau_2$ to a point
lying forward on that geodesic with respect to the canonical orientation.

\subsection{Fuchsian and surface groups}

\noindent A subgroup $\Gamma$ of $\PSL(2,\R)$ is discrete iff it acts
properly discontinuously on the upper half plane by fractional linear
transformations \eqref{fraclin}. In this case $\Gamma$ is called a
{\em Fuchsian group}. By definition, a {\em surface group} is a Fuchsian
group $\Gamma$ without elliptic elements. 

\vspace{2mm}

\paragraph{Remark.}
There exists an uncountable infinity of conjugacy classes of surface
groups.

\subsection{Uniformization of hyperbolic surfaces}

\paragraph{Uniformization to the hyperbolic plane.}
Any oriented smooth hyperbolic surface $(\Sigma,G)$ can be written as
the quotient $\H/\Gamma$ (endowed with the quotiented Poincar\'e
metric), where $\H$ is the hyperbolic plane endowed with the
Poincar\'e metric and $\Gamma$ is a surface group. Notice that the
groups $\Gamma$ and $\pi_1(\Sigma)$ are isomorphic. Let
$\pi_\H:\H\rightarrow \Sigma$ be the projection map (which is known as
the {\em uniformization map}). When $\Sigma$ is endowed with the
complex structure corresponding to the conformal class of $G$ and $\H$
is endowed with its hyperbolic complex structure, the map $\pi_\H$ is
a holomorphic covering map. Conversely, any holomorphic covering map
$\pi_\H:\H\rightarrow \Sigma$ induces a hyperbolic metric on $\Sigma$
which makes $\pi_\H$ into a local isometry. Moreover, $(\Sigma,G)$
determines $\pi_\H$ up to composition with an element of $\PSL(2,\R)$;
this implies that the uniformizing group $\Gamma$ is determined by
$(\Sigma, G)$ up to conjugation inside $\PSL(2,\R)$. Let $z$ be a
local complex coordinate on $\Sigma$. Then $\pi_\H$ has the local
representation $z=z(\tau)$ and we have $\dd s_G^2=\lambda_\Sigma^2|\dd
z|^2$, which gives:
\ben
\label{ulambdaH}
\lambda_\Sigma(z)=\frac{1}{\Im \tau (z)}|\frac{\dd \tau}{\dd z}|~~,
\een
where $\tau(z)$ is a local inverse of $z(\tau)$.

\vspace{2mm}

\paragraph{Uniformization to the hyperbolic disk.}

The Cayley correspondence \eqref{DeltaGamma} identifies Fuchsian
groups $\Gamma$ with the discrete subgroups $\Delta$ of $\PSU(1,1)$.
Composing $\pi_\H$ with the Cayley transformation $f:\mD\rightarrow
\H$ gives a holomorphic covering map $\pi_\mD:\mD\rightarrow \Sigma$
and a presentation $(\Sigma,G)=\mD/\Delta$. The map $\pi_\mD$ has the
local representation $z=z(u)$ and we have $\dd
s_G^2=\lambda_\Sigma^2|\dd z|^2$, which gives:
\ben
\label{ulambdaD}
\lambda_\Sigma(z)=\frac{2}{1-|u(z)|^2}|\frac{\dd u}{\dd z}|~~,
\een
where $u(z)$ is a local inverse of $z(u)$.

\vspace{2mm}

\paragraph{Remark.}
When $(\Sigma,G)$
is given, the highly-nontrivial problem of determining the functions
$z(u)$ (or $z(\tau)$) explicitly is known as the ``explicit
uniformization problem''.

\subsection{Classification of surface groups}

\noindent Let $\bar{\mD}\eqdef \mD\sqcup \partial_\infty \mD$ be the
conformal compactification of $\mD$ (which equals the closed unit
disk). Let $\Lambda_\mD\subset \partial_\infty \mD$ denote the set of
limit points (in the Euclidean topology) of the orbits of $\Delta$ on
$\mD$ and $\Lambda_\H\subset \partial_\infty \H$ denote the limit set
of the corresponding Fuchsian group $\Gamma$. We can of course
identify $\partial_\infty \H$ with $\partial_\infty \mD$ and hence
$\Lambda_\H$ with $\Lambda_\mD$.  The set $\Lambda_\mD$ is closed and
stabilized by the action of $\Delta$ on the complex plane given by
fractional transformations. A classical theorem of Poincar\'e and
Fricke-Klein states that one has the trichotomy
\cite{Katok,Beardon,FrenchelNielsen}:
\begin{itemize}
\item $\Lambda_\mD$ is finite, in which case $\Delta$ and $\Gamma$ are
  called {\em elementary}.
\item $\Lambda_\mD=\partial_\infty \mD$, in which case $\Delta$ and
  $\Gamma$ are called of {\em the first kind}.
\item $\Lambda_\mD$ is a perfect and nowhere-dense\footnote{I.e. a closed
  set with empty interior and without isolated points.} subset of
  $\partial_\infty \mD$, in which case $\Delta$ and $\Gamma$ are
  called of {\em the second kind}.
\end{itemize}
A hyperbolic surface $(\Sigma,G)$ is called {\em elementary}, {\em of
  the first kind} or {\em of the second kind} if its uniformizing
surface group $\Gamma$ (equivalently, $\Delta$) is of the
corresponding type. For an elementary surface group $\Gamma$, the
cardinality of $\Lambda_\mD\simeq \Lambda_\H$ can equal $0$, $1$ or
$2$, namely:
\begin{itemize}
\itemsep 0.0em
\item $\Lambda_\mD=\emptyset$ iff $(\Sigma,G)$ is isometric with the
  hyperbolic disk $\mD$.  In this case, $\Gamma=1$ is the trivial
  group.
\item $\Lambda_\mD=\{u\}$ for some $u\in \partial_\infty \mD$ iff
  $(\Sigma,G)$ is isometric with the hyperbolic punctured
  disk\footnote{The {\em hyperbolic punctured disk} $\mD^\ast$ (also
    known as the {\em parabolic cylinder} \cite{Borthwick}) is the
    non-compact hyperbolic surface obtained by endowing the open unit
    punctured disk $\{u\in \C|0<|u|<1\}$ with its unique Riemannian
    metric of Gaussian curvature equal to $-1$.}  $\mD^\ast$. In this
  case, $\Gamma\simeq \Z$ is the parabolic cyclic group consisting of
  all elements of $\PSL(2,\R)$ which fix the point of $\partial_\infty
  \H$ corresponding to $u$.
\item $\Lambda_\mD=\{u_1,u_2\}$ with $u_1,u_2\in\partial_\infty \mD$
  and $u_1\neq u_2$ iff $\Sigma$ is isometric with the hyperbolic
  annulus\footnote{For each $R>1$, the {\em hyperbolic annulus}
    $\A(R)$ (also known as the {\em hyperbolic cylinder}
    \cite{Borthwick}) is the non-compact hyperbolic surface obtained
    by endowing the open annulus $\{u\in \C|\frac{1}{R}<|u|<R\}$ with
    its unique Riemannian metric of Gaussian curvature equal to
    $-1$. The positive quantity $\mu\eqdef 2\log R$ is known as the
    {\em modulus} of $\A(R)$.}  $\A(R)$ for some $R>1$. In this case,
  $\Gamma\simeq \Z$ is the hyperbolic cyclic group consisting of all
  elements of $\PSL(2,\R)$ which fix both points of
  $\partial_\infty \H$ which correspond to $u_1$ and $u_2$.
\end{itemize}
Moreover, $\Gamma$ (equivalently, $\Delta$) is of the first kind iff
$\area_G(\Sigma)$ is finite (see \cite{Katok}); this happens iff a
fundamental polygon of $\Delta$ has a finite number of vertices and
sides and no free sides, where a side of a fundamental polygon is
called {\em free} if it is a portion of the conformal boundary
$\partial_\infty \mD$ (see Appendix \ref{subsec:fp}). When $\Gamma$
is of the second kind, the complement $\partial_\infty \mD\setminus
\Lambda_\mD\simeq \partial_\infty\H\setminus \Lambda_\H$ is an open
subset of $\rS^1$ and hence it is a countable disjoint union of open
segments (intervals) which we denote by $I_k$ ($k\in \Z_{>0}$); these
are called the {\em intervals of discontinuity} of $\Gamma$.

\subsection{Orientation-preserving isometries}
\label{subsec:Iso}

\noindent For a smooth hyperbolic surface $(\Sigma,G)$ uniformized by the
surface group $\Gamma\subset \PSL(2,\R)$, the group of
orientation-preserving isometries coincides with the group of
biholomorphisms (automorphisms of the underlying complex manifold
$(\Sigma,J)$) and is given by:
\be
\Iso^+(\Sigma,G)=N(\Gamma)/\Gamma~~,
\ee
where:
\be
N(\Gamma)\eqdef \{h\in \PSL(2,\R)~|~h\Gamma h^{-1}=\Gamma\}
\ee
is the normalizer of $\Gamma$ in $\PSL(2,\R)$. When $\Gamma$
is elementary, this group is infinite and isomorphic with $\U(1)$.
When $\Gamma$ is non-elementary, this group is finite and
its cardinality satisfies the bound \cite{Oikawa}:
\ben
\label{IsoBound}
|\Iso^+(\Sigma,G)|\leq 12 (g-1) + 6 n~ , ~~ n>0~,
\een
where $g$ is the genus of $\Sigma$ and $n$ is the number of ends.  For
$n=0$ the previous formula does not apply, being replaced instead by
the Hurwitz bound $|\Iso^+(\Sigma,G)|\leq 84 (g-1)$. Notice that
every finite group arises as the automorphism group of some compact
hyperbolic surface.

\subsection{Fundamental polygons}
\label{subsec:fp}

\noindent Let $\Gamma\subset \PSL(2,\R)$ be a Fuchsian group with limit set
$\Lambda_\H\subset \partial_\infty \H$ and let $\Sigma=\H/\Gamma$ be the
corresponding hyperbolic surface. A subset $\fD_\H\subset \H$ is called a
{\em fundamental domain} for $\Gamma$ if it satisfies the following
three conditions:
\begin{enumerate}[1.]
\itemsep 0.0em
\item $\fD_\H$ is non-empty, open and connected.
\item For any $\gamma\in \Gamma\setminus\{1\}$, we have
  $\gamma(\fD_\H)\cap \fD_\H=0$.
\item We have $\sqcup_{\gamma\in \Gamma}{\overline{\fD_\H}}=\H$, where
  $\overline{\fD_\H}$ denotes the relative closure of $\fD_\H$ in $\H$.
\end{enumerate}
A fundamental domain $\fD_\H$ is called a {\em fundamental polygon} if it
satisfies certain technical conditions for which we refer the reader
to \cite{Katok, Beardon, FrenchelNielsen, MaskitPoincare}. These
conditions state that $\fD_\H$ is a convex polygon which in general
may have an infinite number of vertices and sides, where sides are
either geodesic segments or so-called {\em free sides}. A free side of
$\fD_\H$ is an interval of $\partial_\infty \H$ which is contained in
some interval of discontinuity of $\Gamma$. A {\em free vertex} is a
vertex of $\fD_\H$ which is an endpoint of a free side. An {\em ideal
  vertex} is a vertex of $\fD_\H$ which lies inside the limit set $\Lambda_\H$ of
$\Gamma$. Any Fuchsian group (in particular, any surface group) admits
a fundamental polygon (for example, the so-called Dirichlet polygon
\cite{Beardon, Katok, FrenchelNielsen}). Two distinct sides of a
fundamental polygon $\fD_\H$ are called {\em equivalent} (or {\em
  $\Gamma$-congruent}) if there exists an element $\gamma\in \Gamma$
which maps one side into the other.  This defines an equivalence
relation whose equivalence classes partition the set of all
sides. Each equivalence class of non-free sides contains a single
(unordered) pair of distinct sides, called a {\em Poincar\'e
  pair}. The set of all such equivalence classes is called the {\em
  Poincar\'e pairing} (or {\em side identification}
\cite{MaskitPoincare}) of $\fD_\H$.  Given two $\Gamma$-congruent
non-free sides, there exist exactly two elements which map these sides
into each other and these elements are mutually inverse. Moreover,
picking one of these elements for each pair of $\Gamma$-congruent
non-free sides one obtains a set of generators for $\Gamma$ \cite{Beardon,
  Katok, FrenchelNielsen}.

Since in our case $\Sigma$ has no orbifold points and $\Gamma$ is a
surface group, all vertices of $\fD_\H$ must be ideal vertices or free
vertices.  When $\Gamma$ is of the first kind, we have
$\Lambda_\H=\partial_\infty \H$. In that case, $\fD_\H$ has no free sides
and all its vertices are ideal vertices; thus $\fD_\H$ is an {\em ideal
  polygon}, i.e. a hyperbolic polygon all of whose vertices lie on
$\partial_\infty \H$.

\section{Topologically finite surfaces}
\label{sec:topfinite}

\noindent This appendix reviews certain classical results
concerning topologically finite surfaces.  An oriented surface
$\Sigma$ is called {\em topologically finite} if its fundamental group
$\pi_1(\Sigma)$ is finitely-generated.  This happens iff $\Sigma$ is
homeomorphic (and hence diffeomorphic) with $\hSigma \setminus
\{p_1,\ldots, p_n\}$, where $\hSigma$ is a compact oriented surface
without boundary and $p_1,\ldots,p_n$ is a finite collection of
distinct points of ${\hat \Sigma}$. In this case, $\hSigma$ is
uniquely determined by $\Sigma$ up to diffeomorphism, while the
diffeomorphism class of $\Sigma$ is determined by that of $\hSigma$
and by the number $n$.

\subsection{The end compactification. Genus and Euler characteristic}

\noindent Let $\Sigma$ be a topologically finite surface. Then
$\hSigma$ can be identified with the end (a.k.a. Ker\'ekj\'art\'o-Stoilow)
compactification of $\Sigma$, where $p_1,\ldots, p_n$ correspond to
the Ker\'ekj\'art\'o-Stoilow ideal points. In particular, $\Sigma$ has
exactly $n$ ends, one for each ideal point $p_j$. Moreover, $\Sigma$
is determined up to diffeomorphism by $n$ and by the genus $g$ of
$\hSigma$, which is also called the genus of $\Sigma$. The Euler
characteristic of $\Sigma$ is given by:
\be
\chi(\Sigma)=2-2g-n~~,
\ee
while that of $\hSigma$ is given by:
\be
\chi(\hSigma)=2-2g~~.
\ee
If $G$ is a {\em complete} metric on $\Sigma$ such that
$\area_G(\Sigma)$ is finite, the Gauss-Bonnet theorem
applies\footnote{This is because the boundary contribution from the
  cusp ends vanishes \cite{Rosenberg}.}, giving:
\be
\int_\Sigma K(G) \vol_G=2\pi \chi(\Sigma)~~.
\ee
In particular, $\Sigma$ admits a hyperbolic metric $G$ of finite area
iff $\chi(\Sigma)=-\frac{\area_G(\Sigma)}{2\pi}$ is negative, i.e. iff
$2g+n>2$. In the planar case $g=0$, which requires $n\geq 3$.

\vspace{2mm}

\paragraph{Remark.}
When $g=0$ (i.e. when $\Sigma$ is a planar surface), the end
compactification $\hSigma$ is the unit sphere $\rS^2$.

\subsection{Prolongation of conformal structures and the conformal compactification}
\label{subsec:conf_compactif}

\noindent Let $\Sigma=\hSigma\setminus \{p_1,\ldots, p_n\}$ be a topologically
finite surface.  Consider a partition:
\be
\cP: \{1,\ldots, n\}=\{i_1,\ldots, i_{n_c}\}\sqcup \{j_1,\ldots, j_{n_f}\}~~,
\ee
where $n_c\geq 0$ and $n_f\geq 0$ are natural numbers. Since any
annulus is diffeomorphic with the punctured unit disk, $\Sigma$ is
diffeomorphic with the borderless surface $\Sigma_\cP \eqdef \hSigma
\setminus (\{p_{i_1},\ldots, p_{i_{n_c}}\}\cup \bar{D}_{j_1}\cup
\ldots \cup \bar{D}_{j_{n_f}})$, where $\bar{D}_j$ are closed disks
embedded in $\hSigma$ and centered at the points $p_{j_1},\ldots,
p_{j_{n_f}}$, such that no two closed disks meet each other and no
closed disk meets any of the points $p_{i_1},\ldots, p_{i_{n_c}}$.

Any orientation-compatible complex structure $I$ on $\hSigma$
determines a complete {\em canonical metric} ${\hat G}_I$ on $\hSigma$
as follows \cite{Maskit}:
\begin{enumerate}[1.]
\itemsep 0.0em
\item When $g=0$, $(\hSigma,I)$ is biholomorphic with the complex
  projective plane $\C\P^1$. In this case, ${\hat G}_I$ is the unique
  metric on $\C\P^1$ which has Gaussian curvature $+1$, i.e. the
  metric which makes $\C\P^1$ isometric with the unit round sphere
  $\rS^2$.
\item When $g=1$, $(\hSigma,I)$ is biholomorphic with an elliptic
  curve. In this case, ${\hat G}_I$ is the unique flat metric for
  which a fundamental domain has sides $1$ and $\tau$, where $\tau\in
  \H$ is chosen such that $|\tau|\geq 1$ and $|\Re\tau|\leq
  \frac{1}{2}$.
\item When $g\geq 2$, ${\hat G}_I$ is the unique complete hyperbolic
  metric on $\hSigma$ whose conformal class corresponds to $I$.
\end{enumerate}
By definition, a {\em closed circular disk} in $\hSigma$ relative to
$I$ is a closed subset of $\hSigma$ which has the form:
\be
\bar{D}_I(p;r)\eqdef \{q\in \hSigma| d_I(q,p)\leq r\}~~,
\ee
where $p$ is a point of $\hSigma$ and $d_I$ is the distance function
defined by the canonical metric $\hG_I$. A {\em circle domain} of
$(\hSigma,I)$ is a connected open subset of $\Sigma$ obtained by
removing a finite number $n_c$ of points and a finite number $n_f$ of
closed circular disks from $\hSigma$ (where $n_c+n_f=n$) and endowing
the resulting surface with the restriction of the complex structure
$I$. Notice that any circle domain is diffeomorphic with a surface of
the form $\Sigma_\cP$ for some partition $\cP$ as above.

Let $J$ be an orientation-compatible complex structure on
$\Sigma$. Then it was shown in \cite{Haas,Maskit} that there exists a
unique complex structure ${\hat J}$ on $\hSigma$ such that
$(\Sigma,J)$ is biholomorphic with a circle domain $\Sigma_J$ of
$(\hSigma,{\hat J})$. Moreover, $\Sigma_J$ is uniquely determined up
to a biholomorphism of $(\hSigma,{\hat J})$. We say that ${\hat J}$ is
the {\em prolongation} of $J$ to $\hSigma$. If $\fc$ and ${\hat \fc}$
are the conformal structures defined by $J$ and ${\hat J}$ on $\Sigma$
and $\hSigma$, then we say that ${\hat \fc}$ is the prolongation of
$\fc$ to $\hSigma$.

By definition, the {\em conformal compactification} $\bSigma_J$ of
$\Sigma$ with respect to $J$ is the surface obtained by taking the
closure of $\Sigma_J$ inside $\hSigma$. The topological boundary
$\partial_\infty^J \Sigma=\bSigma_J\setminus \Sigma$ of $\Sigma_J$
consists of $n_c$ isolated points and $n_f$ disjoint simple closed
curves; this is called the {\em conformal boundary} of
$(\Sigma,J)$. In particular, $(\bSigma_J, {\hat J}|_{\bSigma_J})$ is a
bordered Riemann surface with $n_c$ interior marked points and $n_f$
boundary components. When $J$ is of hyperbolic type, one has $n_f=0$
iff the area of $\Sigma$ (computed with respect to the hyperbolic
metric determined by $J$) is finite. When a hyperbolic metric $G$ on
$\Sigma$ is given, we define $\partial_\infty^G\Sigma\eqdef
\partial_\infty^J\Sigma$, where $J$ is the complex structure
determined by $G$. We sometimes write this simply as
$\partial_\infty\Sigma$, when the hyperbolic metric $G$ is understood.

\vspace{2mm}

\paragraph{Remark.}
Suppose that $\Sigma$ is a planar surface, i.e. $g=0$.  In this case,
$\hSigma=\rS^2$ admits a unique orientation-compatible complex
structure $I$ (which makes it biholomorphic with the Riemann sphere
$\C\P^1$) and the result above reduces to the classical statement that
any topologically finite planar Riemann surface is biholomorphic with
a circle domain in the Riemann sphere. 

\vspace{2mm}

\section{Geometrically finite hyperbolic surfaces}
\label{sec:geomfinite}

\noindent This appendix contains a brief review of basic results on
geometrically finite hyperbolic surfaces, i.e. those hyperbolic
surfaces whose underlying 2-manifold is topologically finite. An
important advantage of such surfaces (which makes them especially
useful for applications to cosmology) is that each end of a
geometrically finite hyperbolic surface admits a canonical
neighborhood on which the hyperbolic metric can be brought to one of a
small number of standard forms in semi-geodesic coordinates. A
standard reference for this subject is \cite{Borthwick}.

\subsection{Basics}

\paragraph{Proposition-definition}\cite{Borthwick}
Let $(\Sigma,G)$ be a hyperbolic surface uniformized by the surface
group $\Gamma\subset \PSL(2,\R)$. Then the following statements are
equivalent:
\begin{enumerate}[(a)]
\item $\Sigma$ is topologically finite.
\item $\Gamma$ is finitely-generated.
\item $\Gamma$ admits a fundamental domain which is a convex polygon
  with a finite number of sides (some of which may be free sides).
\end{enumerate}
In this case, one says that $\Gamma$ and $(\Sigma,G)$ are {\em
  geometrically finite}.

\vspace{2mm}

\noindent All elementary hyperbolic surfaces (the Poincar\'e disk, 
the hyperbolic punctured disk and the hyperbolic annuli) are geometrically finite.
Moreover, a result due to Siegel states that any hyperbolic
surface of the first kind (i.e. with finite area) is geometrically finite
(see \cite{Katok}); in particular, all compact hyperbolic
surfaces are geometrically finite. A hyperbolic surface of the second
kind may be either geometrically finite or geometrically infinite.

Elementary hyperbolic surfaces are special in that the isometry type
of their ends can be exceptional. Due to this fact, we start by
discussing non-elementary geometrically finite hyperbolic surfaces,
whose ends admit a uniform classification up to isometry
\cite{Borthwick}.

\subsection{Hyperbolic type of ends of non-elementary geometrically finite hyperbolic surfaces}
\label{sub:isom}

\noindent A non-elementary geometrically finite hyperbolic surface $(\Sigma,G)$
can have only two types of ends, namely {\em cusp} or {\em funnel}
ends. The cusp ends correspond to isolated points of the conformal
boundary $\partial_\infty^G\Sigma$, while the funnel ends correspond
to the simple closed curve components of $\partial_\infty^G\Sigma$.

\vspace{2mm}

\paragraph{Hyperbolic cusps.}

Given a parabolic element $P\in \PSL(2,\R)$, the {\em cusp domain}
$\cC_P\subset \H$ defined by $P$ is the interior of the disk bounded
by the unit horocycle $c_P$ of $P$ (see Appendix
\ref{subsec:el_classif}). The {\em hyperbolic cusp} defined by $P$ is
the quotient $\rC_P=\cC_P/\langle P\rangle$ (endowed with the quotient
metric), where $\langle P\rangle$ denotes the infinite cyclic group
generated by $P$. The cusp defined by any parabolic element $P\in
\PSL(2,\R)$ is isometric with a sub-disk of the hyperbolic punctured
disk $\mD^\ast$ which contains the origin (see Appendix
\ref{subsec:cfcoords}). In particular, the cusp is independent of the
choice of the parabolic element $P$ up to isometry. A hyperbolic cusp
is non-compact but has finite hyperbolic area.

\vspace{2mm}

\paragraph{Hyperbolic funnels.}

Given a hyperbolic element $H\in \PSL(2,\R)$, the {\em funnel domain}
$\cF_H\subset \H$ defined by $H$ is the region bounded by
$\partial_\infty \H$ and by the axis $c_H$ of $H$. The {\em hyperbolic
  funnel} defined by $H$ is the quotient $\rF_H=\cF_H/\langle
H\rangle$ (endowed with the quotient metric), where $\langle H\rangle$
denotes the infinite cyclic group generated by $H$. Up to isometry,
$\rF_H$ depends only on the displacement length $\ell(H)$ of $H$ and
can be identified with the ``inner half'' of the hyperbolic annulus
$\A(R_H)$ of modulus $\mu_H=2\log R_H=\frac{2\pi^2}{\ell(H)}$, where
$R_H$ is the displacement radius of $H$ (see \eqref{Relldef}). The
displacement length $\ell(H)$ is called the {\em width} of the
hyperbolic funnel. A hyperbolic funnel is non-compact and of infinite
hyperbolic area.

\vspace{2mm}

\paragraph{The Nielsen and truncated Nielsen regions.}

Let $\Gamma\subset \PSL(2,\R)$ be a non-elementary finitely-generated
surface group and $\Sigma\eqdef \H/\Gamma$. For each such group, we
define the {\em cusp domains} $\cC_j:=\cC_j(\Gamma)$, {\em funnel domains}
$\cF_k:=\cF_k(\Gamma)$, {\em Nielsen region} $\cN_\Gamma$ and {\em truncated
  Nielsen region} $\cK_\Gamma$ as follows. Let $\Lambda_\H\subset
\partial_\infty \H$ denote the limit set of $\Gamma$. The set of
points of $\partial_\infty \H$ which are fixed by some parabolic
cyclic subgroup of $\Gamma$ is a countable subset of $\Lambda_\H$ called
the set of {\em cusp points of $\Gamma$}. For each cusp point
$\tau_j\in \Lambda_\H$, let $P_j$ be the positive generator of the
parabolic cyclic subgroup of $\Gamma$ which fixes $\tau_j$ and let
$\cC_j\eqdef \cC_{P_j}\subset \H$ be the cusp domain defined
by $P_j$. It can be shown (see \cite[Lemma 2.12]{Borthwick}) that
$\overline{\cC_{j_1}}\cap \overline{\cC_{j_2}}=\emptyset$ for $j_1\neq
j_2$ and that an element $\gamma\in \Gamma$ satisfies
$\gamma(\cC_j)\cap \cC_j\neq \emptyset$ iff $\gamma(\cC_j)=\cC_j$ iff
$\gamma$ belongs to the parabolic cyclic group $\langle P_j\rangle$ 
generated by $P_j$. Distinguish the cases:
\begin{enumerate}[A.]
\itemsep 0.0em
\item When $\Gamma$ is of the first kind, we have
$\Lambda_\H=\partial_\infty\H$. In this case, define:
\be
\cN_\Gamma\eqdef \H~~,~~\cK_\Gamma\eqdef \H\setminus \left(\sqcup_j \cC_j\right)~~.
\ee
\item When $\Gamma$ is of the second kind, the complement
  $\partial_\infty\H\setminus \Lambda_\H$ is a countable union of
  intervals of discontinuity of $\Gamma$. For each such interval
  $I_k$, let $H_k$ be the generator of the hyperbolic cyclic group
  consisting of all hyperbolic elements of $\PSL(2,\R)$ which fix the
  two endpoints of $I_k$ and let $\cF_k\eqdef \cF_{H_k}\subset \H$
  denote the funnel domain defined by $H_k$. It can be shown (see
  \cite[Lemma 2.12]{Borthwick}) that the closure of $\cF_k$ does not
  meet any $\cF_{k'}$ with $k'\neq k$ and also does not meet the union
  of the cusp regions $\cup_j \cC_j$. Define:
\be
\cN_\Gamma\eqdef \H\setminus \left(\sqcup_j \cF_j\right)~~,~~\cK_\Gamma\eqdef \cN_\Gamma\setminus \left(\sqcup_j \cC_j\right)~~.
\ee
It is easy to see the set $\cN_\Gamma$ is geodesically convex,
being the geodesic convex hull of $\Lambda_\H$.
\end{enumerate}

\vspace{2mm}

\paragraph{The canonical decomposition of $\Sigma$.}

\vspace{2mm}

\paragraph{Definition.}
The {\em convex core} of $\Sigma$ is the bordered surface:
\be
N_\Sigma\eqdef \cN_\Gamma/\Gamma~~,
\ee
endowed with the quotient hyperbolic metric. The {\em compact core} of
$\Sigma$ is the bordered compact surface:
\be
K_\Sigma\eqdef \cK_\Gamma/\Gamma~~,
\ee
endowed with the quotient hyperbolic metric. The {\em cusp regions}
$\rC_1,\ldots, \rC_{n_c}$ of $\Sigma$ are the distinct regions of
$\Sigma$ obtained from the sequence of hyperbolic cusps $\cC_j/\langle
P_j\rangle$ under the full quotient by $\Gamma$.  The {\em funnel
  regions} $\rF_1,\ldots, \rF_{n_f}$ of $\Sigma$ are the distinct
regions of $\Sigma$ obtained from the sequence of hyperbolic funnels
$\cF_k/\langle H_k\rangle$ under the full quotient by $\Gamma$. The
quotient of $\H$ by $\Gamma$ identifies those hyperbolic cusps
$\cC_{j_1}/P_{j_1}$ and $\cC_{j_2}/P_{j_2}$ for which $P_{j_1}$ is
conjugate to $P_{j_2}$ inside $\Gamma$, and similarly for hyperbolic
funnels. 

\vspace{2mm}

\noindent We have:
\be
N_\Sigma=K_\Sigma\sqcup \rC_1\sqcup \ldots \sqcup \rC_{n_c}~~.
\ee
The convex core $N_\Sigma$ is the largest geodesically convex subset
of $\Sigma$. 

\vspace{2mm}

\paragraph{Theorem.}\cite{Borthwick}
A non-elementary geometrically finite hyperbolic surface $(\Sigma,G)$
admits a disjoint union decomposition $\Sigma = N_\Sigma \sqcup \rF_1
\sqcup \ldots \sqcup \rF_{n_f}=K_\Sigma \sqcup \rC_1 \sqcup \ldots \sqcup
\rC_{n_c} \sqcup \rF_1 \sqcup \ldots \sqcup \rF_{n_f}$.
 
\vspace{2mm}

\noindent In particular, each end of sigma corresponds to a cusp or
a funnel region, being either a cusp end or a funnel
  end. We have:
\be
n=n_c+n_f~~,
\ee
where $n$ is the total number of ends. Moreover:
\begin{itemize}
\item $(\Sigma,G)$ has no funnel ends ($n_f=0$) iff it is of the
  first kind, i.e. iff $(\Sigma,G)$ has finite area.
\item $(\Sigma,G)$ has no cusp ends ($n_c=0$) iff the convex core
  $N_\Sigma$ is compact.
\item $(\Sigma,G)$ is compact iff it has no ends
  ($n=0$).
\end{itemize}

\begin{figure}[H]
\centering
\includegraphics[width=.75\columnwidth]{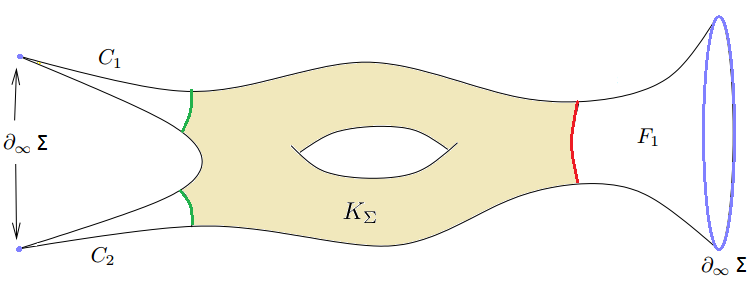}
\caption{A geometrically finite hyperbolic surface with two cusp ends 
  and one funnel end, where $K_\Sigma$ indicates the compact
  core. In this example, the ideal boundary consists of $n=3$ points
  while the conformal boundary $\partial_\infty \Sigma$ consists of
  two points (the blue dots corresponding to the cusp ends) 
and a circle (the blue circle corresponding to the funnel end). 
The two green lines are the horocycles which separate the compact core $K_\Sigma$
from the two cusp regions $C_1$ and $C_2$, while the red line is the geodesic 
which separates $K_\Sigma$ from the funnel region $F_1$.
\label{fig:Surface}}
\end{figure}

\noindent A geometrically finite hyperbolic surface with two cusp ends
and one funnel end is sketched in Figure \ref{fig:Surface}.

\subsection{Hyperbolic pants decompositions of the convex core}
\label{subsec:hyppants}

\noindent Recall that a (topological) {\em pair of pants} $X$ is a
genus zero surface with boundary which is diffeomorphic with a sphere
with three holes, where the boundaries of the holes (each of which is
diffeomorphic with a circle and is called a {\em cuff}) are the
boundary components of $X$. Such a surface has genus zero and 3 ends,
thus $\chi(X)=-1$. In particular, the fundamental group $\pi_1(X)$ is
free on two generators. A {\em hyperbolic pants metric} is a
hyperbolic metric $G$ on $X$ for which each boundary component is a
geodesic. When endowed with such a metric, the Riemannian manifold
with boundary $(X,G)$ is called a {\em hyperbolic pair of pants}. This
has area $2\pi$ and the hyperbolic metric $G$ is uniquely determined
by the hyperbolic lengths (called the {\em cuff lengths})
$\ell_1,\ell_2,\ell_3$ of the boundary components. A cuff length
vanishes iff the corresponding cuff is collapsed to a point; such a
generated pair of pants is called {\em tight}. The non-negative real
numbers $\ell_1,\ell_2$ and $\ell_3$ determine $(X,G)$ up to isometry,
so we shall use the notation $X(\ell_1,\ell_2,\ell_3)$ for the
latter. The uniformizing group of $X(\ell_1,\ell_2,\ell_3)$ is a free
group on two generators. Moreover, $X(\ell_1,\ell_2,\ell_3)$ can be
obtained by gluing two copies of a right angled hyperbolic hexagon
$\cH\subset \H$ with three alternating sides of lengths $\ell_1/2$,
$\ell_2/2$ and $\ell_3/2$, where the gluing is performed along the
other three sides. Such a hyperbolic hexagon is uniquely-determined by
$\ell_1,\ell_2$ and $\ell_3$ up to the action of $\PSL(2,\R)$. The
following is a fundamental result in Teichm\"uller theory \cite{Buser,
  Borthwick}:

\vspace{2mm}

\paragraph{Theorem.}
Let $(\Sigma,G)$ be a non-elementary geometrically finite hyperbolic
surface. Then the convex core $N_\Sigma$ of $(\Sigma,G)$ can be decomposed as
a finite union of hyperbolic pairs of pants $X_1,\ldots, X_m$, where
$m=-\chi(\Sigma)=n+2g-2$ and the hyperbolic pants metric of $X_j$ is
given by restricting $G$:
\be
N_\Sigma=X_1\cup \ldots \cup X_{n+2g-2}~~.
\ee

\noindent The hyperbolic pairs of pants in the theorem are separated by geodesics
of $(\Sigma,G)$. One can order these such that $X_1,\ldots, X_{n_c}$
are tight pairs of pants, each of which has precisely one vanishing
cuff length; this isolates each cusp end of $\Sigma$ within a tight
pair of pants. Notice that the hyperbolic pants decomposition of
$(\Sigma,G)$ is not unique.

\subsection{Cusp and funnel coordinates}
\label{subsec:cfcoords}

\noindent Consider a geometrically finite hyperbolic surface
$\Sigma=\H/\Gamma$, where $\Gamma\subset \PSL(2,\R)$ is a surface
group with limit set $\Lambda_\H\subset \partial_\infty \H$. Recall
that $\partial_\infty \H\setminus \Lambda_\H$ is a countable union of
intervals of discontinuity $I_k$. Since $\partial_\infty \H$ and
$\Lambda_\H$ are stabilized by the action of $\Gamma$, this action
permutes the intervals $I_k$.

\vspace{2mm}

\paragraph{Ideal vertices and free sides of a fundamental polygon.}
Let $\fD_\H$ be a fundamental polygon for $\Gamma$.
The Euclidean closure of $\fD_\H$ touches $\partial_\infty \H$ at a
finite number of ideal vertices and along a
finite number of free sides. Each ideal vertex
lies in $\Lambda_\H$ while each free side is contained in one of the
intervals $I_k$. Two ideal vertices or two free sides of $\fD_\H$ are
called equivalent if they are related by the action of an
element of $\Gamma$. This defines equivalence relations on the sets of
ideal vertices and free sides of $\fD_\H$, which partition these sets
into equivalence classes called {\em ideal vertex cycles} and {\em
  free side cycles}.  Equivalent vertices project through $\pi_\H$ to
the same isolated point of the conformal boundary $\partial_\infty
\Sigma$, while equivalent free sides project to the same circle
component of $\partial_\infty \Sigma$. It follows that cusp points and
circle components of $\partial_\infty\Sigma$ are respectively in
bijection with ideal vertex cycles and free side cycles.

\vspace{2mm}

\begin{figure}[H]
\centering 
\includegraphics[width=.7\columnwidth]{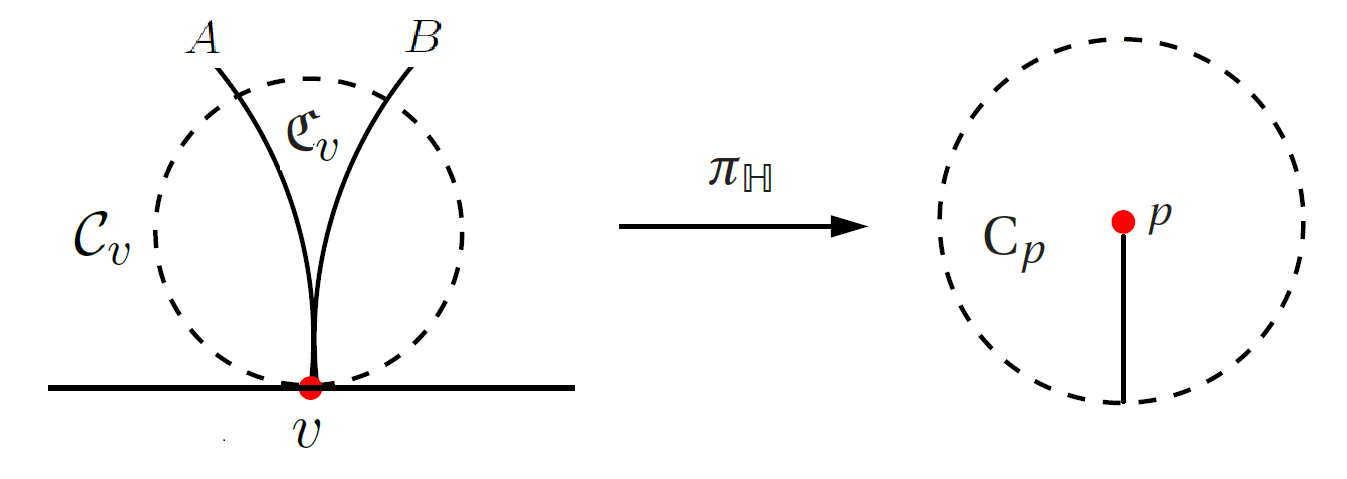}
\caption{Two sides $A$ and $B$ of a fundamental polygon $\fD_\H$
  meeting at an ideal vertex $v\in \partial_\infty \H$ and the unit
  horocycle of the parabolic generator $P_v$. The relative cusp
  neighborhood $\fC_v$ of $v$ in $\fD_\H$ is the region lying between
  the sides and the horocycle (represented by the dashed circle). 
The holomorphic covering $\pi_\H$ maps
  the cusp domain $\cC_v$ bounded by the unit horocycle to the
  hyperbolic cusp region $\rC_p$ of $\Sigma$, which can be identified
  with a punctured disk of radius $e^{-2\pi}$, centered at the
  corresponding ideal point $p=\pi_\H(v) \in \hSigma$ (shown in
  red). The covering map identifies the two sides of $\fD_\H$ which
  meet at $v$.}
\label{fig:cuspcoord}
\end{figure}

\paragraph{Cusp coordinates.}
At each ideal vertex $v$ of the fundamental polygon, two sides of
$\fD_\H$ (which are hyperbolic geodesic segments of $\H$) meet with a
vanishing angle, being paired by a parabolic element $P_v\in
\Gamma$. This element is a generator of the parabolic cyclic group
consisting of all elements of $\PSL(2,\R)$ which fix $v$, and we can 
take it to be the positive generator of that group. The {\em
  relative cusp neighborhood} of $v$ in $\fD_\H$ is the intersection:
\be
\fC_v \eqdef \fD_\H\cap \cC_v
\ee
of $\fD_\H$ with the cusp domain $\cC_v\eqdef \cC_{P_v}$ of $P_v$. It
can be shown that no vertex of $\fD_\H$ except $v$ meets the Euclidean
closure of $\fC_v$. The holomorphic covering map $\pi_\H$ maps $\cC_v$
onto the cusp region $\rC_p=\cC_v/\langle P_v\rangle$ of $\Sigma$,
which is a holomorphically embedded disk punctured at the
corresponding ideal point $p=\pi_\H(v)\in \hSigma\setminus \Sigma$
(see Figure \ref{fig:cuspcoord}). There exists a unique element
$T_v\in \PSL(2,\R)$ such that $T_vP_vT_v^{-1}$ equals the translation
$\tau\rightarrow \tau+1$. The local holomorphic coordinate $z_p$ on
$\rC_p$ which identifies the latter with the punctured disk:
\ben
\label{dotD}
\dot{D}\eqdef \{z\in \C\,|~0<|z|< e^{-2\pi}\}
\een
is called a {\em local cusp coordinate} for $\Sigma$ near $p$ and is
given by\footnote{We have $T_v(v)=\infty$ and $T_v(\cC_v)=\{\tau'\in \H~|~\Im
  \tau'>1\}$.}:
\ben
\label{cuspcoord}
z_p=e^{2\pi i T_v\tau} ~~ (\tau\in \cC_v)~.
\een
Notice that $z_p$ depends only on the ideal point $p$. Indeed, if $v'$
is another ideal vertex of $\fD_\H$ such that
$\pi_\H(v')=\pi_\H(v)=p$, then we have $v'=\gamma v$ for some
$\gamma\in \Gamma$ and hence $P_{v'}=\gamma P_v\gamma^{-1}$ and
$\cC_{v'}=\gamma \cC_v$.  Thus $T_{v'}=T_v\gamma^{-1}$ and
$T_{v'}\tau'=T_v(\gamma^{-1}\tau')=T_v\tau$ for any
$\tau'=\gamma\tau\in \cC_{v'}$, where $\tau\in \cC_v$. With respect to
the coordinate $z_p$, the ideal point $p$ corresponds to $z_p=0$ and
the restriction of the hyperbolic metric $G$ to the cusp region
$\rC_p$ of $\Sigma$ corresponds to the restriction of the hyperbolic
metric of the punctured disk $\mD^\ast$ to the punctured disk \eqref{dotD}.

\vspace{2mm}

\begin{figure}[H]
\centering 
\includegraphics[width=.75\columnwidth]{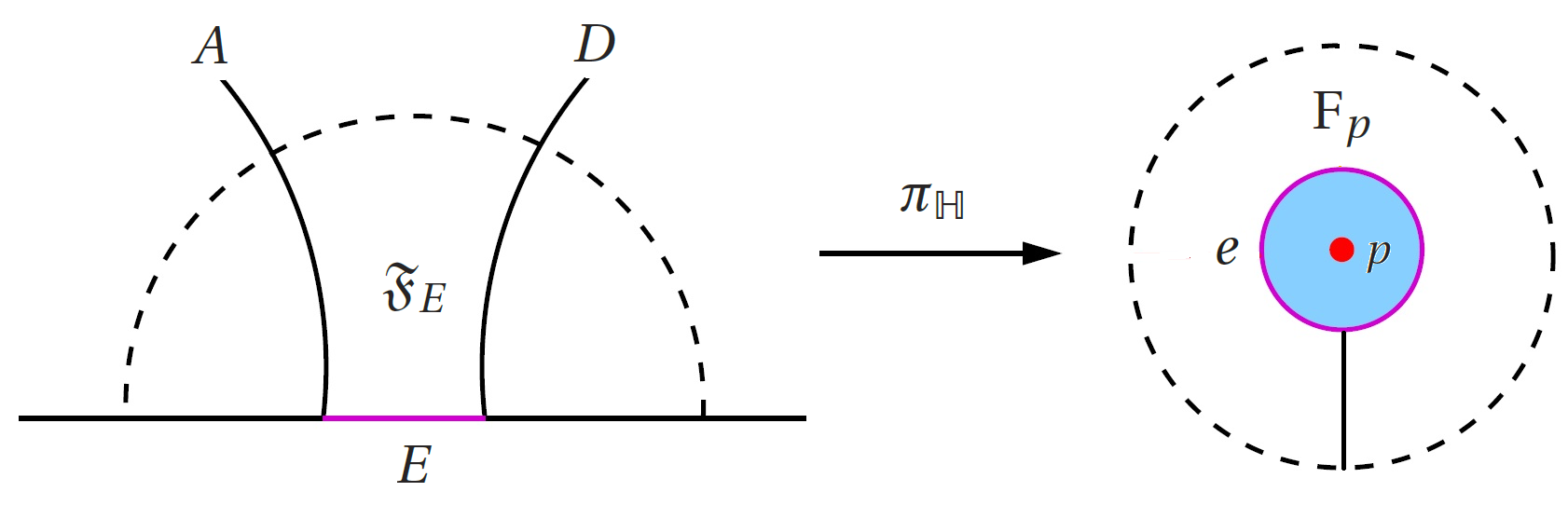}
\caption{A free side $E$ of a fundamental polygon is shown in
  magenta. The dashed half-circle is the axis of the hyperbolic
  transformation $H$ which pairs the two non-free sides $A$ and $D$ of
  $\fD_\H$ which meet the endpoints of $E$; this axis meets
  $\partial_\infty \H$ at the endpoints of the interval of
  discontinuity $I_E$ which contains $E$. The relative funnel
  neighborhood $\fF_E$ of $E$ in $\fD_\H$ is the region lying between
  $E$, these two non-free sides and the dashed half-circle. The
  holomorphic covering $\pi_\H$ maps $E$ to the inner circle
  (continuous line $e$ of radius $1/R_E$) of the annulus $A_{R_E}$ which
  corresponds to the funnel region $\rF_p$. The dashed boundary of the
  funnel neighborhood is mapped to the outer circle (radius $1$) of
  this annuls, while the sides $A$ and $D$ of the fundamental polygon
  are identified by the projection. The ideal point $p$ corresponding
  to the funnel end is depicted in red. The shadowed disk shown in
  blue is {\em not} part of $\Sigma$ or of its conformal
  compactification $\partial_\infty^G\Sigma$.}
\label{fig:funnelcoord}
\end{figure}

\paragraph{Funnel coordinates.}
The ends of each free side $E$ of $\fD_\H$ lie on two non-free sides
of $\fD_\H$ (which are hyperbolic geodesic segments in $\H$). These
meet $\partial_\infty \H$ orthogonally, being paired by a hyperbolic
element $H_E\in \Gamma$. Let $e\eqdef \pi_\H(E)$ denote the circle
component of the conformal boundary $\partial_\infty^G \Sigma$ which
corresponds to the free side $E$ and let $p\in \hSigma\setminus
\Sigma$ be the corresponding ideal point. Let $I_E$ be the unique
interval of discontinuity of $\Gamma$ which contains $E$. Then $H_E$
is a generator of the hyperbolic cyclic subgroup of $\Gamma$
consisting of all elements of $\PSL(2,\R)$ which fix each of the
endpoints of $I_E$, and we can take $H_E$ to be the positive generator
of this group. The {\em relative funnel neighborhood} of $E$ in
$\fD_\H$ is defined as the intersection:
\be
\fF_E\eqdef \fD_\H\cap \cF_E
\ee
of $\fD_\H$ with the funnel domain $\cF_E\eqdef \cF_{H_E}$ of
$H_E$. It can be shown that no vertex of $\fD_\H$ except the two
endpoints of $E$ meets the Euclidean closure of $\fF_E$. Let
$\ell_E\eqdef \ell(H_E)$ and $R_E\eqdef
R_{H_E}=e^{\frac{\pi^2}{\ell_E}}$ be the displacement length and
displacement radius of $H_E$ (see \eqref{elldef} and
\eqref{Relldef}). The holomorphic covering map $\pi_\H$ projects
$\cF_E$ onto the funnel region $\rF_p$ of $\Sigma$ corresponding to
$p$, which is a holomorphically embedded open annulus, whose inner
boundary corresponds to $e$ (see Figure \ref{fig:funnelcoord}). There
exists a unique element $T_E\in \PSL(2,\R)$ such that $T_E H_E
T_E^{-1}$ is the dilation $\tau \rightarrow e^{\ell_E} \tau$. The
local holomorphic coordinate $z_p$ on $\rF_p$ which identifies the
latter with the open annulus:
\ben
\label{ARE}
A_{R_E}\eqdef \{z\in \C \, \big|\,\frac{1}{R_E}<|z|< 1\}
\een
is called a {\em local funnel coordinate} on $\Sigma$ near $p$ and is
given by\footnote{ We have $T_E(E)=\{\tau'\in \H~|~\Re \tau'=0\}$ and
  $T_E(\cF_E)=\{\tau'\in \H~|~\Re\tau'>0\}$.}:
\ben
\label{funnelcoord}
z_p=e^{\frac{2\pi i}{\ell_E} \log(T_E\tau)}~~(\tau\in \cF_E)~~.
\een
Again notice that $z_p$ depends only on the ideal point $p$. With
respect to this coordinate, $e$ corresponds to the circle of radius
$|z_p|=\frac{1}{R_E}$ and the restriction of the hyperbolic metric $G$
to the funnel region $\rF_p$ of $\Sigma$ corresponds to the
restriction of the hyperbolic metric of $\A(R_E)$ to the annulus
\eqref{ARE}.

\subsection{Ends of elementary hyperbolic surfaces}

\noindent Elementary hyperbolic surfaces can have special types of hyperbolic
ends, known as the {\em plane} and {\em horn} ends. The plane end is
the only end of the Poincar\'e disk $\mD$, while the horn end is one
of the two hyperbolic ends of the hyperbolic punctured disk $\mD^\ast$
(the other being a cusp end). On the other hand, a hyperbolic annulus
$\A(R)$ has two ends, both of which are hyperbolic funnels.  
More details about these ends and some applications of elementary
hyperbolic surfaces to cosmology can be found in \cite{elem,modular}.

\subsection{Explicit form of the hyperbolic metric on a canonical neighborhood of each end}
\label{subsec:endmetric}

\noindent The ends of a geometrically finite hyperbolic surface admit
semi-geodesic coordinate neighborhoods on which the hyperbolic
metric takes a canonical form \cite{Borthwick,BorthwickRev}:

\begin{enumerate}

\item ({\bf cusp ends}) The cusp region $\rC_p$ of $\Sigma$ corresponding to a cusp
  ideal point $p\in \hSigma\setminus \Sigma$ is isometric with
  $\R_{>0}\times \R/(2\pi\Z)$, equipped with the following metric:
\be
\dd s_G^2 = \dd r^{2} + e^{-2 r} \frac{\dd \theta^{2}}{(2\pi)^{2}}~~.
\ee
On this region, $r>0$ and $\theta\in (0,2\pi)$ are semi-geodesic
coordinates. The boundary of this region relative to $\Sigma$ is a
horocycle of length $1$ placed at $r=0$, while the ideal point $p$
corresponds to $r\rightarrow +\infty$.

\item ({\bf funnel ends}) The funnel region $\rF_p$ of $\Sigma$ corresponding to a funnel
  ideal point $p\in \hSigma\setminus \Sigma$ is isometric with
  $\R_{>0}\times \R/(2\pi\Z)$, equipped with the following metric:
\be
\dd s_G^2 = \dd r^{2} + \ell_p^{2} \cosh(r)^{2}
\frac{\dd\theta^{2}}{(2\pi)^{2}}~~,
\ee
where $\ell_p$ is the width of the funnel.  On this region, $r>0$ and
$\theta\in (0,2\pi)$ are semi-geodesic coordinates. The boundary of
this region relative to $\Sigma$ is a closed geodesic of length $\ell_p$ placed
at $r=0$, while the ideal point $p$ (and the associated circle of the
conformal boundary $\partial_\infty^G \Sigma$) correspond to
$r\rightarrow +\infty$.

\item ({\bf horn end}) A canonical punctured neighborhood in\footnote{Recall that the
  hat denotes the end compactification.} $\widehat{\mD^\ast}$ of the
  ideal point corresponding to the horn end of $\mD^\ast$ is isometric
  to $\R_{>0}\times \R/(2\pi\Z)$, equipped with the following metric:
\be
\dd s_G^2 = \dd r^{2} + e^{2 r} \frac{\dd\theta^{2}}{(2\pi)^{2}}~~.
\ee
On such a neighborhood, $r>0$ and $\theta\in (0,2\pi)$ are
semi-geodesic coordinates. The boundary of such a neighborhood
relative to $\mD^\ast$ is a horocycle of length $1$ placed at $r=0$.
The ideal point (and the associated circle of the conformal boundary
$\partial_\infty^G \Sigma$) corresponds to $r\rightarrow +\infty$. The
only hyperbolic surface which admits a horn end is the hyperbolic
punctured disk $\mD^\ast$.

\item ({\bf plane end}) A canonical punctured neighborhood in $\hat{\mD}$ of
  the ideal point corresponding to the plane end of $\mD$ is isometric
  to $\R_{>0}\times \R/(2\pi\Z)$, equipped with the following metric:
\be
\dd s_G^2 = \dd r^{2} + \sinh(r)^{2} \dd\theta^{2}~~.
\ee
On such a neighborhood, $r>0$ and $\theta\in (0,2\pi)$ are
semi-geodesic coordinates. The ideal point (and the associated circle
of the conformal boundary $\partial_\infty^G \Sigma$) corresponds to
$r\rightarrow +\infty$. The only hyperbolic surface which admits a
plane end is the Poincar\'e disk $\mD$.
\end{enumerate}

\

\noindent  The previous list implies the following asymptotic behavior
for $r\gg 1$, near an ideal point $p\in \hSigma\setminus \Sigma$:
\ben
\label{asmetric}
\dd s_G^2=\dd r^2+ \left(\frac{c_p}{4\pi}\right)^2 e^{2\epsilon_p r} \dd \theta^2~~,
\een
where $\theta\in (0,2\pi)$ and\footnote{The sign factor $\epsilon_p$
  should not be confused with the slow-roll parameter $\epsilon$.}:
\beqan
\label{Cdef}
&& c_p=\threepartdefmod{\ell_p~,~}{funnel~end}{2\pi~,~}{plane~end}{2~,~}{horn~or~cusp~end}~~,\\
&& \epsilon_p=\twopartdefmod{+1}{~funnel,~plane~or~horn~end}{-1}{~cusp~end}~~.\nn
\eeqan

\noindent A funnel, plane or horn end will be called a {\em flaring
  end}. Flaring ends are characterized by the fact that the size of
horocycles in $(\Sigma,G)$ grows exponentially towards the end
($\epsilon_p=+1$); such an end corresponds to a circle boundary component
of the conformal compactification of $(\Sigma,G)$. The only
non-flaring ends are cusp ends, each of which corresponds to a point
added when passing to the conformal compactification; for such ends,
the length of horocycles in $(\Sigma,G)$ tends to zero as one approaches
the ideal point.

\end{document}